\documentclass[
    aps,prb,twocolumn,superscriptaddress,floatfix,10pt,amssymb,amsfont,amsmath,
    nofootinbib,float,floatfix
]{revtex4-1}

\usepackage[T1]{fontenc}
\usepackage[utf8]{inputenc}
\usepackage{helvet}
\usepackage{dsfont}
\usepackage[american]{babel}
\usepackage[autostyle=try,english=american]{csquotes}
\usepackage{xfrac}
\usepackage{microtype}
\usepackage{verbatim}
\usepackage{color}
\usepackage{graphicx}
\usepackage[pdftex, plainpages=false]{hyperref}
\usepackage[retainorgcmds]{IEEEtrantools}
\usepackage{todonotes}
\usepackage{array}
\usepackage{tabularx}
\usepackage{booktabs}

\hypersetup{%
	bookmarks=true,         
	unicode=true,           
	pdftoolbar=true,        
	pdfmenubar=true,        
	pdffitwindow=false,     
	pdfstartview={FitH},    
	pdftitle={Topological conformal defects with tensor networks},
	pdfauthor={M. Hauru, W. W. Ho, G. Evenbly, D. Gaiotto, G. Vidal},
	pdfsubject={},          
	pdfcreator={},          
	pdfproducer={pdfLaTeX}, 
    pdfkeywords={ising model} {tensor networks} {conformal} {defects},
	pdfnewwindow=true,      
	colorlinks,
	citecolor=black,
	filecolor=black,
	linkcolor=black,
	urlcolor=black
}

\DeclareMathOperator{\Tr}{Tr}

\newcommand{\bra}[1]{\langle #1|}
\newcommand{\ket}[1]{|#1\rangle}

\newcommand{\unity}{\mathds{1}}

\newcommand{\Naturals}{\mathbb{N}}
\newcommand{\Integers}{\mathbb{Z}}
\renewcommand{\Re}{\operatorname{Re}}
\renewcommand{\Im}{\operatorname{Im}}

\newcommand{\cO}{\mathcal{O}}

\newcommand{\bV}{\mathbb{V}}

\newcommand{\dg}{^{\dagger}}

\newcommand{\znet}{Z}
\newcommand{\trm}{M}
\newcommand{\tro}{T}
\newcommand{\vs}{m}
\newcommand{\iters}{s}
\newcommand{\hs}{n}
\newcommand{\trms}{l}
\newcommand{\cdf}{\gamma}

\begin{document}

\title{Topological conformal defects with tensor networks}

\author{Markus Hauru}
\email{markus@mhauru.org}
\affiliation{Perimeter Institute for Theoretical Physics, Waterloo, Ontario N2L 2Y5, Canada}
\affiliation{Department of Physics and Astronomy, University of Waterloo, Waterloo, Ontario,
Canada, N2L 3G1.}

\author{Glen Evenbly}
\affiliation{Department of Physics and Astronomy, University of California, Irvine, CA 92697-4575
USA}

\author{Wen Wei Ho}
\affiliation{Department of Theoretical Physics, University of Geneva, 24 quai Ernest-Ansermet, 1211
Geneva, Switzerland}

\author{Davide Gaiotto}
\affiliation{Perimeter Institute for Theoretical Physics, Waterloo, Ontario N2L 2Y5, Canada}

\author{Guifre Vidal}
\affiliation{Perimeter Institute for Theoretical Physics, Waterloo, Ontario N2L 2Y5, Canada}

\date{\today}

\begin{abstract}
The critical two-dimensional classical Ising model on the square lattice has two topological
conformal defects: the $\mathbb{Z}_2$ symmetry defect $D_{\epsilon}$ and the Kramers-Wannier
duality defect $D_{\sigma}$.
These two defects implement antiperiodic boundary conditions and a more exotic form of twisted
boundary conditions, respectively.
On the torus, the partition function $Z_{D}$ of the critical Ising model in the presence of a
topological conformal defect $D$ is expressed in terms of the scaling dimensions $\Delta_{\alpha}$
and conformal spins $s_{\alpha}$ of a distinct set of primary fields (and their descendants, or
conformal towers) of the Ising conformal field theory.
This characteristic conformal data $\{\Delta_{\alpha}, s_{\alpha}\}_{D}$ can be extracted from
the eigenvalue spectrum of a transfer matrix $M_{D}$ for the partition function $Z_D$.
In this paper, we investigate the use of tensor network techniques to both represent and
coarse-grain the partition functions $Z_{D_\epsilon}$ and $Z_{D_\sigma}$ of the critical Ising
model with either a symmetry defect $D_{\epsilon}$ or a duality defect $D_{\sigma}$.
We also explain how to coarse-grain the corresponding transfer matrices $M_{D_\epsilon}$ and
$M_{D_\sigma}$, from which we can extract accurate numerical estimates of $\{\Delta_{\alpha},
s_{\alpha}\}_{D_{\epsilon}}$ and $\{\Delta_{\alpha}, s_{\alpha}\}_{D_{\sigma}}$.
Two key new ingredients of our approach are (i) coarse-graining of the defect $D$, which applies to
any (i.e.\ not just topological) conformal defect and yields a set of associated scaling dimensions
$\Delta_{\alpha}$, and (ii) construction and coarse-graining of a generalized translation operator
using a local unitary transformation that moves the defect, which only exist for topological
conformal defects and yields the corresponding conformal spins $s_{\alpha}$.
\end{abstract}

\maketitle

\section{Introduction}
\label{sec:introduction}

\raggedbottom  

A conformal defect is a universality class of critical behavior at the junction of two critical
systems.
Relevant examples include point impurities, interfaces and boundary phenomena in critical 1D
quantum systems, as well as line defects, interfaces and boundaries in critical 2D classical
systems~\cite{oshikawa_boundary_1997,difrancesco_conformal_1997,henkel_conformal_1999,affleck_quantum_2008,quella_reflection_2007}.
A \emph{topological} conformal defect in a conformal field theory (CFT) is a particular type of
conformal defect that is totally transmissive~\cite{quella_reflection_2007} and can be deformed
without affecting the value of correlators as long as it is not taken across a field insertion.
It can also be regarded as defining a form of twisted boundary conditions for that CFT\@.

The goal of this paper is to investigate the use of tensor network techniques to describe
topological conformal defects in microscopic lattice models.
For simplicity, we analyze the two-dimensional critical Ising model, working mostly with the 2D
classical statistical spin system, but also repeatedly connecting to the (1+1)D quantum spin chain.
The critical Ising model turns out to have two non-trivial topological conformal defects: a
symmetry defect $D_{\epsilon}$ and a duality defect $D_{\sigma}$~\cite{petkova_generalised_2001}.
The symmetry defect $D_{\epsilon}$ relates to the global $\mathbb{Z}_2$ spin-flip symmetry of the
Ising model and implements antiperiodic boundary conditions, whereas the duality defect
$D_{\sigma}$ relates to the Kramers-Wannier self-duality of the critical Ising model and can be
thought of as implementing some form of twisted boundary conditions.

\subsection{Defects and transfer matrices}
\begin{figure}[htbp]
    \centering
    \includegraphics[width=1.0\linewidth]{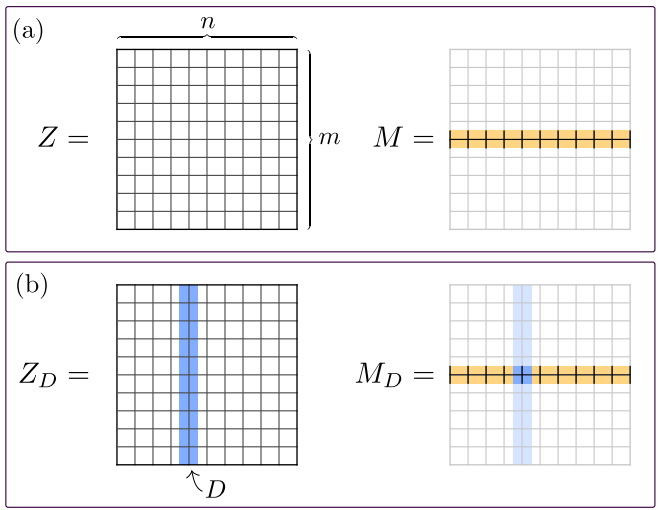}
    \caption{%
       (a) Partition function $Z$ on a square lattice made of $n \times m$ sites with periodic
       boundary conditions in both directions (a torus), and the corresponding transfer matrix $M$,
       such that $Z = \Tr \left( M^{m} \right)$.
       (b) Partition function $Z_D$ on the same torus, where the defect $D$ implements some form of
       boundary conditions, and the generalized transfer matrix $M_D$, such that $Z_D = \Tr \left(
       (Z_D)^m \right)$.
    }\label{fig:Z_and_M}
\end{figure}

Consider the statistical partition function $Z$ of the critical 2D classical Ising model on a
square lattice with periodic boundary conditions (that is, on a torus), made of $m\times n$ sites.
By the operator-state correspondence of CFT~\cite{difrancesco_conformal_1997}, this partition
function is expressed in terms of the scaling dimensions $\Delta_{\alpha}$ and conformal spins
$s_{\alpha}$ of some specific set of scaling operators $\phi_{\alpha}$, namely, those in the
conformal towers of the three local primary fields of the Ising CFT: the identity,
energy-density and spin primaries.
We can extract $\{\Delta_{\alpha},s_{\alpha}\}$ from the spectrum of eigenvalues of a transfer
matrix $M$ for the partition function $Z$, fulfilling $Z = \Tr \left( M^m\right)$ [see
Fig.~\ref{fig:Z_and_M}(a)].

\flushbottom  

In the presence of a topological conformal defect $D$, the modified partition function $Z_{D}$ is
expressed now in terms of the scaling dimensions and conformal spins $\{\Delta_{\alpha},
s_{\alpha}\}_{D}$ of other conformal towers, corresponding to a different set of primary fields.
We can again extract $\{\Delta_{\alpha},s_{\alpha}\}_{D}$ from the spectrum of eigenvalues of a
modified transfer matrix $M_D$ for the partition function $Z_D$, fulfilling $Z_D = \Tr \left(
(M_D)^m\right)$ [see Fig.~\ref{fig:Z_and_M}(b)].

In this paper, we build tensor network representations of the transfer matrices $M_{D_{\epsilon}}$
and $M_{D_{\sigma}}$ for the two topological conformal defects $D_{\epsilon}$ and $D_{\sigma}$ of
the critical Ising model and explain how to extract accurate estimates of the corresponding scaling
dimensions and conformal spins $\{\Delta_{\alpha},s_{\alpha}\}_{D_{\epsilon}}$ and
$\{\Delta_{\alpha},s_{\alpha}\}_{D_{\sigma}}$, which we regard as a characterization of these
defects.
These estimates are obtained by first coarse-graining and then diagonalizing the transfer matrices
$M_{D_{\epsilon}}$ and $M_{D_{\sigma}}$.

We emphasize that on a sufficiently small $m\times n$ torus, say in the range $n \sim 10-20$, one
can already diagonalize the transfer matrix $M_{D}$ for defect $D$, and thus obtain numerical
estimates of $\{\Delta_{\alpha}, s_{\alpha}\}_D$, by using exact diagonalization techniques.
Why do we then need to use tensor networks?
As we will review, these estimates for $\{\Delta_{\alpha}, s_{\alpha}\}_D$ are affected by
non-universal, finite-size corrections, which diminish with growing $n$.
The merit of tensor network techniques is then merely that, through proper coarse-graining, they
allow us to consider a much larger $n$ than exact diagonalization techniques, thus producing more
accurate numerical estimates.
However, the coarse-graining of the tensor network introduces \emph{truncation errors}, which must
be kept under check and effectively limit the size $n$ that can be reliably considered.

In the absence of defects, the use of tensor networks to obtain more accurate estimates of the
conformal data $\{\Delta_{\alpha}, s_{\alpha}\}$ by diagonalizing a transfer matrix $M$ for the
partition function $Z$ was proposed and demonstrated in
Ref.~\onlinecite{gu_tensorentanglementfiltering_2009}.
In this paper, we generalize the proposal of Ref.~\onlinecite{gu_tensorentanglementfiltering_2009}
to the presence of defects, focusing on topological conformal defects for concreteness.
This requires several new steps, which we list here.
The first two steps, which apply to generic line defects on a 2D classical partition function
(equivalent to a point defect in a 1D quantum model), are: (i) encoding of the defect $D$ as a 1D
tensor network, which upon insertion in the 2D tensor network for the clean partition function $Z$
produces a tensor network for the defect partition function $Z_D$ as well as a tensor network for
the corresponding defect transfer matrix $M_D$; (ii) coarse-graining of the partition function
$Z_D$ / transfer matrix $M_{D}$.
Diagonalization of the coarse-grained transfer matrix $M_{D}$ will yield the scaling dimensions
$\Delta_{\alpha}$ associated to defect $D$.
Steps~(i) and~(ii) apply to a generic type of conformal defect.
Topological conformal defects are special in that we can extract additional conformal data, namely,
the conformal spins $s_{\alpha}$, by following two additional new steps: (iii) identification of a
local unitary transformation that moves the location of the topological defect, so as to be able to
define a generalized translation operator $T_{D}$ that commutes with the transfer matrix $M_D$;
(iv) coarse-graining of the generalized translation operator $T_D$ and diagonalization of the
product $T_D M_D$ of coarse-grained versions of the translation operator $T_D$ and the transfer
matrix $M_D$.

The main result of this paper is the proposal of the steps~(i) and~(iv) described above to extract
accurate estimates of the $\{\Delta_{\alpha}, s_{\alpha}\}_D$ associated to a topological conformal
defect, together with a thorough demonstration of the approach for the symmetry defect
$D_{\epsilon}$ and duality defect $D_{\sigma}$ of the critical Ising model.
We also propose steps (i)-(ii) to extract accurate estimates of the $\Delta_{\alpha}$ associated to
a generic (i.e.\ non-topological) conformal defect $D$, which we demonstrate with specific
examples.

\subsection{Structure of this paper}
Sections~\ref{sec:ising_model}--\ref{sec:topological_defects} are mostly devoted to discussing
background material, whereas Secs.~\ref{sec:D_epsilon}-\ref{sec:discussion} contain our main
results.

In Sec.~\ref{sec:ising_model}, we review the Ising model in the absence of a defect.
This includes the 2D classical Ising model on the square lattice, the 1D quantum Ising model, and
the Ising CFT that effectively describes the previous two lattice models at criticality.
We introduce the partition function $Z$ and transfer matrix $M$, and relate the eigenvalue spectrum
of $M$ to the conformal data $\{\Delta_{\alpha}, s_{\alpha}\}$ of the three local primary fields of
the Ising model.
In Sec.~\ref{sec:tensor_networks}, we review the use of tensor networks to represent the
partition function $Z$ and transfer matrix $M$, and of coarse-graining algorithms for tensor
networks, which allow us to obtain estimates of $\{\Delta_{\alpha}, s_{\alpha}\}$ with smaller
finite-size errors than is possible with exact diagonalization.
In Sec.~\ref{sec:topological_defects}, we review the properties of the topological defects
$D_{\epsilon}$ and $D_{\sigma}$ of the critical Ising model, including their associated field
content and their fusion rules.

Sections~\ref{sec:D_epsilon} and~\ref{sec:D_sigma} analyze the symmetry defect $D_{\epsilon}$ and
the duality defect $D_{\sigma}$, respectively.
On the lattice, the topological character of a defect appears to be related to the existence of a
local unitary transformation that changes the location of the defect.
We identify such local unitary transformations for $D_{\epsilon}$ and $D_{\sigma}$, which also
allow us to fuse these defects and confirm the expected fusion rules.
We express the partition functions $Z_{D_\epsilon}$ and $Z_{D_\sigma}$ of the critical Ising model
in terms of transfer matrices $M_{D_\epsilon}$ and $M_{D_\sigma}$, and propose generalized
translation operators $T_{D_\epsilon}$ and $T_{D_\sigma}$, which commute with $M_{D_\epsilon}$ and
$M_{D_\sigma}$, respectively.
Working with a tensor network representation of all the above objects, we then consider a
coarse-graining transformation for the products $M_{D_{\epsilon}}T_{D_{\epsilon}}$ and
$M_{D_{\sigma}}T_{D_{\sigma}}$, whose eigenvalue spectra yield the conformal data
$\{\Delta_\alpha, s_{\alpha}\}_{D_{\epsilon}}$ and $\{\Delta_\alpha, s_{\alpha}\}_{D_{\sigma}}$.

It turns out that, from a tensor network perspective, the two topological conformal defects of the
Ising model are very different.
Indeed, as we will see in Sec.~\ref{sec:D_epsilon}, the symmetry defect $D_{\epsilon}$ can be
readily incorporated into a tensor network by employing $\mathbb{Z}_2$-symmetric
tensors~\cite{singh_tensor_2011}, in which case $D_{\epsilon}$ is represented by a simple unitary
matrix $V$ acting on a bond index.
This implies, in particular, that the defect transfer matrix $M_{D_\epsilon}$ and translation
operator $T_{D_\epsilon}$ are obtained from the clean transfer matrix $M$ and translation operator
$T$ by insertions of $V$.
This also applies to the coarse-grained version of $M_{D_\epsilon}$ and $T_{D_\epsilon}$, and we
conclude that one can extract the new set of $\{\Delta_{\alpha}, s_{\alpha}\}_{D_{\epsilon}}$ with
remarkably little effort by recycling the coarse-grained tensor networks used for the critical
Ising model without a defect. On the other hand, we will see in Sec.~\ref{sec:D_sigma} that
representing the duality defect $D_{\sigma}$ requires using different tensors altogether (as does a
generic conformal defect), which one needs to explicitly coarse-grain.

In Sec.~\ref{sec:generic} we then briefly discuss the case of a generic (i.e.\ non-topological)
conformal defect $D$, for which we can also extract the scaling dimensions $\{\Delta_{\alpha}\}_D$
by diagonalizing a transfer matrix $M_D$ for the defect partition function $Z_D$, and demonstrate
the performance of the approach for a known continuous family of conformal defects of the critical
Ising model.

Section~\ref{sec:discussion} concludes the paper with a discussion of the present tensor network
approach, including its extension to defects in other critical lattice models, a discussion of
different coarse-graining transformations one can use, and a comparison to other tensor network
approaches (based on explicitly realizing scale invariance of the tensor network under
coarse-graining) that can also be used to extract conformal data.
Finally, several appendices contain technical discussions, as well as a study of the $\Integers_3$
symmetry defects of the three-level Potts model.

\subsection{Source code}
The numerical results we present were obtained using a Python 3 implementation of the algorithms we
describe.
The source code is available at \url{https://arxiv.org/src/1512.03846/anc}, licensed under the MIT
License, a permissive free software license.
It can be used to reproduce our results and as the ultimate reference on details of the
algorithms we describe here.

\section{Critical Ising model}
\label{sec:ising_model}
In this section, we review the critical Ising model on the lattice, both the 2D classical partition
function and the 1D quantum spin chain.
We also review their continuum limit, the Ising conformal field theory.
Universal properties of the phase transition (conformal data) can be numerically estimated using
exact diagonalization.
The accuracy of these numerical estimates is limited by non-universal, finite-size corrections.

\subsection{Classical partition function}
The classical Ising model is defined by its Hamiltonian $K = -\sum_{\langle i,j \rangle} \sigma_i
\sigma_j$, where $i$ and $j$ label sites on the lattice and $\sigma_i$ is a classical spin variable
on site $i$ that can take the values $\pm 1$.
$\sum_{\langle i,j \rangle}$ is a sum over nearest-neighbor pairs of sites.
The partition function at inverse temperature $\beta$ is
\begin{IEEEeqnarray}{c}
    \label{eq:Z_definition}
    Z =  \sum_{\{\sigma\}} e^{-\beta K}
    = \sum_{\{\sigma\}} e^{\beta \sum_{\langle i,j \rangle} \sigma_i \sigma_j},
\end{IEEEeqnarray}
where $\sum_{\{\sigma\}}$ is a sum over all the spin configurations.
Here, we consider that the spins inhabit the sites of a square lattice with periodic boundaries in
both directions (a torus).
The Ising Hamiltonian $K$ is invariant under a simultaneous flip of all the spins so that if we map
$\sigma_i \mapsto - \sigma_i$ for all sites $i$ the Hamiltonian remains unchanged: $K \mapsto
K$.
Flipping a spin twice recovers the original configuration, so the spin-flip symmetry is a global,
internal $\Integers_2$ symmetry.  This symmetry is preserved at high temperatures (low $\beta$) and
spontaneously broken at low temperatures (large $\beta$).
At $\beta = \frac{1}{2}\log(1+\sqrt{2})$ there is a second-order phase transition that separates
the low temperature symmetry-breaking, ordered phase from the high temperature disordered phase.

The Ising model on the square lattice also has a order-disorder duality called the Kramers-Wannier
duality which states that a low temperature Ising model is equivalent to a high temperature model
on the dual lattice.
At the critical point, the model is self-dual under this duality map.~\cite{baxter_exactly_1982}

On a torus made of $\vs \times \hs$ spins we can write down a transfer matrix $\trm$ such that $Z =
\Tr(\trm^\vs)$.
The transfer matrix $\trm$ (which is made of $2^\hs \times 2^\hs$ entries) is associated to a row
of $\hs$ spins.
Many questions about the model can be answered in terms of the eigenvalue spectrum of $\trm$.
We omit here the explicit form of $\trm$, which we will later build using tensor networks.

\subsection{Quantum spin chain}
The one-dimensional quantum Ising model is defined by the Hamiltonian
\begin{IEEEeqnarray}{c}
    H(h) = -\left(\sum_{i=1}^{\hs}
    \sigma_i^z \sigma_{i+1}^z + h \sum_{i=1}^{\hs} \sigma_i^x \right).
\end{IEEEeqnarray}
On every site $i$ there is a quantum spin and $\sigma^z$ and $\sigma^x$ are Pauli matrices.
The parameter $h$ is the strength of a transverse field.

The Hamiltonian is symmetric under a global spin-flip or, in other words,
\begin{IEEEeqnarray}{c}
    \Sigma^x H \left( \Sigma^x \right)\dg = H,
\end{IEEEeqnarray}
where $\Sigma^x = \bigotimes_{i=1}^\hs \sigma_i^x$, and there are again two phases:
a symmetry-breaking phase for small transverse magnetic field, and a disordered phase for large
magnetic field, with a continuous quantum phase transition at $h=1$.

The Kramers-Wannier duality for the quantum model equates it with another spin chain living on the
dual lattice with a similar Hamiltonian but with external field~\cite{grimm_spin12_1993}
$\tilde{h} = \frac{1}{h}$.
As in the classical model, the spin chain is Kramers-Wannier self-dual at the critical point.

The partition function of the quantum model for inverse temperature $\beta_Q$ is $Z_Q = \Tr\left(
e^{-\beta_Q H} \right)$.
A transfer matrix for this partition function can be written as $\trm_Q \equiv
e^{-\frac{\beta_Q}{\vs} H}$, so that $Z_Q = \Tr\left( (\trm_Q)^\vs \right)$.
Through the standard classical-quantum mapping, $\trm_Q$ also corresponds to a transfer matrix for
the partition function of a classical two-dimensional Ising model~\cite{sachdev_quantum_2011-1}.
However, this classical dual is not the isotropic Ising model that we discussed above, but an
extremely anisotropic one with very different couplings in different directions.
The anisotropic and isotropic classical models are, nevertheless, in the same universality class,
and hence the universal properties of the quantum and the classical Ising models are the same.
To extract these universal properties we will mostly concentrate on studying the classical model.

\subsection{Ising CFT}
As we have mentioned above, the classical square lattice Ising model has a critical point at $\beta
= \frac{1}{2}\log(1+\sqrt{2})$ and the Ising spin chain has a quantum critical point at $h=1$.
The continuum limits of both of these critical points are described by the Ising CFT\@.
The universal properties of the phase transition are captured by the conformal data of this CFT\@.

Consider the torus formed by parameterizing the two coordinates $(x,y)$ of the plane by a complex
variable $w \equiv x + i y$ and by identifying the points $w$, $w+2\pi$ and $w + 2\pi\tau$, for
$\tau = \tau_1 + i \tau_2$ a complex \emph{modular parameter} that defines the shape of the torus.
A purely imaginary modular parameter $\tau= i \tau_2$ produces a torus consisting of a rectangle
with periodic boundaries.

On a torus defined by a complex modular parameter $\tau$, the partition function of a CFT is
\begin{IEEEeqnarray}{rCl}
    \label{eq:Z_CFT1}
    \nonumber
        Z_{\mathrm{CFT}} &=& \Tr\left( e^{-2\pi \tau_2 (L_0 + \overline{L}_0 - \frac{c}{12})}
        e^{2\pi i \tau_1 (L_0 - \overline{L}_0)} \right) \\
        &=& \Tr\left( e^{-2\pi \tau_2 H_{\mathrm{CFT}}} e^{2\pi i \tau_1 P}  \right).
\end{IEEEeqnarray}
Here $L_0$ and $\overline{L}_0$ are the Virasoro generators and $H_{\mathrm{CFT}} = L_0 +
\overline{L}_0 - \frac{c}{12}$ and $P = L_0 - \overline{L}_0$ are the Hamiltonian and momentum
operators that generate translations in the directions $\Im(w)$ and $\Re(w)$, respectively, while
$c$ is the central charge.~\cite{ginsparg_applied_1988}

The scaling operators $\phi_\alpha$ of the CFT are eigenoperators of dilations on an infinite
plane.
The operator-state correspondence identifies them with states $\ket{\phi_\alpha}$ that are the
eigenstates of $L_0$ and $\overline{L}_0$:
$L_0 \ket{\phi_\alpha} = h_\alpha \ket{\phi_\alpha}$ and $\overline{L}_0 \ket{\phi_\alpha} =
\overline{h}_\alpha \ket{\phi_\alpha}$.
$h_\alpha$ and $\overline{h}_\alpha$ are known as the holomorphic and antiholomorphic conformal
dimensions of $\phi_\alpha$.
In terms of the eigenvalues of $L_0$ and $\overline{L}_0$ we can rewrite the partition function as
\begin{IEEEeqnarray}{rCl}
    \label{eq:Z_CFT2}
    Z_{\mathrm{CFT}} &=& \sum_{\alpha} e^{-2\pi\tau_2(h_{\alpha} + \bar{h}_{\alpha} - \frac{c}{12})
    + 2\pi i \tau_1 (h_{\alpha} - \bar{h}_{\alpha})}, \\
    &=&\sum_{\alpha} e^{-2\pi\tau_2(\Delta_\alpha - \frac{c}{12}) + 2\pi i \tau_1 s_\alpha},
\end{IEEEeqnarray}
where $\Delta_\alpha = h_\alpha + \overline{h}_\alpha$ and $s_\alpha = h_\alpha -
\overline{h}_\alpha$ are known as the scaling dimension and conformal spin of $\phi_\alpha$,
respectively.~\cite{ginsparg_applied_1988}
The scaling operators come in conformal towers built up from the primary operators.
If $h_p$ and $\overline{h}_p$ are the conformal dimensions of a primary operator, then the scaling
operators in its conformal tower have conformal dimensions of the form $h = h_p + k$ and
$\overline{h} = \overline{h}_p + l$, where $k,l \in \Naturals$.~\cite{ginsparg_applied_1988}

The Ising CFT is a conformal field theory of central charge $c=\frac{1}{2}$.
For a unitary $c=\frac{1}{2}$ CFT the conformal dimensions $h$ and $\bar{h}$ of the primaries can
take the values $0$, $\frac{1}{2}$ and $\frac{1}{16}$.~\cite{difrancesco_conformal_1997}
However, not all the possible combinations of these values of $h$ and $\bar{h}$ are realized as
local primary operators in the CFT\@.
The Ising CFT only includes the three ``diagonal'' primary operators that have $h = \overline{h}$.
They are called the identity $\unity$ for $(0,0)$, the energy density $\epsilon$ for $(\frac{1}{2},
\frac{1}{2})$ and the spin $\sigma$ for $(\frac{1}{16}, \frac{1}{16})$.
Because of the $\Integers_2$ symmetry of the model, the conformal towers come with a parity (a
$\Integers_2$ charge).
This parity is $+1$ for $\unity$ and $\epsilon$, and $-1$ for $\sigma$.

We will see later in Sec.~\ref{sec:topological_defects} that the non-diagonal combinations of
$h$ and $\overline{h}$ are relevant to the discussion of topological conformal defects of the
Ising CFT.

\subsection{Extracting the universal data}
For a critical, classical lattice model that has a CFT as its continuum limit, the partition
function on an $\vs \times \hs$ torus (corresponding to $\tau = i\frac{\vs}{\hs}$) can be written
as~\cite{cardy_operator_1986}
\begin{IEEEeqnarray}{c}
    \label{eq:Z_spectrum}
    Z = \sum_\alpha e^{2 \pi \frac{\vs}{\hs} \left( \frac{c}{12} - \Delta_\alpha \right) + \vs\hs f
    + \cO\left( \frac{\vs}{\hs^\gamma} \right)}, \quad \gamma > 1.
\end{IEEEeqnarray}
The sum is again over scaling operators.
Equation~\eqref{eq:Z_spectrum} only differs from Eq.~\eqref{eq:Z_CFT2} in two non-universal terms.
One non-universal term is the free energy term $\vs\hs f$, where $f$ is the free energy per
site at the thermodynamic limit.
The second is the subleading finite-size corrections $\cO\left( \frac{\vs}{\hs^\gamma} \right)$,
which become negligible for a large torus.
The transfer matrix $\trm$ corresponding to a row of $\hs$ sites of the lattice so that $Z =
\Tr\left( \trm^{\vs}\right)$ then has eigenvalues
\cite{cardy_operator_1986}
\begin{IEEEeqnarray}{c}
    \label{eq:cardys_formula0}
    \lambda_\alpha = e^{ 2 \pi\frac{1}{\hs} \left( \frac{c}{12} - \Delta_\alpha \right)
    + \hs f + \cO\left( \frac{1}{\hs^\gamma} \right)}.
\end{IEEEeqnarray}

\begin{figure}[htbp]
    \centering
    \includegraphics[width=1.0\linewidth]{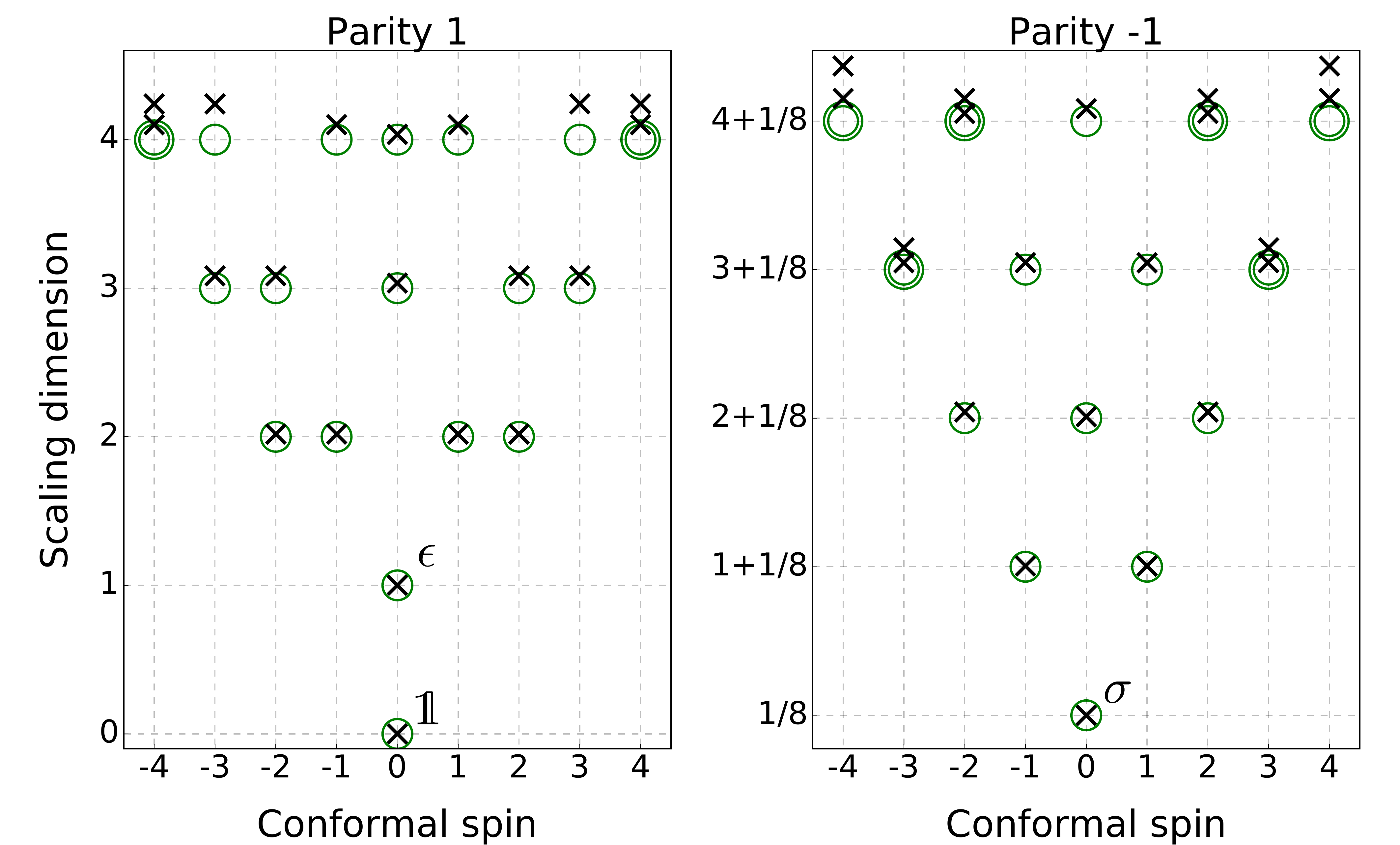}
    \caption{%
        The scaling dimensions (vertical axis) and conformal spins (horizontal axis) of the first
        scaling operators of the Ising CFT obtained from exact diagonalization of a transfer matrix
        of $\hs=18$ sites.
        The scaling operators are divided by their parity, i.e.\ their eigenvalue under the
        $\Integers_2$ symmetry operator $\Sigma^x$ that commutes with the transfer matrix.
        The crosses mark the numerical values that can be compared with the circles that are
        centered at the exact values.
        Several concentric circles denote the degeneracy $N_\alpha$ of that $(\Delta_\alpha,
        s_\alpha)$ pair.
        The primary fields \emph{identity}~$\mathbb{I}$, \emph{spin}~$\sigma$ and \emph{energy
        density}~$\epsilon$ appear at the basis of their three conformal towers.
    }\label{fig:ed_ising_results}
\end{figure}

More generally, if the transfer matrix $\trm$ corresponds instead to $\trms$ rows of $\hs$ sites,
so that now $Z = \Tr\left( \trm^{\frac{\vs}{\trms}}\right)$, then its eigenvalues
are~\cite{cardy_operator_1986}
\begin{IEEEeqnarray}{c}
    \label{eq:cardys_formula}
    \lambda_\alpha = e^{2 \pi \frac{\trms}{\hs} \left( \frac{c}{12} - \Delta_\alpha \right)
    + \trms \hs f + \cO\left( \frac{\trms}{\hs^\gamma} \right)}.
\end{IEEEeqnarray}
Thus, if we manage to diagonalize a transfer matrix for large $\hs$, the subleading, non-universal
corrections will become negligible, whereas we can extract $f$ by varying $\trms$ and $\hs$ while
keeping $\frac{\trms}{\hs}$ fixed.
We can then rescale $\trm$ by $e^{\trms \hs f}$ (or equivalently we can rescale the Boltzmann
weights $e^{\beta \sigma_i \sigma_j}$ in the partition function $Z$) to get rid of the free energy
term in the eigenvalue spectrum.
From now on we will always assume the transfer matrix $\trm$ has been rescaled in this way, so that
its spectrum is
\begin{IEEEeqnarray}{c}
    \label{eq:cardys_formula_no_f}
    \lambda_\alpha \approx e^{2 \pi \frac{\trms}{\hs} \left( \frac{c}{12} - \Delta_\alpha \right)}.
\end{IEEEeqnarray}
The equality is approximate because we have left out the subleading, non-universal, finite-size
terms.
Each of the eigenvalues $\lambda_\alpha$ then tells us the value of $\frac{c}{12} - \Delta_\alpha$
for one of the scaling operators $\phi_\alpha$.
The scaling dimension $\Delta_\unity$ of the identity operator is always $0$ and for a unitary CFT
this is the smallest scaling dimension possible.
Thus we can obtain $c$ from the largest eigenvalue $\lambda_0$ and the rest of the
$\lambda_\alpha$'s give us the rest of the scaling dimensions.\footnote{%
    In a finite system $\trm$ only has a finite number of eigenvalues whereas there is an infinite
    number of scaling operators.
    However, we do observe that at least the largest $\lambda_\alpha$'s correspond to the scaling
    operators with smallest $\Delta_\alpha$, see Fig.~\ref{fig:ed_ising_results}.
}

For a critical quantum spin chain of $\hs$ spins the spectrum of the Hamiltonian
is~\cite{cardy_operator_1986}
\begin{IEEEeqnarray}{c}
    \label{eq:H_spectrum}
    E_\alpha = a+ b\left[ 2 \pi \frac{1}{\hs} \left( \frac{c}{12} - \Delta_\alpha \right)
    + \hs f + \cO\left( \frac{1}{\hs^\gamma} \right) \right]
\end{IEEEeqnarray}
where $a$ and $b$ are non-universal constants and $\gamma > 1$ as before.
We could thus extract all the same critical data by diagonalizing the quantum Hamiltonian $H$
instead of the transfer matrix $M$ of the classical partition function $Z$.
In this work we choose, however, to work mostly with the transfer matrix.

For a translationally invariant lattice model, such as the Ising model, the transfer matrix $\trm$
commutes with the translation operator $\tro = e^{\frac{2 \pi i}{\hs} P}$ that implements a
discrete translation by one lattice site.
Each eigenstate of $\tro$, with eigenvalue $e^{\frac{2 \pi i}{\hs} p_\alpha}$, has well-defined
momentum $p_\alpha$.
As discussed above, the momentum operator is $P = L_0 - \overline{L}_0$ and thus the momentum
$p_\alpha$ corresponds to a conformal spin $s_\alpha = h - \overline{h}$.
Hence we can diagonalize $\tro$ and $\trm$ simultaneously to obtain both the scaling dimensions
$\Delta_\alpha$ and the conformal spins $s_\alpha$ for the scaling operators $\phi_\alpha$.
In fact, we can get away with even less work by diagonalizing only the product $\tro \cdot \trm$,
which corresponds to the transfer matrix on a torus with a modular parameter $\tau$ that has real
part $\tau_1 = \Re(\tau) = 1/n$ and imaginary part $\tau_2 = \Im(\tau) = l/n$.
The eigenvalues of $\tro \cdot \trm$ are the products of the eigenvalues of $\tro$ and $\trm$,
\begin{IEEEeqnarray}{c}
    \label{eq:cardys_formula_momentum_no_f}
    \tilde{\lambda}_\alpha = \lambda_\alpha \cdot e^{\frac{2 \pi i}{\hs} p_\alpha}
    \approx e^{2 \pi\frac{\trms}{\hs} \left( \frac{c}{12} - \Delta_\alpha \right)
    + \frac{2 \pi i}{\hs} s_\alpha
    },
\end{IEEEeqnarray}
where we have again scaled away the free energy contribution and ignored the subleading
finite-size corrections.
The real part $\Re(\log \tilde{\lambda}_\alpha) = 2 \pi\frac{\trms}{\hs} \left( \frac{c}{12} -
\Delta_\alpha \right)$ then yields the scaling dimension $\Delta_\alpha$ and the imaginary part
$\Im(\log \tilde{\lambda}_\alpha) = \frac{2 \pi i}{\hs} s_\alpha$ the conformal spin $s_\alpha$.
Note that because of the periodicity of $e^{\frac{2 \pi i}{\hs} s_\alpha}$, the spin can only be
determined modulo $\hs$, a point we will come back to later.

Thus to obtain scaling dimensions and conformal spins $\{\Delta_{\alpha},s_{\alpha}\}$ of the CFT
numerically we can construct $\tro \cdot \trm$ for a finite but sufficiently large system and
diagonalize it using an exact diagonalization algorithm.
Results obtained in this manner are shown in Fig.~\ref{fig:ed_ising_results}.
They clearly show the structure of the conformal towers coming out correctly and the accuracy of
the estimates of $\Delta_{\alpha}$ for the operators with lowest scaling dimensions is remarkably
good.
However, for operators with higher scaling dimensions the numerical estimates start to deteriorate
significantly.
This is due to the subleading finite-size corrections, which are still large at the system size
$\hs=18$ that we used here.

Unfortunately if we try to push for larger systems the computations quickly become prohibitively
expensive because the dimension of the transfer matrix grows as $2^\hs$ and the cost of exact
diagonalization grows as the third power of this dimension.
To diminish the effect of these non-universal corrections, we can describe the system using tensor
networks and apply tensor network coarse-graining algorithms to reach large system sizes, as
we review in the next section.

\section{Tensor networks}
\label{sec:tensor_networks}
In this section, we first review how tensor networks can be used to express the partition function
$Z$ and its transfer matrix $M$.
We then describe how a coarse-graining algorithm for tensor networks can be used to analyze larger
systems than with exact diagonalization, as first proposed and demonstrated in
Ref.~\onlinecite{gu_tensorentanglementfiltering_2009}.
By diagonalizing a transfer matrix $M$ corresponding to a large number $n$ of sites we can reduce
very significantly the errors, due to finite-size corrections, in the estimates of scaling
dimensions and conformal spins $\{\Delta_{\alpha}, s_{\alpha}\}$.
This comes at the price of introducing truncation errors.

\subsection{Tensor network representation}

\begin{figure*}[htbp]
	\centering
	\includegraphics[width=1.0\linewidth]{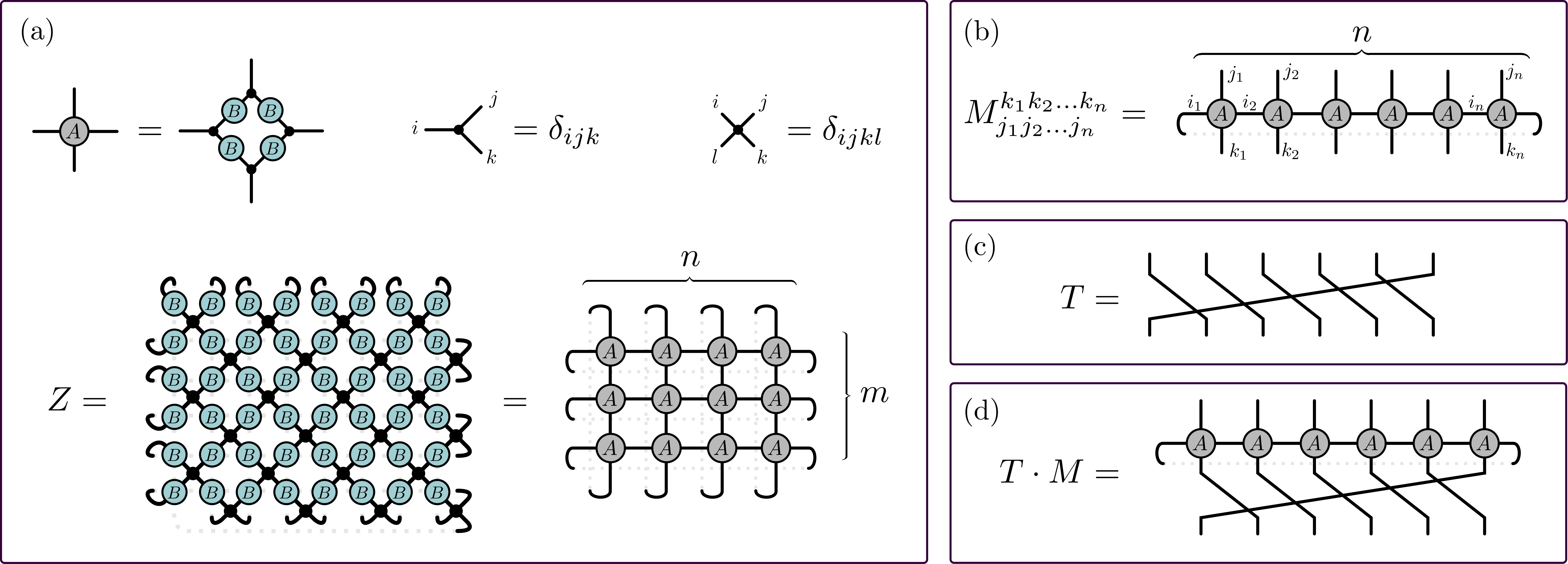}
	\caption{%
        In this figure and throughout the paper we use the usual graphical tensor network language
        where tensors are represented by various shapes and their indices by legs coming out from
        them.
        A leg connecting two tensors is summed over.
        (a)
        The partition function $Z$ of a classical two dimensional lattice model on a torus as a
        tensor network, first using the Boltzmann weights $B_{ij} = e^{- \beta E_{ij}}$ and then in
        terms of the tensor $A_{ijkl} = B_{ij} B_{jk} B_{kl} B_{li}$.
        We call the network on the right $\znet_{\hs,\vs}(A)$.
        $\delta_{ijk}$ and $\delta_{ijkl}$ are three- and four-way Kronecker deltas that fix all
        their indices to have the same value.
        (b)
        The transfer matrix $\trm$ as a tensor network.
        (c)
        The one-site translation operator $\tro$.
        (d)
        The translation operator composed with the transfer matrix.
    }\label{fig:Z_and_T_collection}
\end{figure*}

In terms of the Boltzmann weights $B_{ij} = e^{\beta \sigma_i \sigma_j}$ the partition function is
\begin{IEEEeqnarray}{c}
    \label{eq:Z_definition_B}
    Z = \sum_{\{\sigma\}} \prod_{\langle i,j \rangle} B_{ij},
\end{IEEEeqnarray}
with the sum and the product being over all spin configurations and all nearest-neighbor pairs,
respectively.
Periodic boundary conditions in both directions are again assumed.
Such a partition function $Z$ can be written as a tensor network in several ways.
We will be using the network shown in Fig.~\ref{fig:Z_and_T_collection}(a) on the right.
The first network in Fig.~\ref{fig:Z_and_T_collection}(a) is a straight-forward translation of
Eq.~\eqref{eq:Z_definition_B} into the graphical tensor network notation.
In it for every spin there is a four-index Kronecker delta $\delta_{ijkl}$.
Each of the four indices connects to one of the neighboring spins through the matrix $B$.
On the right this is then rewritten in terms of tensor $A_{ijkl} = B_{ij} B_{jk} B_{kl} B_{li}$
that encodes the interactions around a plaquette of spins.
Every index of $A$ corresponds to one spin.
We will be working with this latter network, which we denote $\znet_{\hs,\vs}(A)$.
Notice that here $m$ and $n$ label the number of rows and columns of tensors $A$, not of spins,
with each tensor accounting for two spins.
However, all the expressions in the preceding section can be seen to remain valid due to the
isotropy of the original spin model.
We also note that, alternatively, one can build a single tensor $A$ for each spin following the
construction in Ref.~\onlinecite{orus_itebd_2008}.

From Fig.~\ref{fig:Z_and_T_collection}(a) it is clear that we can write $Z =
\Tr\left(\trm^\vs\right)$ where the transfer matrix is as in Fig.~\ref{fig:Z_and_T_collection}(b),
or in other words
\begin{IEEEeqnarray}{c}
    \trm_{j_1 j_2 \dots j_\hs}^{k_1 k_2 \dots k_\hs}
    = \sum_{i_1, i_2, \dots, i_\hs} \prod_{\alpha=1}^\hs
    A_{i_\alpha j_\alpha i_{\alpha+1} k_\alpha}.
\end{IEEEeqnarray}
Here all the $i_\alpha$ indices are summed over and $i_{\hs+1}$ is identified with $i_1$.
When writing $\Tr\left( \trm^\vs \right)$ we have interpreted $\trm$ as a linear map from $\bV_j$
to $\bV_k$, where $\bV_j$ (respectively $\bV_k$) is the tensor product of the vector spaces of the
indices $j_\alpha$ (respectively $k_\alpha$).

Implementing a lattice translation in the network is straight-forward, and shown in
Fig.~\ref{fig:Z_and_T_collection}(c).
In Fig.~\ref{fig:Z_and_T_collection}(d) is the operator $\tro \cdot \trm$, which we want to
diagonalize in order to extract universal data of a phase transition (see previous section).

It should be noted here that when we translate between network diagrams and equations, our
convention is that reading an equation from left to right corresponds to reading a diagram from
either left to right or bottom to top, but never right to left or top to bottom.

\begin{figure}[htbp]
	\centering
	\includegraphics[width=0.6\linewidth]{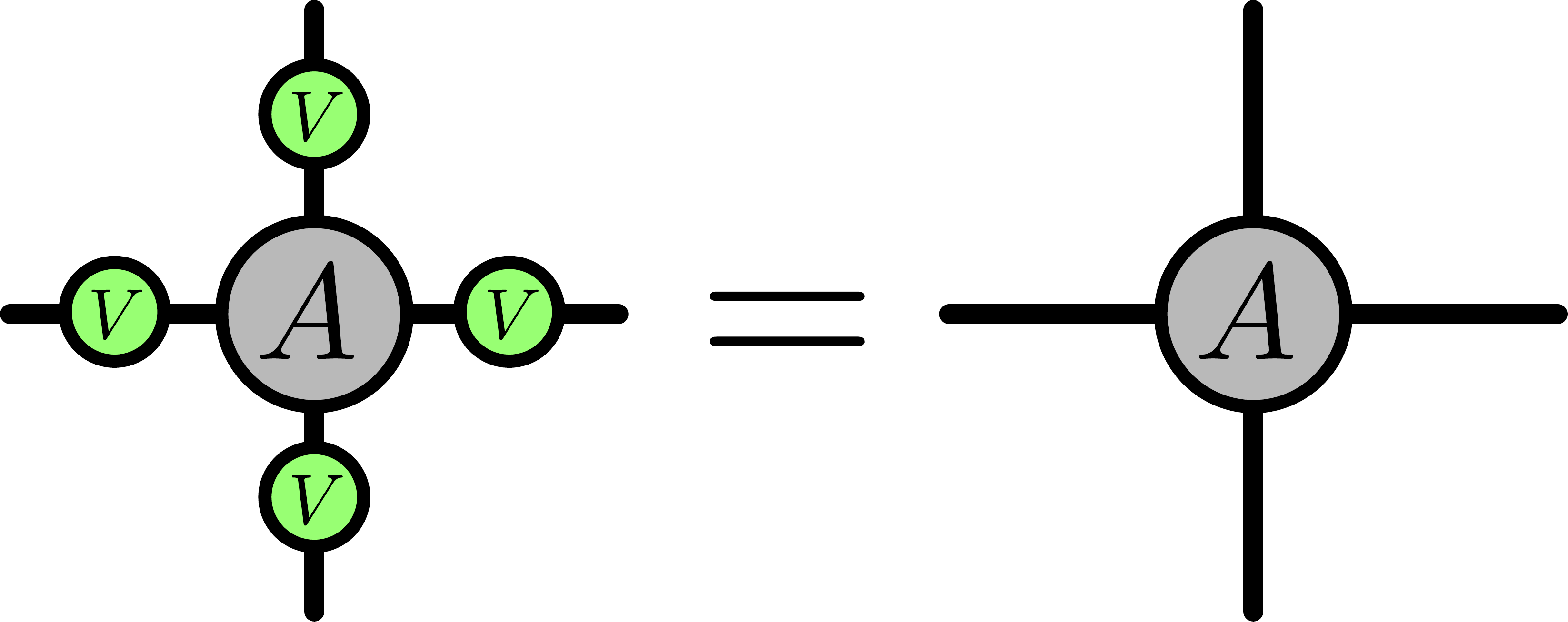}
	\caption{%
        The invariance of tensor $A$ under the symmetry transformation $V$.
    }\label{fig:A_invar}
\end{figure}

The $\Integers_2$ symmetry of the Ising model plays an important role in the tensor network
representation, as we will see later when we consider a system with a topological defect.
For a model with a global internal symmetry, the symmetry can be made manifest in the tensors
themselves.
This is covered in length in
Refs.~\onlinecite{singh_tensor_2010,singh_tensor_2011,singh_tensor_2012}.
For the present discussion it suffices to know that for the Ising model, we use tensors that
fulfill the identity in Fig.~\ref{fig:A_invar}, namely, tensors that are left unchanged if we
apply a spin-flip matrix $V$ on each of the indices.
The spin-flip matrix $V$ here is nothing but $\sigma^x$, but we call it $V$ for consistency with
the case where it acts on a coarse-grained index.
As explained in Appendix~\ref{app:tnr}, in general $V$ is some unitary representation of the
non-trivial element of the symmetry group $\Integers_2$.
In other words it is a unitary matrix such that $V^2 = \unity$.
We call a tensor that obeys the invariance property of Fig.~\ref{fig:A_invar} a $\Integers_2$
invariant tensor.
The vector space attached to each index of a $\Integers_2$ invariant tensor can be understood as
the direct sum of two subspaces, one for each parity $\pm $ (i.e.\ each $\Integers_2$ charge).
In this way, we can attach a parity to each eigenvalue/vector of the transfer matrix and, by
extension, to the corresponding scaling operators $\phi_{\alpha}$ (discussed in the previous
section).

\subsection{Coarse-graining}
A tensor network coarse-graining transformation maps a network like $\znet_{\hs,\vs}(A)$ [see
Fig.~\ref{fig:Z_and_T_collection}(a)] to a smaller network $\znet_{\hs', \vs'}(A')$ that is of
the same form but consists of tensors $A'$.
Each $A'$ corresponds to a small region of the original network $\znet_{\hs,\vs}(A)$ and
describes longer length scale features of the system.
For concreteness we consider a coarse-graining where each $A'$ corresponds to four of the original
tensors $A$ and $\hs' = \frac{\hs}{2}$ and $\vs' = \frac{\vs}{2}$, i.e.\ the coarse-graining has
scaled the linear size of the system by a factor of $\frac{1}{2}$.

Ideally, this coarse-graining would be such that the original network $\znet_{\hs, \vs}(A)$ and
the coarse-grained $\znet_{\frac{\hs}{2}, \frac{\vs}{2}}(A')$ would contract to exactly the same
value: the partition function $Z$.
In practice small errors, called truncation errors, need to be introduced to keep the computational
cost from growing too large, and thus the two networks contract to only approximately the same
value.

\begin{figure}[htbp]
	\centering
	\includegraphics[width=1.0\linewidth]{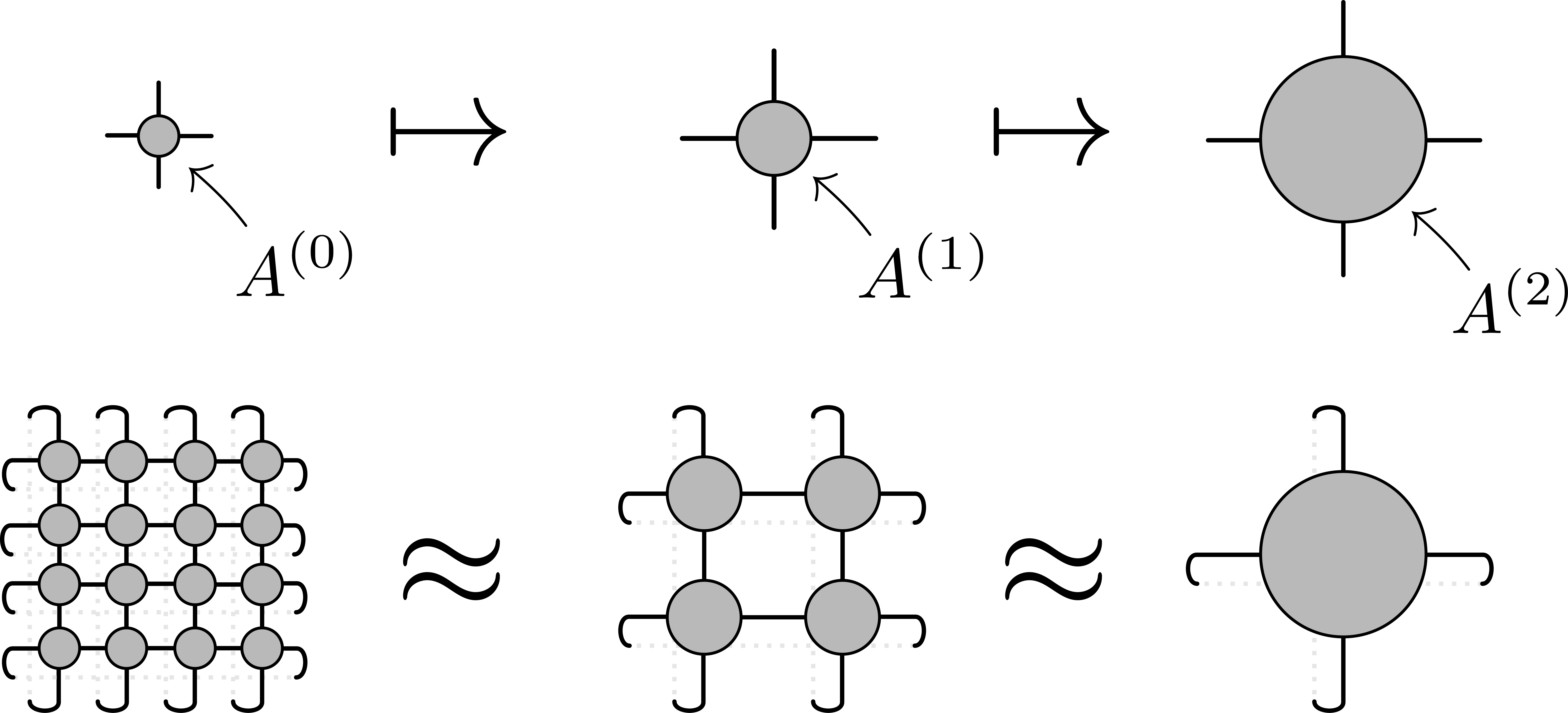}
	\caption{%
        Repeating a coarse-graining produces a series of tensors $A^{(\iters)}$ and corresponding
        networks that all contract to approximately the same value.
        We think of each $A^{(\iters)}$ as representing a local patch of the system at a different
        length scale.
        With a $2 \times 2 \mapsto 1$ coarse-graining like the one we consider, a network
        $\znet_{2^k, 2^k}(A)$ can be coarse-grained to a single tensor in $k$ steps.
    }\label{fig:TNR_iteration}
\end{figure}

We can use such a coarse-graining repeatedly to produce a series of tensors $A^{(0)} \mapsto
A^{(1)} \mapsto \dots \mapsto A^{(\iters)}$ such that $A^{(0)} \equiv A$ and each $A^{(\iters)}$
represents $4^{\iters}$ of the original tensors $A^{(0)}$.
For each tensor $A^{(\iters)}$ the network $\znet_{\frac{\hs}{2^s}, \frac{\vs}{2^s}}\left(
A^{(\iters)} \right)$ is an approximate representation of the original network $\znet_{\hs,
\vs}(A^{(0)})$.
This is illustrated in Fig.~\ref{fig:TNR_iteration}.
As first proposed and demonstrated in Ref.~\onlinecite{gu_tensorentanglementfiltering_2009},
we can then use the coarse-grained tensors $A^{(\iters)}$ to produce a transfer matrix $M$
representing many spins and extract $\{\Delta_{\alpha},s_{\alpha}\}$ with smaller finite-size
corrections.
We emphasize that although Ref.~\onlinecite{gu_tensorentanglementfiltering_2009} described
this approach in the context of a particular coarse-graining scheme, namely the \emph{tensor
entanglement-filtering renormalization} (TEFR) method, it can be used with any coarse-graining
scheme that accurately preserves the partition function $Z$ or transfer matrix $M$.

A key role in any coarse-graining scheme is played by the dimension $\chi$ of the indices of
$A^{(\iters)}$, called the bond dimension of the network.
The bond dimension controls both the computational cost of the coarse-graining, which grows as a
power of $\chi$, and the truncation errors introduced at each coarse-graining step, which decrease
with growing $\chi$.
For the purpose of estimating $\{\Delta_{\alpha}, s_{\alpha}\}$, a useful coarse-graining scheme is
then one where a sufficiently small $\chi$ (leading to a sufficiently small computational cost) can
be kept over several coarse-graining steps (so that large system sizes can be considered, reducing
the finite-size corrections) while at the same time keeping the truncation errors sufficiently
small, so that they do not significantly affect the numerical estimates.
Thus, for a fixed bond dimension $\chi$ (that is, for a fixed computational cost per
coarse-graining step), the best numerical estimates are obtained by applying a number of iterations
$s$ such that the finite-size corrections and truncation errors are of the same magnitude, and
their cumulative effect on the results is at a minimum.

The specific coarse-graining scheme that we use in this paper is called \emph{tensor network
renormalization}~(TNR)~\cite{evenbly_tensor_2015-2}.
It is based on inserting approximate partitions of unity into the network, consisting of isometric
and unitary tensors, that can be optimized to minimize the truncation error.
We will not explain the details of the algorithm in this paper, but refer the reader
to Refs.~\onlinecite{evenbly_algorithms_2015} and~\onlinecite{evenbly_tensor_2015-2}.
However, an outline of the algorithm can be found in Appendix~\ref{app:tnr}.
Similar results to the ones obtained in this paper with TNR could equally well be obtained (perhaps
with less accuracy or increased computational cost) using any coarse-graining that acts
sufficiently locally, such as the simpler tensor renormalization group
algorithm~\cite{levin_tensor_2007}. This point is elaborated further in the discussion section.

\begin{figure}[htbp]
	\centering
	\includegraphics[width=1.0\linewidth]{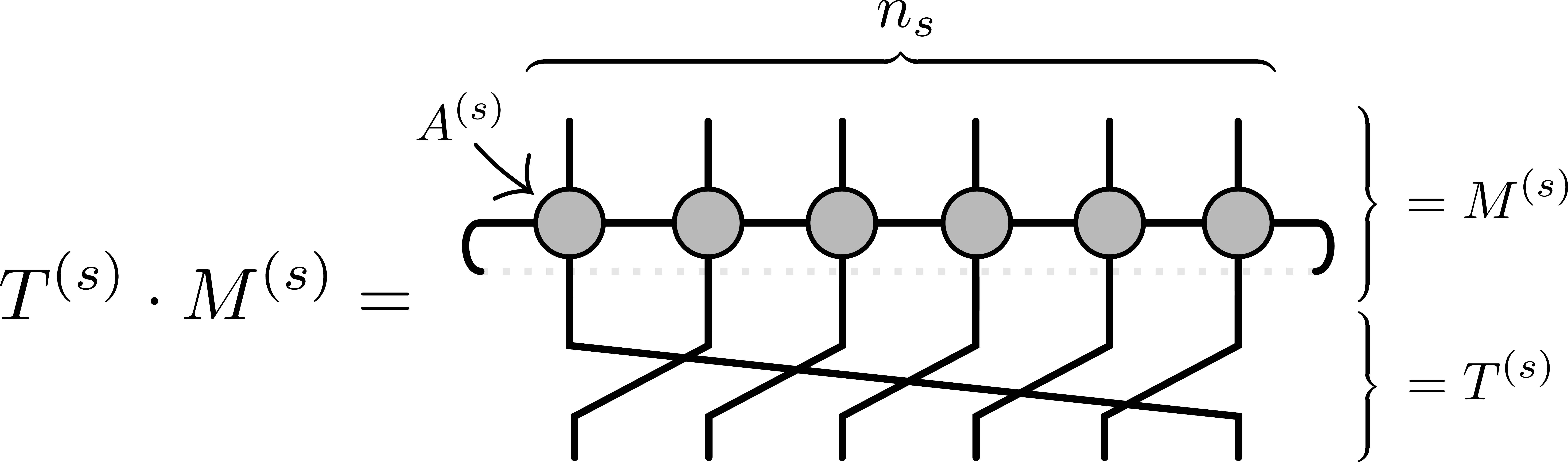}
	\caption{%
        The coarse-grained transfer matrix and translation operator.
    }\label{fig:cT_T_m}
\end{figure}

Applying a $2 \times 2 \mapsto 1$ coarse-graining transformation $s$ times, a transfer matrix of
$2^\iters \times (2^\iters \cdot \hs_\iters)$ tensors $A$ can be coarse-grained down to the
transfer matrix  $\trm^{(\iters)}$ in Fig.~\ref{fig:cT_T_m}, consisting of a row of $n_s$ tensors
$A^{(s)}$.
The computational cost thus scales logarithmically in system size.
Interpreted as a matrix, $\trm^{(\iters)}$ has dimensions $\chi^{\hs_s} \times \chi^{\hs_s}$, and
can be diagonalized for sufficiently small values of $n_s$ and $\chi$.
We diagonalize $\trm^{(\iters)}$ simultaneously with a translation operator $\tro^{(\iters)}$, also
shown in Fig.~\ref{fig:cT_T_m}.
$\tro^{(\iters)}$ is a translation by $2^\iters$ sites in the original system and its eigenvalues
yield the conformal spins modulo $\hs_\iters$, as explained in Appendix~\ref{app:tnr}.
In Appendix~\ref{app:coarse_momenta} we show how to perform a final coarse-graining step on the
composite operator $\tro^{(\iters)} \cdot \trm^{(\iters)}$ to raise the periodicity of the
conformal spins to $2\hs_s$.

\subsection{Numerical results}
\begin{figure}[htbp]
	\centering
    \includegraphics[width=1.0\linewidth]{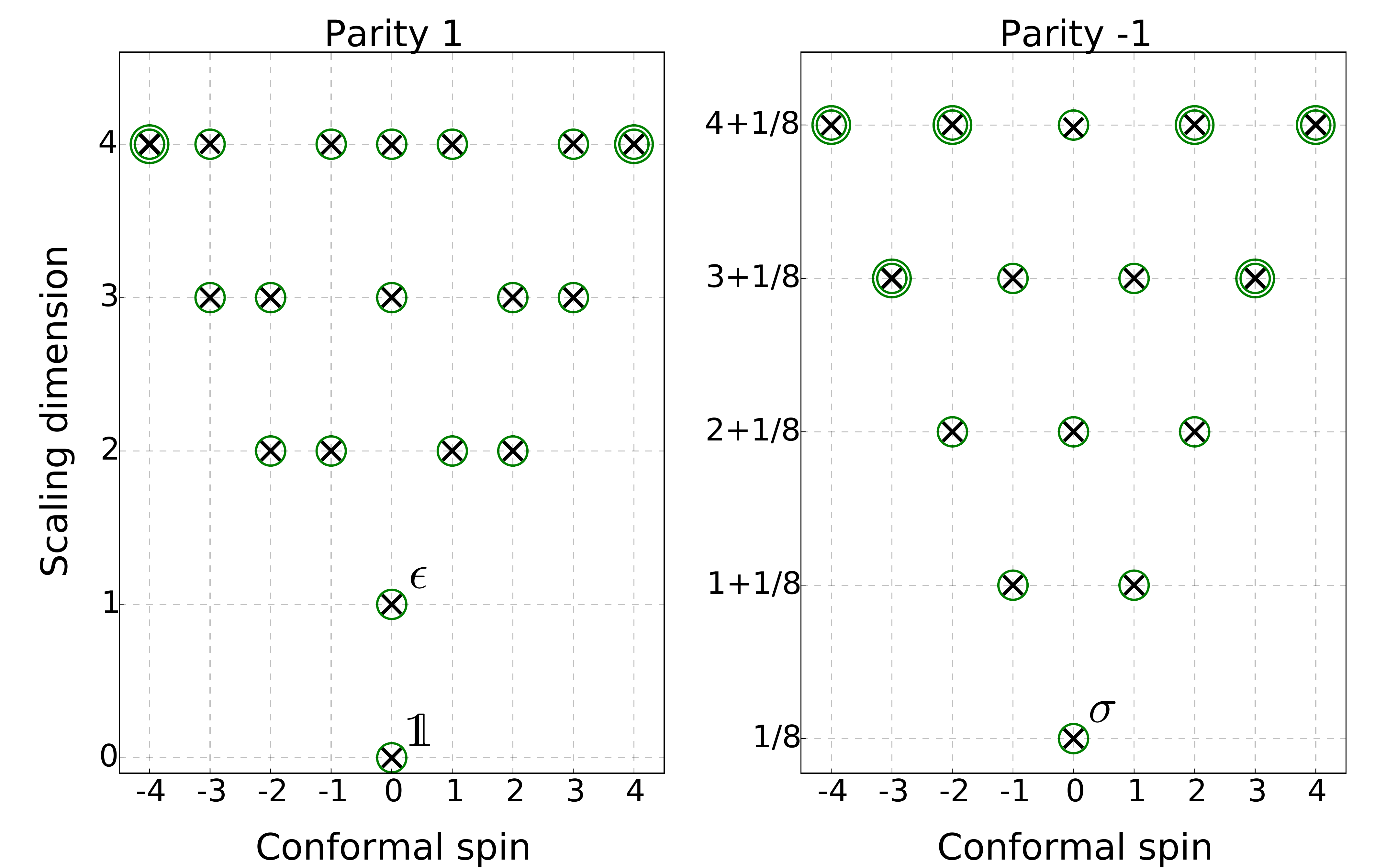}
	\caption{%
        By coarse-graining and diagonalizing a transfer matrix for the classical square lattice
        Ising model we have obtained scaling dimensions (vertical axis) and conformal spins
        (horizontal axis) of the first few scaling operators with lowest dimensions in the Ising
        CFT\@.
        The crosses mark the numerical values that should be compared with the circles that are
        centered at the exact values.
        The scaling operators are divided according to their $\Integers_2$ charge, that is
        their parity under a global spin-flip.
        Several concentric circles denote the degeneracy $N_\alpha$ of that $(\Delta_\alpha,
        s_\alpha)$ pair.
        Although it is not clear from the figure, these degeneracies also come out correctly.
    }\label{fig:D_unity_results}
\end{figure}

In Fig.~\ref{fig:D_unity_results}, we show scaling dimensions and conformal spins
$\{\Delta_{\alpha}, s_{\alpha}\}$ obtained by diagonalizing a transfer matrix that has been
coarse-grained using TNR\@.
To obtain the scaling dimensions we have coarse-grained a transfer matrix of $4$ tensors
$A^{(\iters)}$ for $\iters = 7$ (corresponding to $2\times 4 \times 2^{7}\times 2^7=2^{17} \approx
130\,000$ spins) using bond dimensions $\chi' = 14$ and $\chi = 28$ (the TNR scheme we use has two
relevant bond dimensions, see Appendix~\ref{app:tnr}).
For the conformal spins a slightly larger system was used, as explained in
Appendix~\ref{app:coarse_momenta}.
The numerical results are in excellent agreement with the exact values even higher up in the
conformal towers, in contrast with the exact diagonalization results in
Fig.~\ref{fig:ed_ising_results}.
Table~\ref{tab:primary_data_unity} shows a comparison of the numerical values to the exact ones for
the primary fields.
For the central charge we obtain $c=0.500091$ where the exact value would be $\frac{1}{2}$.

\begin{table}[tb]
    \begin{tabular}{cccccc}
        Primary & $\left(h, \overline{h}\right)$
        & $\Delta_\text{TNR}$ & $\Delta_\text{exact}$
        & $s_\text{TNR}$ & $s_\text{exact}$\\%
        \midrule

        $\unity$ & $\left(0,0\right)$
        & \textendash & $0$
        & $0$ & 0\\%

        $\epsilon$ & $\left(\sfrac{1}{2},\sfrac{1}{2}\right)$
        & $1.000256$ & $1$
        & $0$ & $0$\\%

        $\sigma$ & $\left(\sfrac{1}{16},\sfrac{1}{16}\right)$
        & $0.125109$ & $0.125$
        & $0$ & $0$\\%
    \end{tabular}
    \caption{%
        The scaling dimensions and conformal spins of the primary fields of the Ising CFT obtained
        using TNR, contrasted with the exact values.
        No numerical value for the scaling dimension of the identity operator is provided because
        we extract the central charge $c$ by assuming that $\Delta_\unity = 0$ exactly.
        The central charge we get is $c=0.500091$ whereas the exact one is $c=\frac{1}{2}$.
        The conformal spins we obtain are exactly zero without any numerical errors, because
        we know that the possible eigenvalues of the translation operator for a four site system
        are $\pm 1$ and $\pm i$, which yields the possible conformal spins $-1$, $0$, $1$ and $2$.
    }\label{tab:primary_data_unity}
\end{table}

\subsection{Multi-scale entanglement renormalization ansatz}

\begin{figure}[htbp]
	\centering
    \includegraphics[width=0.667\linewidth]{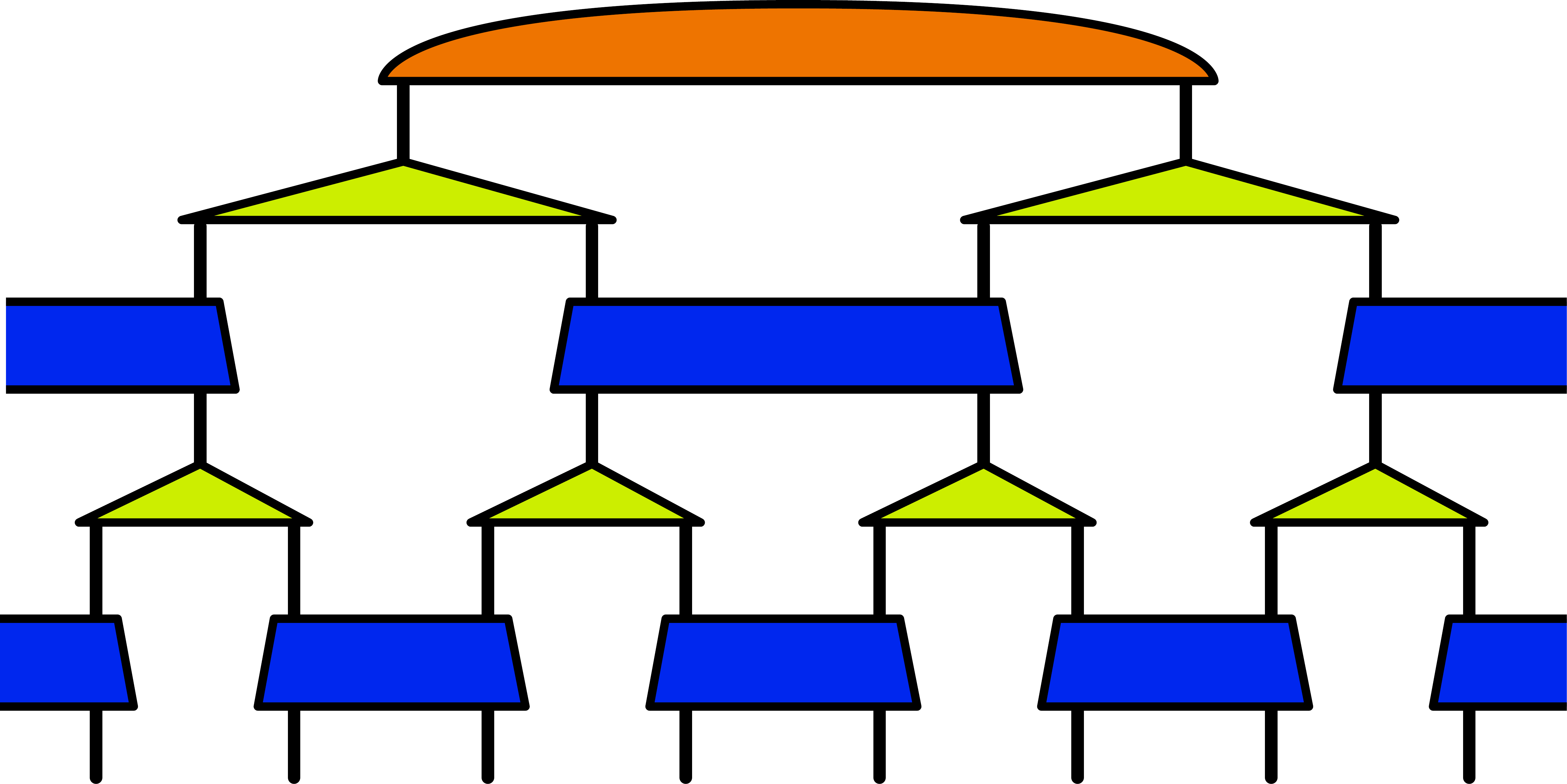}
	\caption{%
        A MERA for a state of eight spins in a system with periodic boundaries.
        Such a network for the ground state of a spin chain can be obtained by applying TNR to a
        tensor network describing the Euclidean path integral of the quantum
        Hamiltonian~\cite{evenbly_tensor_2015-1}.
        The tensors in the network are the same unitaries and isometries as used in
        Appendix~\ref{app:tnr}.
        Even though our notation does not reflect it, the unitaries and isometries on different
        layers generally differ from each other.
    }\label{fig:MERA_threelayer}
\end{figure}

We conclude this section by recalling that, as explained in
Ref.~\onlinecite{evenbly_tensor_2015-1}, if we apply TNR to a tensor network representing the
Euclidean path integral $e^{-\beta_Q H}$ of a quantum Hamiltonian $H$, we obtain a multi-scale
entanglement renormalization ansatz (MERA) for the ground state of $H$.
Such a MERA, shown in Fig.~\ref{fig:MERA_threelayer}, is built of tensors (called disentanglers
and isometries) that are produced during the coarse-graining of the Euclidean path integral.
We have observed empirically that the partition function of the 2D classical Ising model that we
study is also the Euclidean path integral of the 1D quantum model.
It then follows that the disentanglers and isometries produced during the coarse-graining of the
classical partition function $Z$ can be put together into a MERA that represents the ground state
of the Hamiltonian $H$ for the 1D quantum Ising model.
This observation will extend to the case of topological defects, discussed in
Secs.~\ref{sec:D_epsilon} and~\ref{sec:D_sigma}.

\section{Topological defects}
\label{sec:topological_defects}
In this section, we review the topological conformal defects (often referred to simply as
topological defects) of the Ising CFT\@.
On a torus these defects can be thought of as different boundary conditions, and their presence
modifies the operator content of the partition function.
The Ising CFT has two different non-trivial topological defects, which we introduce in this section
and whose lattice realization will be analyzed in the next two sections.

As stated earlier in Eq.~\eqref{eq:Z_CFT1}, the partition function of a CFT on a torus can be
written as
\begin{IEEEeqnarray}{c}
    Z_{\mathrm{CFT}} = \Tr\left( e^{-2\pi \tau_2 H_{\mathrm{CFT}}} e^{2\pi i \tau_1 P}  \right).
\end{IEEEeqnarray}
We consider now a twisted partition function $Z_{D}$ of the form\footnote{%
    We have dropped here the previous distinction between the partition function $Z$ of a lattice
    model and $Z_\mathrm{CFT}$ of the CFT\@:
    Here by $Z_D$ we mean a field theory partition function, whereas later we will use the same
    symbol to refer to its lattice realization.
    Context should make clear which one we are referring to.
}
\begin{IEEEeqnarray}{c}
    \label{eq:Z_DCFT}
    Z_{D} = \Tr\left( D e^{-2\pi \tau_2 H_{\mathrm{CFT}}} e^{2\pi i \tau_1 P}  \right).
\end{IEEEeqnarray}
$D$ is the twist operator, which can be thought of as implementing a special type of boundary
condition on the torus.
If $D$ commutes with all the generators of the Virasoro algebra it is called a topological
conformal defect (topological defect for simplicity).
A twist operator can be seen as a line defect that loops around the torus.
If the defect is topological the loop can be freely deformed without affecting correlation
functions in the system as long as the defect is not moved across a field insertion.
The conformality of the defect also means that it is invariant under scale
transformations.~\cite{frohlich_duality_2007,petkova_generalised_2001}

The twisted partition function $Z_D$ for a topological defect can be written as a sum of terms
corresponding to scaling dimensions and conformal spins $\{\Delta_{\alpha}, s_{\alpha}\}_{D}$,
similarly as for the non-twisted $Z$ in Eq.~\eqref{eq:Z_CFT2}.
However, the $\{\Delta_{\alpha}, s_{\alpha}\}_{D}$ present in the sum are in general different from
those of the non-twisted $Z$.~\cite{frohlich_duality_2007,petkova_generalised_2001}

For the Ising CFT, all possible topological defects can be built as linear combinations of three
defects which we denote $D_\unity$, $D_\epsilon$ and $D_\sigma$.
They are known as the \emph{simple} defects of the Ising CFT and are related to the same
irreducible representations of the Virasoro algebra as the primary fields $\unity$, $\epsilon$ and
$\sigma$, hence the names.~\cite{petkova_generalised_2001,frohlich_kramerswannier_2004}
The $D_\unity$ defect is the trivial defect or rather the lack of any defect where the twist
operator is just the identity.
The partition function for it, $Z_{D_{\unity}} = Z$, has been the topic of the last two sections of
the paper.
$D_\epsilon$ is also known as the ($\Integers_2$) symmetry defect and $D_\sigma$ is called the
(Kramers-Wannier) duality defect.

The operators present in the twisted partition functions $Z_{D_{\epsilon}}$ and
$Z_{D_{\sigma}}$ come organized in conformal towers built on top of primary operators, each of
which is identified with conformal dimensions $(h,\overline{h})$ that may take values $0$,
$\frac{1}{2}$ and $\frac{1}{16}$, just like for the operators of the non-twisted $Z_{D_{\unity}}$.
The combinations of $h$ and $\overline{h}$ that are present in each partition function are shown in
Table~\ref{tab:operator_content}.
Note that together the three partition functions include all the possible pairs $(h,
\overline{h})$.~\cite{petkova_generalised_2001}

{\setlength{\tabcolsep}{0.4em}\renewcommand{\arraystretch}{1.3}%
\begin{table}[htbp]
    \hfill
    \begin{tabular}{c|ccc}
        $Z_{D_{\unity}}$ & 0 & $\frac{1}{2}$ & $\frac{1}{16}$\\%
        \hline
        0 & $\unity$ & &\\%
        $\frac{1}{2}$ & & $\epsilon$ &\\%
        $\frac{1}{16}$ & & & $\sigma$
    \end{tabular}
    \hfill
    \begin{tabular}{c|ccc}
        $Z_{D_\epsilon}$ & 0 & $\frac{1}{2}$ & $\frac{1}{16}$\\%
        \hline
        0 & & $\psi$ &\\%
        $\frac{1}{2}$ & $\overline{\psi}$ & &\\%
        $\frac{1}{16}$ & & & $\mu$
    \end{tabular}
    \hfill
    \begin{tabular}{c|ccc}
        $Z_{D_{\sigma}}$ & 0 & $\frac{1}{2}$ & $\frac{1}{16}$\\%
        \hline
        0 & & & $X$ \\%
        $\frac{1}{2}$ & & & $Y$\\%
        $\frac{1}{16}$ & $\overline{X}$ & $\overline{Y}$ &
    \end{tabular}
    \hfill{}
    \caption{%
        The primary operators with conformal dimensions $(h, \overline{h})$ included in the Ising
        partition functions with different defects in them.
        The horizontal axis is $h$, the vertical one is $\overline{h}$.
        The operators in $Z_{D_{\unity}}$ and $Z_{D_{\epsilon}}$ have established names shown in
        the table, with the $Z_{D_{\unity}}$ ones already familiar to us from earlier in the paper.
        The ones included in $Z_{D_{\sigma}}$ we denote $X$, $\overline{X}$, $Y$ and $\overline{Y}$
        in the absence of more a established convention.
    }\label{tab:operator_content}
\end{table}}

Consider bringing two defects next to each other so that they effectively behave as one defect.
This gives rise to fusion rules for topological defects.
The fusion rules of the topological defects of the Ising model are the same as the fusion rules of
the primary operators, namely
\begin{IEEEeqnarray}{rCl}
    D_\epsilon \times D_\epsilon &=& D_\unity\\%
    D_\sigma \times D_\epsilon &=& D_\sigma\\%
    D_\sigma \times D_\sigma &=& D_\unity + D_\epsilon.
\end{IEEEeqnarray}

The next two sections are devoted to analyzing the two non-trivial defects $D_\epsilon$ and
$D_\sigma$ on the lattice.
Similarly as we have done in the two preceding sections for the trivial defect $D_\unity$, we will
first study the realization of each defect in the quantum and classical lattice models, then
discuss how to represent them using tensor networks, and finally study how to coarse-grain these
networks.
We will then diagonalize transfer matrices for the twisted partition functions and
extract the scaling dimensions and conformal spins of the scaling operators.
In the process we will also end up discussing how these defects can be moved around and how their
fusion rules manifest in the lattice models.

\section{Symmetry defect $D_{\epsilon}$}
\label{sec:D_epsilon}
In this section, we review how to realize, on the lattice, the symmetry defect $D_\epsilon$ of the
Ising CFT in the classical and quantum Ising models and present a way of implementing
$D_{\epsilon}$ in a tensor network.
We discuss how to coarse-grain this tensor network representation and use it to numerically
evaluate the scaling dimensions and conformal spins $\{\Delta_{\alpha}, s_{\alpha}\}_{D_\epsilon}$
of the operators in the twisted partition function $Z_{D_{\epsilon}}$.

\subsection{Lattice representation}
The symmetry defect $D_\epsilon$ is directly related to the $\Integers_2$ spin-flip symmetry of the
Ising model.
Consider the classical  Ising model on a square lattice and draw a closed loop around a connected
region of spins.
If one flips the spins encircled by the loop, then most of the terms $-\sigma_i \sigma_j$ in the
Hamiltonian are unaffected, but along the loop there is a string of nearest-neighbor pairs where
one of the spins is flipped and the other one is not and their couplings become $\sigma_i
\sigma_j$.
This string is the classical lattice realization of the $D_\epsilon$ defect.

We are interested in a partition function $Z_{D_{\epsilon}}$ where such a defect forms a
non-contractible loop around a torus.
$Z_{D_{\epsilon}}$ is said to have antiperiodic boundary conditions due to how most of the spins
are coupled ferromagnetically but along the boundary the coupling is antiferromagnetic.

Analogously, in the quantum spin chain the $D_\epsilon$ defect is realized by changing the sign
of one of the nearest-neighbor terms to obtain
\begin{IEEEeqnarray}{c}
    \label{eq:H_epsilon}
    H_{D_\epsilon}
    = -\left(\sum_{i=1}^{\hs-1} \sigma_i^z \sigma_{i+1}^z - \sigma_\hs^z \sigma_1^z
    + \sum_{i=1}^{\hs} \sigma_i^x \right).
\end{IEEEeqnarray}

Let us concentrate on the quantum case for a moment and consider moving and fusing $D_\epsilon$
defects.
In the Hamiltonian~$\ref{eq:H_epsilon}$ the defect is located on the coupling between spins $\hs$
and $1$.
If we conjugate the Hamiltonian with $\sigma_\hs^x$, we effectively move the defect by one site:
\begin{IEEEeqnarray}{rCll}
    \sigma_\hs^x H_{D_\epsilon} \sigma_\hs^x
    &=& -\Bigg( & \sum_{i=1}^{\hs-2} \sigma_i^z \sigma_{i+1}^z + \sum_{i=1}^{\hs} \sigma_i^x\\%
    &&&  - \sigma_{\hs-1}^z \sigma_\hs^z + \sigma_\hs^z \sigma_1^z \Bigg).
\end{IEEEeqnarray}
$\sigma_\hs^x$ is unitary and Hermitian and conjugating by it does not affect the spectrum of the
Hamiltonian.
Using $\sigma^x$ to move defects one can easily check that taking a Hamiltonian with two
$D_\epsilon$ defects and moving them to the same site gives the usual Ising Hamiltonian $H$.
This is the fusion rule $D_\epsilon \times D_\epsilon = D_\unity$.

Similarly in the classical model a $D_\epsilon$ defect can be moved by flipping a spin that is next
to it.
This is because flipping a spin changes all its couplings from ferromagnetic to antiferromagnetic
or vice versa.
Such a spin-flip is only a matter of relabeling a degree of freedom and does not affect the
partition function.
Now consider a system with two $D_\epsilon$ defects parallel to each other and move one of them
until there is only a string of spins between the two defects.
Flipping the spins between the defects changes all the antiferromagnetic couplings to ferromagnetic
and we are left with the usual Ising model --- a $D_\unity$ defect.

\subsection{Tensor network representation}
\begin{figure}[tbp]
	\centering
	\includegraphics[width=1.0\linewidth]{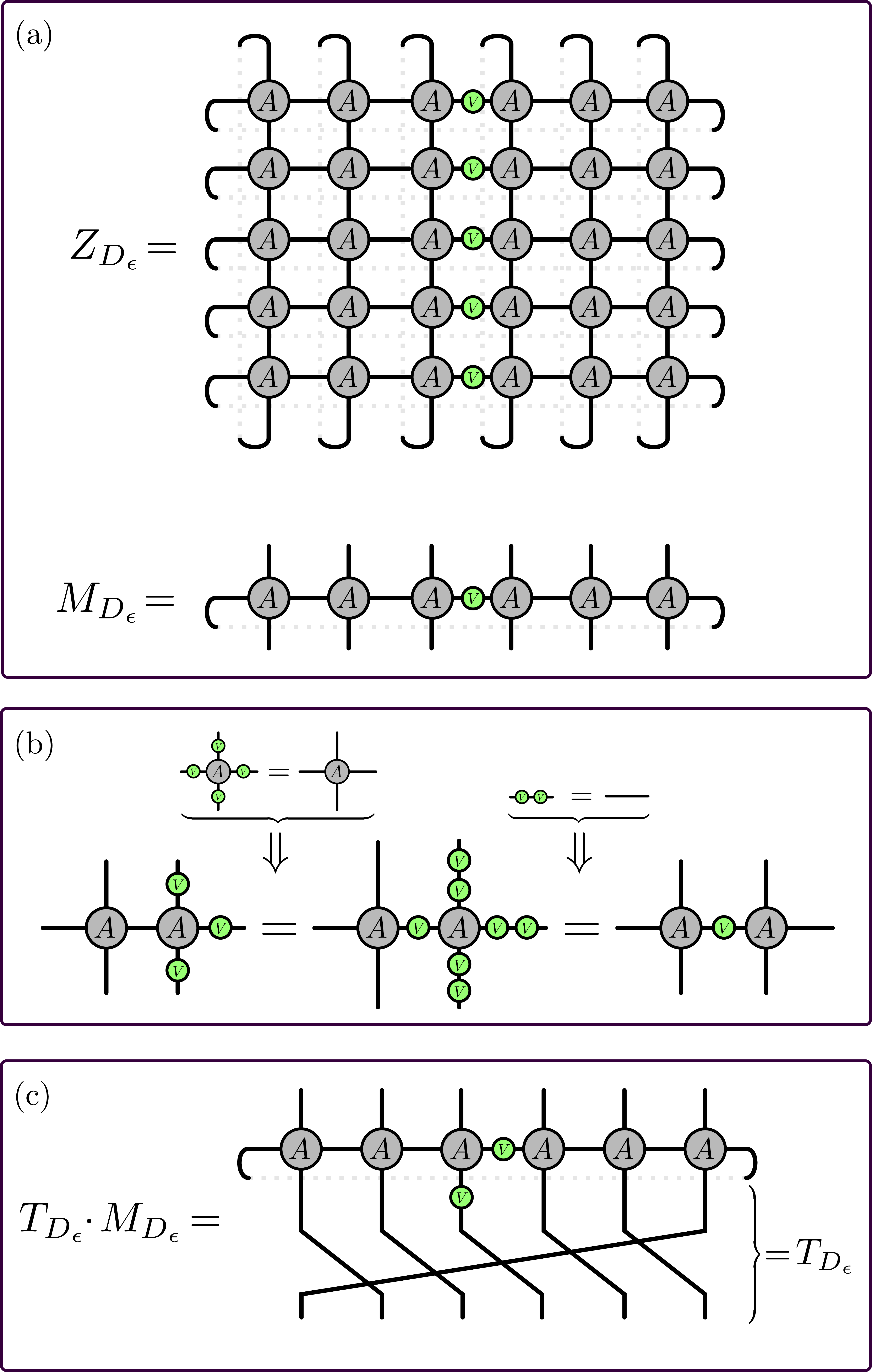}
	\caption{%
        (a)
        The tensor networks for the partition function $Z_{D_{\epsilon}}$ and its transfer matrix
        $\trm_{D_\epsilon}$, that fulfills $Z_{D_{\epsilon}} = \Tr \big( \trm_{D_\epsilon}^m
        \big)$.
        $V$ is the Pauli matrix $\sigma^x$ that flips a spin.
        (b)
        The invariance property of $A$ and the fact that $V^2 = \unity$  imply that conjugating
        the tensor next to the defect from above and below with $V$ moves the defect by one site.
        (c)
        The operator $\tro_{D_\epsilon}\cdot\trm_{D_\epsilon}$ whose diagonalization produces the
        conformal data $\{\Delta_{\alpha}, s_{\alpha}\}_{D_{\epsilon}}$ for $Z_{D_{\epsilon}}$.
        Here, $\tro_{D_\epsilon}$ is the translation operator that commutes with the transfer
        matrix $\trm_{D_\epsilon}$.
    }\label{fig:D_eps_collection}
\end{figure}

In a tensor network $\znet_{\hs,\vs}(A)$ that represents a partition function $Z_{D_{\unity}}$ [see
Fig.~\ref{fig:Z_and_T_collection}(a)], every bond corresponds to one classical spin.
If we multiply one of the legs $i$ of a tensor $A_{ijlk}$ with a spin-flip matrix $V$ this makes
the corresponding spin $i$ couple antiferromagnetically to its neighbors on one side.
Thus, the network for $Z_{D_{\epsilon}}$ is as depicted in Fig.~\ref{fig:D_eps_collection}(a),
where the defect lives on a string of bonds.
The same figure shows the transfer matrix $\trm_{D_\epsilon}$ for this twisted partition function.

Just as in the system without a defect, by diagonalizing $\trm_{D_\epsilon}$ we can extract the
scaling dimensions of the operators in $Z_{D_{\epsilon}}$.
For the conformal spins we would need to diagonalize $\trm_{D_\epsilon}$ simultaneously with the
translation operator.
However, the usual lattice translation $\tro$ does not commute with $\trm_{D_\epsilon}$ because the
translation moves the defect by one lattice site.
We can move the defect back to where it was by conjugating the tensor $A$ that is next to the
defect with $V$ from above and below, as shown in Fig.~\ref{fig:D_eps_collection}(b).

\begin{figure}[htbp]
    \centering
    \includegraphics[width=1.0\linewidth]{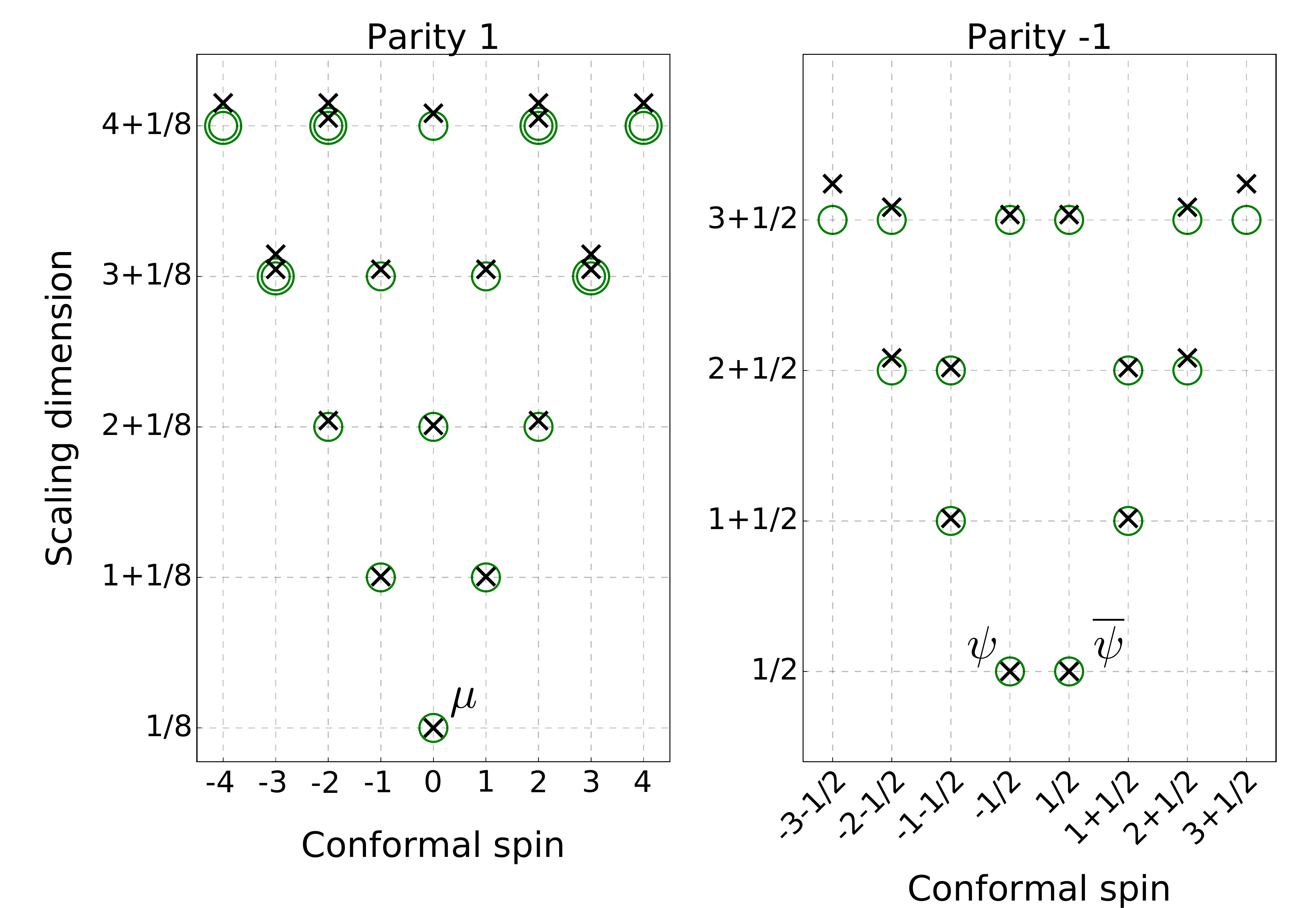}
    \caption{%
        The scaling dimensions (vertical axis) and conformal spins (horizontal axis) of the first
        scaling operators in $Z_{D_{\epsilon}}$ obtained from exact diagonalization of a transfer
        matrix of $\hs=18$ sites.
        The scaling operators are again divided by their $\Integers_2$ charge.
        The crosses mark the numerical values that can be compared with the circles that are
        centered at the exact values.
        Several concentric circles denote the degeneracy $N_\alpha$ of that $(\Delta_\alpha,
        s_\alpha)$ pair.
    }\label{fig:ed_eps_results}
\end{figure}

Thus the operator $\tro_{D_\epsilon} = V \tro$ commutes with $\trm_{D_\epsilon}$.
We call $\tro_{D_\epsilon}$ the generalized translation operator for the $D_\epsilon$ defect.
It is the notion of translation under which the partition function with a $D_\epsilon$ defect is
translation invariant.~\cite{gehlen_ashkin-teller_1987,chaselon_toroidal_1989}
$\tro_{D_\epsilon} \cdot \trm_{D_\epsilon}$, shown in Fig.~\ref{fig:D_eps_collection}(c), has a
spectrum of the form in Eq.~\eqref{eq:cardys_formula_momentum_no_f} from which the conformal
spins and scaling dimensions of the operators in $Z_{D_{\epsilon}}$ can be obtained.
Figure~\ref{fig:ed_eps_results} shows estimates for $\{ \Delta_\alpha, s_\alpha \}$ obtained
numerically by diagonalizing a transfer matrix consisting of $\hs=18$ tensors.
The structure of the bottom of the conformal towers can be clearly recognized, but the accuracy of
the estimates deteriorates quickly as we look at larger scaling dimensions.
We shall see below that again increasing the system size by using a coarse-graining algorithm will
greatly improve these estimates.

\subsection{Coarse-graining}
\begin{figure}[htbp]
	\centering
	\includegraphics[width=1.0\linewidth]{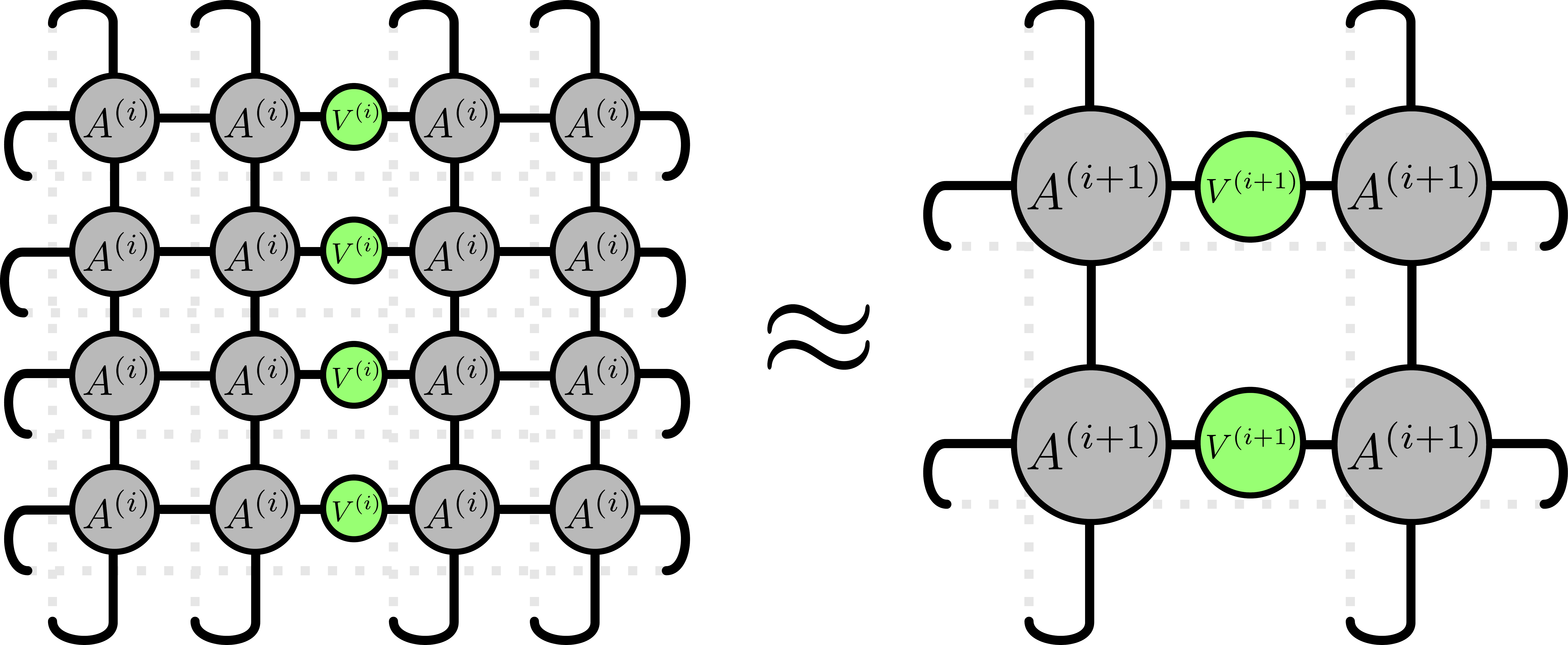}
	\caption{%
        Coarse-graining a $D_\epsilon$ defect produces a similar defect at the next scale.
        $A^{(i)}$ and $A^{(i+1)}$ are the same tensors that we obtain when coarse-graining a system
        without a defect and $V^{(i)}$ and $V^{(i+1)}$ are the spin-flip matrices of the bonds they
        are on.
    }\label{fig:TNR_epsilon_iteration2}
\end{figure}

Coarse-graining the symmetry defect is trivial when we use a coarse-graining scheme that is based
on using $\Integers_2$ invariant tensors:
By utilizing the invariance of all the tensors involved we can move the string of spin-flip
matrices $V$ to the next scale without any additional numerical effort.
In other words, as is shown in Fig.~\ref{fig:TNR_epsilon_iteration2}, a network
$\znet_{\hs,\vs}(A)$ that has $V$ matrices on a string of bonds coarse-grains into a network
$\znet_{\frac{\hs}{2},\frac{\vs}{2}}(A')$ with matrices $V'$ on a string of bonds.
$A'$ is the same tensor that we obtain when coarse-graining a network without a defect.
$V'$'s are the spin-flip matrices of the bonds of the coarse-grained network, i.e.\ the unitary
representations of the non-trivial element of $\Integers_2$ under which $A'$ is $\Integers_2$
invariant.
Similarly, the generalized translation operator $\tro_{D_\epsilon}$ coarse-grains into a
translation operator $\tro_{D_\epsilon}'$ at next scale with a $V'$ on one of the legs.
How, exactly, this happens when using TNR is explained in Appendix~\ref{app:tnr}.
However, this also holds for other coarse-graining schemes, provided that $\Integers_2$ symmetric
tensors are used.

The coarse-graining of $Z_{D_{\epsilon}}$ or $\trm_{D_{\epsilon}}$ can be iterated.
We then diagonalize an operator $\tro_{D_\epsilon}\cdot\trm_{D_\epsilon}$ similar to that in
Fig.~\ref{fig:D_eps_collection}(c), but with $A^{(\iters)}$'s and $V^{(\iters)}$'s instead of $A$'s
and $V$'s.
The form of $V^{(\iters)}$ is determined by the $\Integers_2$ invariance property of
$A^{(\iters)}$.
Notice that if we had already coarse-grained an equivalent network without defects, then we already
have the tensors $A^{(\iters)}$ and $V^{(\iters)}$ and the only additional computational work
required is the diagonalization of $\tro_{D_\epsilon}\cdot\trm_{D_\epsilon}$.

\subsection{Numerical results}
\begin{figure}[htbp]
	\centering
	\includegraphics[width=1.0\linewidth]{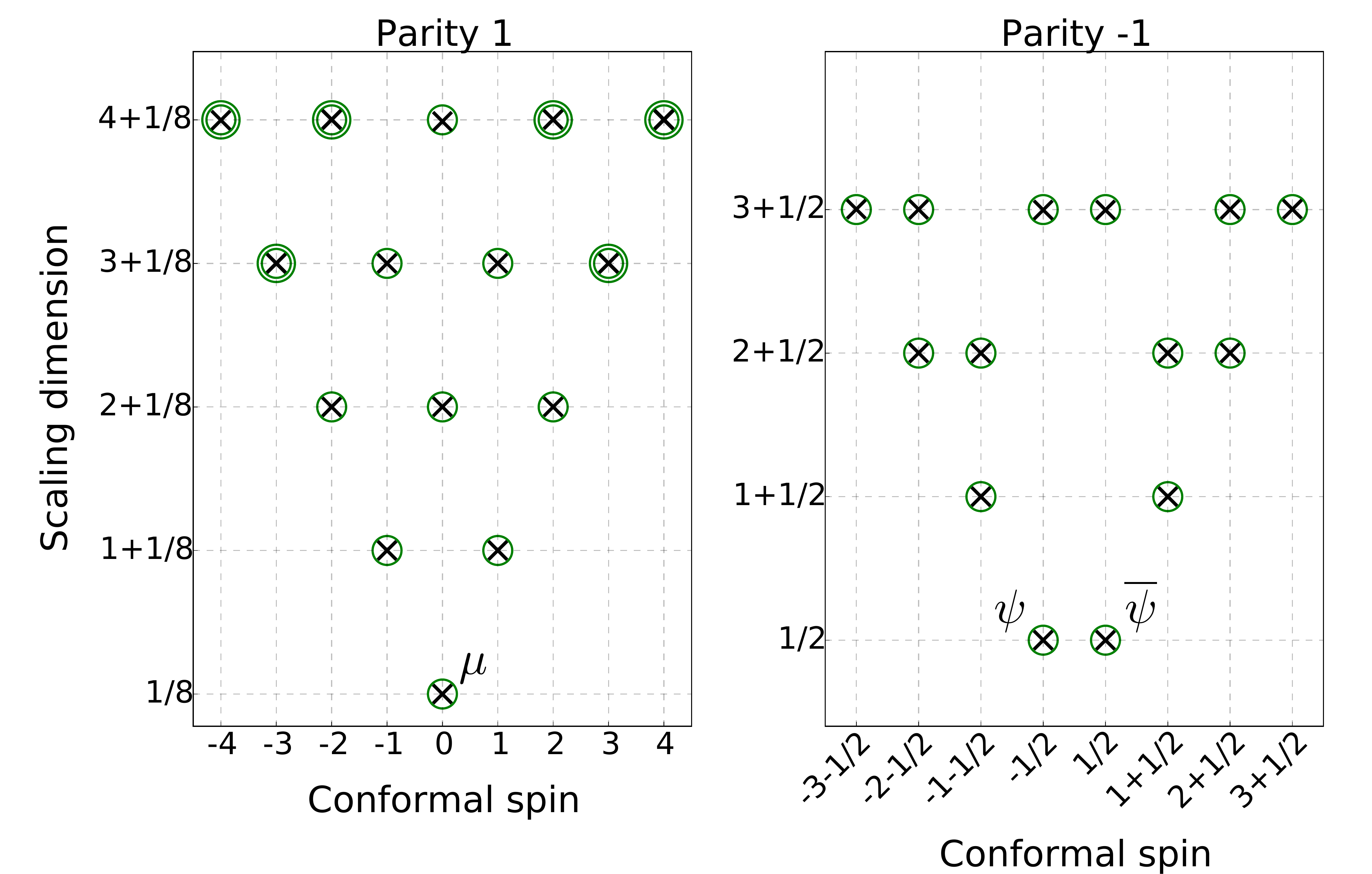}
	\caption{%
        The scaling dimensions (vertical axis) and conformal spins (horizontal axis) of the first
        scaling operators of the two-dimensional classical Ising model with a $D_\epsilon$ defect
        as obtained with TNR\@.
        The crosses mark the numerical values and the circles mark the exact values.
        The scaling operators are divided according to their $\Integers_2$ charge, that is
        their parity under a global spin-flip.
        Several concentric circles denote the degeneracy $N_\alpha$ of that $(\Delta_\alpha,
        s_\alpha)$ pair.
    }\label{fig:D_eps_results}
\end{figure}

As mentioned earlier in Sec.~\ref{sec:topological_defects}, the primary fields present in
$Z_{D_{\epsilon}}$ are $\mu$, $\psi$ and $\overline{\psi}$ with the conformal dimensions
$(\frac{1}{16}, \frac{1}{16})$, $(\frac{1}{2}, 0)$ and $(0,
\frac{1}{2})$~\cite{petkova_generalised_2001}.
$\mu$ has parity $+1$, $\psi$ and $\overline{\psi}$ have parity $-1$.
Numerical results for the scaling dimensions and conformal spins of these primaries obtained with
the TNR method are shown in Table~\ref{tab:primary_data_epsilon}.
Similar values for some of the first descendants are shown in Fig.~\ref{fig:D_eps_results}.
The results are in excellent agreement with the exact results even higher up in the conformal
towers.
The parameters used to obtain these results are the same as for the $D_\unity$ case:
A transfer matrix of $2^7 \times (4 \times 2^7)$  tensors $A^{(0)}$ (corresponding to $2 \times 4
\times 2^7 \times 2^7 = 2^{17}$ spins) coarse-grained with bond dimensions $\chi' = 14$ and $\chi =
28$ for the scaling dimensions and a transfer matrix twice as wide and with one additional
coarse-graining step for the conformal spins, as explained in Appendix~\ref{app:coarse_momenta}.

\begin{table}[htbp]
    \begin{tabular}{cccccc}
        Primary & $\left(h, \overline{h}\right)$
        & $\Delta_\text{TNR}$ & $\Delta_\text{exact}$
        & $s_\text{TNR}$ & $s_\text{exact}$\\%
        \midrule

        $\mu$ & $\left(\sfrac{1}{16},\sfrac{1}{16}\right)$
          & $0.1249287$ & $0.125$ & $10^{-16}$ & $0$\\%
        $\psi$ & $\left(\sfrac{1}{2},0\right)$
          & $0.5000704$ & $0.5$ & $\phantom{+}0.4999847$ & $\phantom{+}0.5$\\%
        $\overline{\psi}$ & $\left(0,\sfrac{1}{2}\right)$
          & $0.5000704$ & $0.5$ & $-0.4999847$ & $-0.5$\\%
    \end{tabular}
    \caption{%
        The scaling dimensions $\Delta$ and conformal spins $s$ for the primaries of
        $Z_{D_{\epsilon}}$ as obtained with TNR compared with the exact values.
        Note that one could easily see that the conformal spins must be half-integers in the parity
        $-1$ sector and integers in the parity $+1$ sector by observing that
        $\left(\tro_{D_\epsilon}\right)^n = \Sigma^x$, the $\Integers_2$ symmetry operator.
        However, we choose to present the numerical values for the conformal spins, including the
        small numerical errors, to demonstrate the accuracy of our method.
    }\label{tab:primary_data_epsilon}
\end{table}

\subsection{MERA}
\begin{figure}[htbp]
	\centering
    \includegraphics[width=1.0\linewidth]{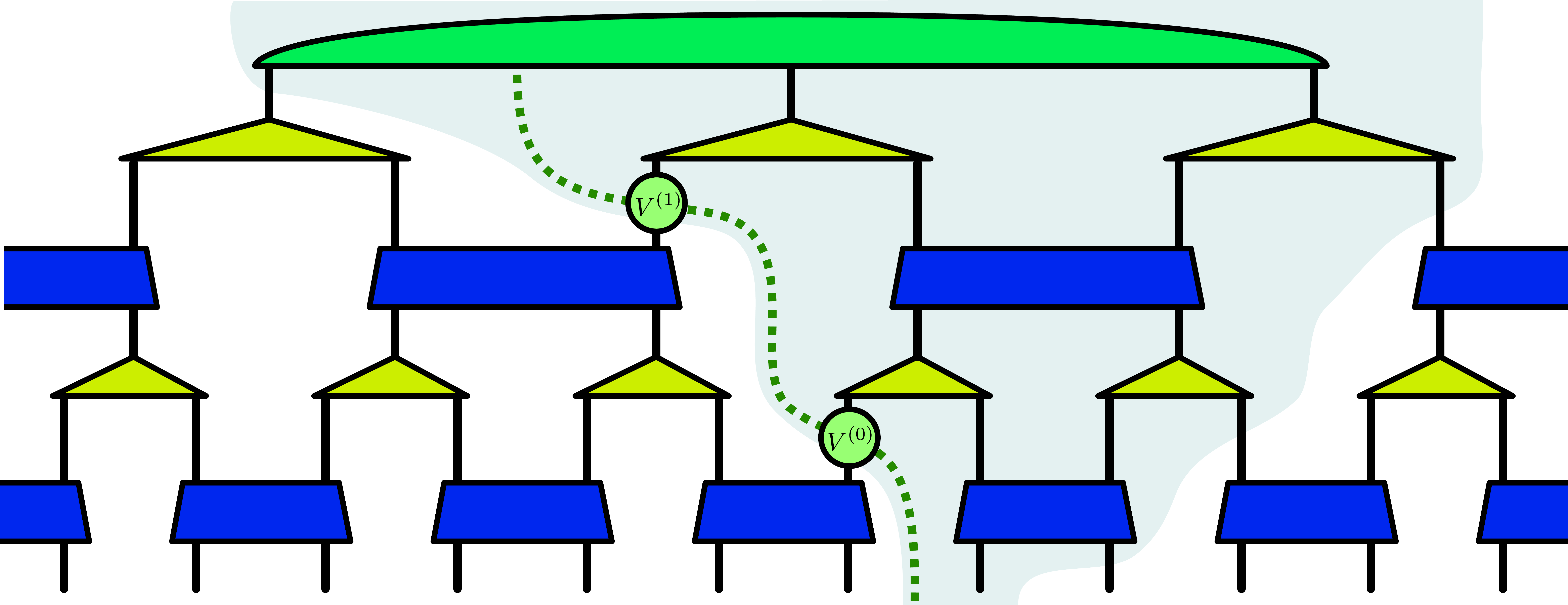}
	\caption{%
        A MERA for the ground state of the quantum spin chain that has a symmetry defect in it,
        produced by coarse-graining the tensor network for $Z_{D_{\epsilon}}$ with TNR\@.
        Compared to the MERA without a defect [Fig.~\ref{fig:MERA_threelayer}] the only
        differences here are the spin-flip matrices and the different top tensor.
        Like the $V$'s, the new top tensor can also be obtained without additional numerical work.
        The dotted green line traces the path of the defect through the coarse-graining.
        Note how at every layer of the MERA the defect is situated between two sites.
        We could have chosen the defect to take different paths through the network by choosing
        slightly different coarse-graining schemes (see Appendix~\ref{app:tnr} for details on the
        scheme we use).
        Shown in pale blue is the causal cone of the site that is situated next to the defect.
        The dotted green line of the defect follows the edge of this causal cone.
        This is a special case of the more general situation where a defect may affect the tensors
        within, and only within, its causal cone.
        We shall encounter the more general situation in the next section, specifically
        in Fig.~\ref{fig:MERA_threelayer_sigma}.
    }\label{fig:MERA_threelayer_epsilon}
\end{figure}

By using TNR to coarse-grain the tensor network for $Z_{D_{\epsilon}}$ we get a MERA for the ground
state of the quantum model with a $D_\epsilon$ defect.
As shown in Fig.~\ref{fig:MERA_threelayer_epsilon} it is like the defectless MERA, but with a
spin-flip matrix $V^{(\iters)}$ at every layer and a different top tensor.
Note that this is a special case of an impurity MERA, discussed in
Ref.~\onlinecite{evenbly_algorithms_2014}.

\section{Duality defect $D_{\sigma}$}
\label{sec:D_sigma}
In this section, we first review the realization, on a quantum spin chain, of the Kramers-Wannier
duality defect $D_\sigma$ of the Ising CFT, and how one can move the defect with a local unitary
operator.
We then construct a tensor network representation of the duality defect $D_{\sigma}$ for the
classical partition function.
We discuss how the defect can be coarse-grained following a strategy that is common to any type of
line defect since, in contrast to the symmetry defect $D_{\epsilon}$, we are not able to
incorporate the duality defect $D_{\sigma}$ into the bond index.
Finally, we present numerical results for the scaling dimensions and conformal spins
$\{\Delta_{\alpha}, s_{\alpha}\}_{D_{\sigma}}$ of the scaling operators in the twisted partition
function $Z_{D_{\sigma}}$.

\subsection{Lattice representation}
The duality defect $D_\sigma$ is related to the Kramers-Wannier self-duality of the critical Ising
model in a manner similar to how the symmetry defect $D_\epsilon$ is related to the $\Integers_2$
symmetry of the model.
To the best of our knowledge, its explicit realization on the classical partition function was not
known (although Ref.~\onlinecite{chui_integrable_2001} reports related work).
For the quantum spin chain, the $D_\sigma$ defect is realized in the
Hamiltonian~\cite{grimm_spin12_1993}
\begin{IEEEeqnarray}{c}
    H_{D_\sigma} = -\left(\sum_{i=1}^{\hs-1} \sigma_i^z \sigma_{i+1}^z
                          + \sum_{i=1}^{\hs-1} \sigma_i^x
                          + \sigma_\hs^y \sigma_1^z \right).
\end{IEEEeqnarray}
Note that, in addition to the new term involving $\sigma^y$, the one-site term $\sigma^x$ is
missing from the $\hs$th site.
We say that in this Hamiltonian the defect is on site $\hs$.

As before for the $D_\epsilon$ defect, we need a way to move the $D_\sigma$ defect from one site to
the next.
The quantum Ising model can be mapped to a theory of free Majorana fermions with a Jordan-Wigner
transformation.
In the Majorana fermion picture the $D_\sigma$ defect corresponds to one fermion missing from the
chain.
There it is then clear what moving the defect means.
Translating this back to the spin chain language gives the two-site unitary
operator~\cite{grimm_spin12_1993} (acting here on sites $1$ and $\hs$)
\begin{IEEEeqnarray}{c}
    \label{eq:U_sigma}
    U_{D_\sigma} =
    \left[ \left(R_z^{\frac{\pi}{4}} \right)_\hs
    \otimes
    \left( R_y^{\frac{\pi}{4}} R_x^{\frac{\pi}{4}} \right)_1 \right]
    \text{CZ}_{1,\hs}.
\end{IEEEeqnarray}
Here $R^\alpha_a = e^{i \alpha \sigma^a} = \unity \cos(\alpha) + i \sigma^a \sin(\alpha)$ with
$\sigma^a$'s being the Pauli matrices and $\text{CZ}_{1,\hs}$ is a controlled-$Z$ gate
$\ket{0}\bra{0}_\hs \otimes \unity_1 + \ket{1}\bra{1}_\hs \otimes \sigma^z_1$.
Which site is considered the control qubit for $\text{CZ}$ does not matter because $\text{CZ}$ is
symmetric under swapping of the two sites.
$U_{D_\sigma}$ moves the defect in the sense that
\begin{IEEEeqnarray}{c}
    U_{D_\sigma} H_{D_\sigma} U_{D_\sigma}\dg =
    -\left(\sum_{i=2}^{\hs} \sigma_i^z \sigma_{i+1}^z
                          + \sum_{i=2}^{\hs} \sigma_i^x
                          + \sigma_1^y \sigma_2^z \right),
\end{IEEEeqnarray}
which is like $H_{D_\sigma}$ but with the defect now on site $1$.
We will be referring to $U_{D_\sigma}$, $H_{D_\sigma}$ and $U_{D_\sigma}\dg$ without mentioning
which site the defect is on and which sites $U_{D_\sigma}$ operates on as this should be clear from
the context.

Next, we briefly investigate how the fusion rules come about in the quantum Hamiltonian.
We take a Hamiltonian that has a $D_\epsilon$ defect on one site and a $D_\sigma$ defect on
another.
By moving either one (or both) of the defects by conjugating with $U_{D_\sigma}$ or $V$ we can
bring both of the two defects to site $\hs$.
The resulting Hamiltonian is
\begin{IEEEeqnarray}{c}
    H_{D_\epsilon \times D_\sigma} = -\left(\sum_{i=1}^{\hs-1} \sigma_i^z \sigma_{i+1}^z
                          + \sum_{i=1}^{\hs-1} \sigma_i^x
                          - \sigma_\hs^y \sigma_1^z \right).
\end{IEEEeqnarray}
This is related to $H_{D_\sigma}$ by conjugation with the unitary $\sigma_\hs^z$.
Therefore it is the same defect $D_{\sigma}$ up to a local change of basis, thus demonstrating the
fusion rule $D_\epsilon \times D_\sigma = D_\sigma$.

Next consider a Hamiltonian with two $D_\sigma$ defects on different sites, such as
\begin{IEEEeqnarray}{c}
    -\left(\sum_{i\neq 1,4} \sigma_i^z \sigma_{i+1}^z + \sum_{i \neq 1,4} \sigma_i^x
                          + \sigma_1^y \sigma_2^z + \sigma_4^y \sigma_5^z\right).
\end{IEEEeqnarray}
If we use conjugation by $U_{D_\sigma}$ to bring the defect on site $1$ to site $4$ we get the
Hamiltonian
\begin{IEEEeqnarray}{c}
    \label{eq:H_sigmatimessigma}
    H_{D_\sigma \times D_\sigma} = -\left(\sum_{i \neq 3,4} \sigma_i^z \sigma_{i+1}^z
                          + \sum_{i \neq 4} \sigma_i^x
                          + \sigma_3^z \sigma_4^x \sigma_5^z \right).
\end{IEEEeqnarray}
This is like the usual Ising Hamiltonian of $\hs-1$ spins but now with an extra spin (the $4$th
one) that is otherwise decoupled, but controls the coupling between spins $3$ and $5$.
$H_{D_\sigma \times D_\sigma}$ is invariant under $\sigma_4^x H_{D_\sigma \times D_\sigma}
\sigma_4^x$ and we can decompose it into two sectors according to the parity of the $4$th spin.
In the $+1$ sector the coupling between spins $3$ and $5$ is ferromagnetic and we have the usual
Ising chain of $\hs-1$ spins.
In the parity $-1$ sector the coupling is antiferromagnetic and we get the Hamiltonian for the
Ising model with $D_\epsilon$ defect.
This is the sense in which $D_\sigma \times D_\sigma = D_\unity + D_\epsilon$ in the quantum
Ising model.

\subsection{Tensor network representation}
\begin{figure}[tbp]
	\centering
	\includegraphics[width=1.0\linewidth]{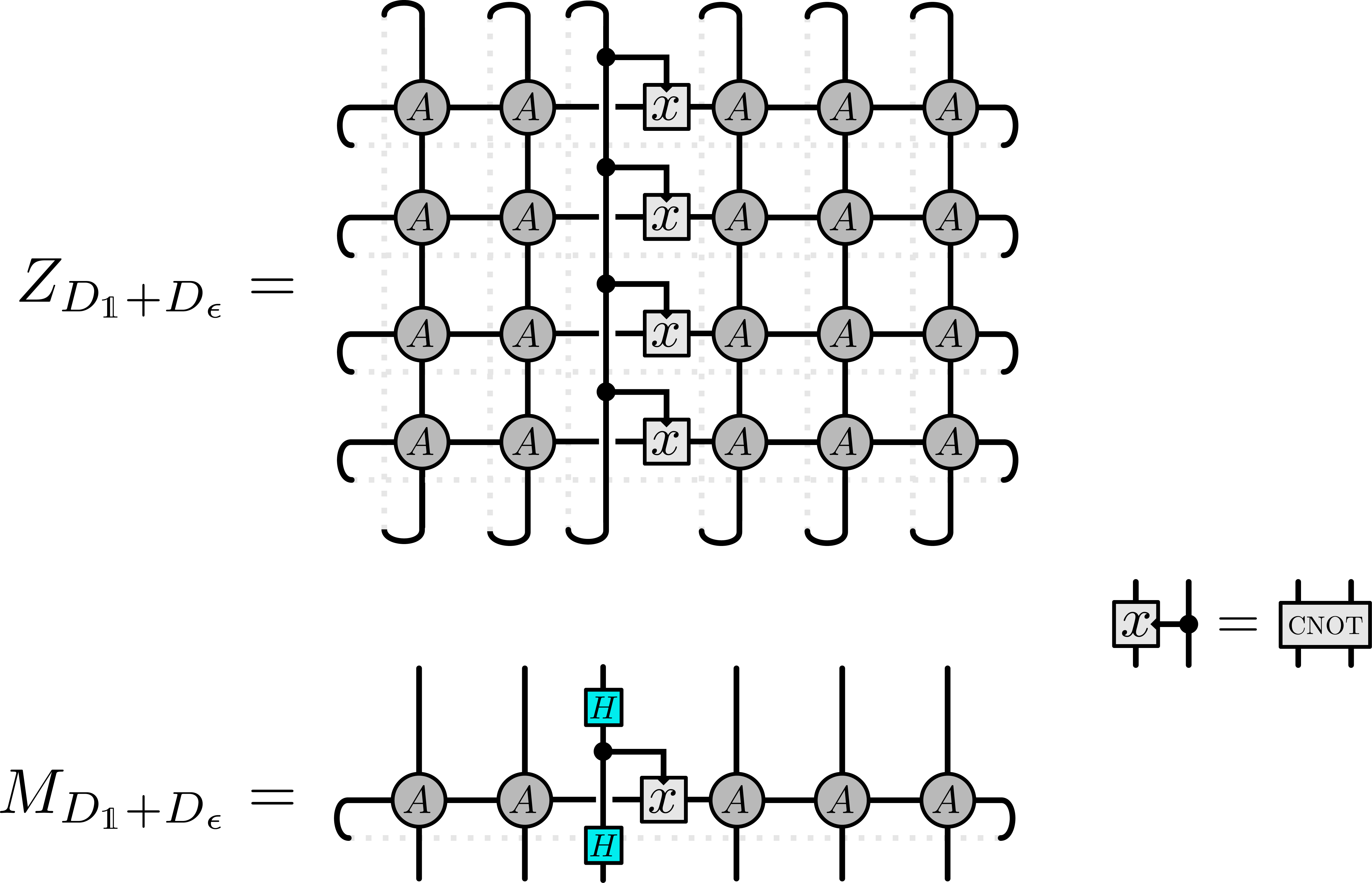}
	\caption{%
        The partition function $Z_{D_\unity + D_\epsilon}$ and its transfer matrix $\trm_{D_\unity
        + D_\epsilon}$ that is the classical equivalent of $H_{D_\sigma \times D_\sigma}$.
        $\text{CNOT}$ is a controlled-NOT gate $\ket{0}\bra{0} \otimes \unity + \ket{1}\bra{1}
        \otimes \sigma^x$ with the dot marking the control qubit.
        $H = \frac{1}{2}(\sigma^x + \sigma^z)$ is the Hadamard matrix that transforms between the
        spin basis and the parity eigenbasis.
        Because $H^2 = \unity$ the presence of the Hadamards in the transfer matrix does not affect
        the partition function but it ensures the transfer matrix is spin-flip invariant.
    }\label{fig:Z_and_T_D_unity_plus_D_eps}
\end{figure}

\begin{figure*}[htbp!]
	\centering
	\includegraphics[width=1.0\linewidth]{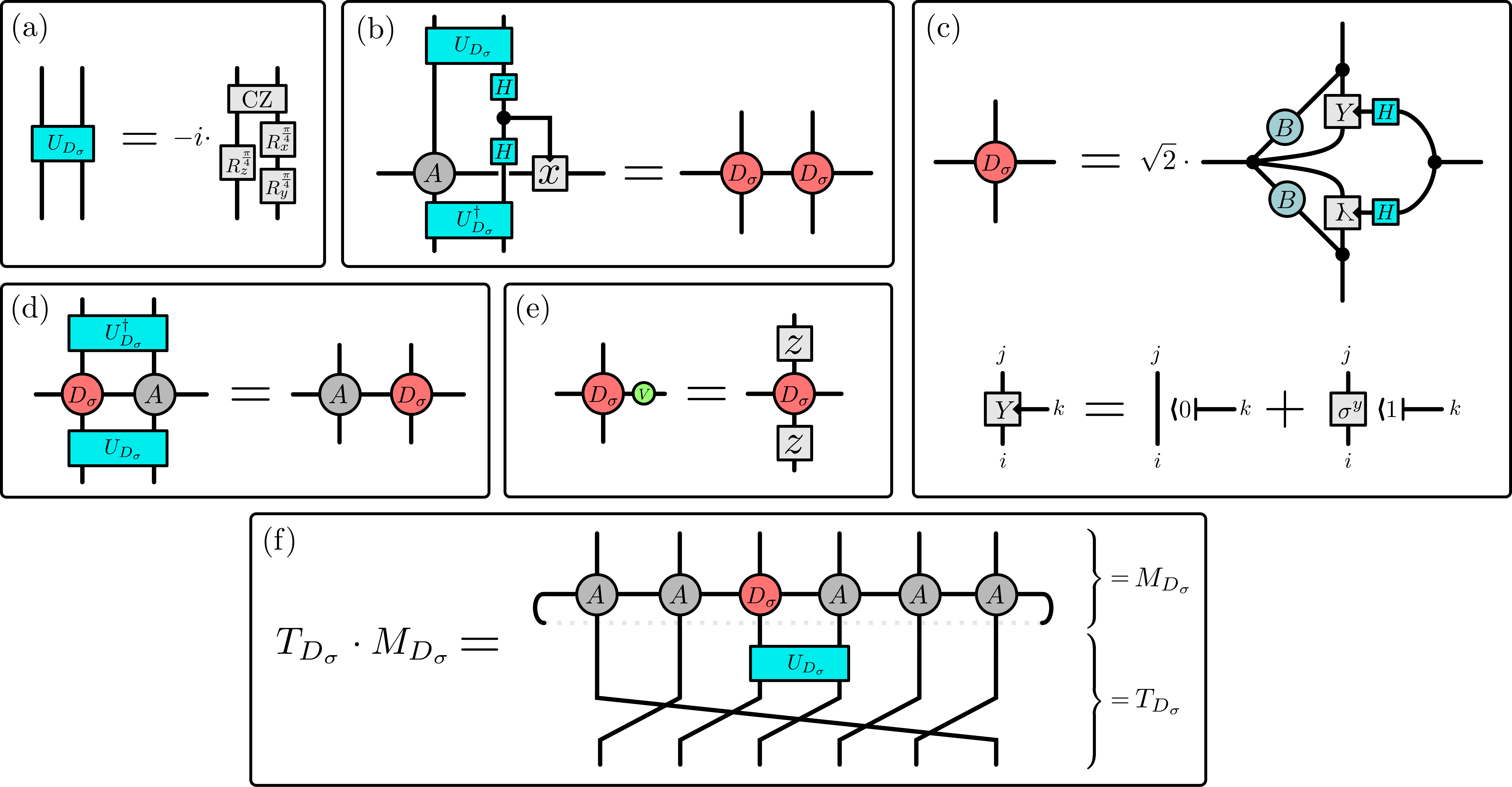}
    \caption[]{
        (a)
        The unitary that moves the $D_\sigma$ defect by one site.
        Here $R^\alpha_a = e^{i \alpha \sigma^a} = \unity \cos(\alpha) + i \sigma^a
        \sin(\alpha)$ with $\sigma^a$'s being the Pauli matrices.
        $\text{CZ}$ denotes a controlled-$Z$ gate $\ket{0}\bra{0} \otimes \unity + \ket{1}\bra{1}
        \otimes \sigma^z$.
        Compared to the $U_{D_\sigma}$ in Eq.~\eqref{eq:U_sigma} this one has an additional
        factor of $-i$.
        We are in fact free to multiply $U_{D_\sigma}$ with an arbitrary phase, because to move a
        defect we multiply it with $U_{D_\sigma}$ and $U_{D_\sigma}\dg$ and the phases cancel.
        Because of this phase freedom in $U_{D_\sigma}$, when extracting the conformal spins from
        the eigenvalues of the translation operator [see
        Eq.~\eqref{eq:cardys_formula_momentum_no_f}] the conformal spins are only determined up
        to an additive constant.
        The phase has been chosen here so that the conformal spins come out correctly, which relies
        on us knowing at least one of the exact conformal spins of the operators in
        $Z_{D_{\sigma}}$.
        (b)
        Taking the tensor network equivalent of having two $D_\sigma$ defects on the same site and
        moving one of them away with $U_{D_\sigma}$ yields two copies of the same tensor contracted
        with each other.
        We call this tensor $D_\sigma$ and identify it as the tensor representing a single duality
        defect.
        Its explicit form is shown in~(c).
        (c)
        The $D_\sigma$ tensor obtained using the procedure in~(b).
        $B_{ij} = e^{\beta \sigma_i \sigma_j}$ are the Boltzmann weights of the Ising model, $H$
        is the Hadamard operator $H = \frac{1}{\sqrt{2}}(\sigma^z + \sigma^x)$, and the dots are
        Kronecker $\delta$'s that fix all their indices to have the same value.
        Note how the tensor marked with a $Y$, defined at the bottom as $\unity \otimes \bra{0} +
        \sigma^y \otimes \bra{1}$, is almost like a controlled-$Y$ gate.
        \raisebox{\depth}{\scalebox{1}[-1]{$Y$}} denotes the same tensor but with the indices $i$
        and $j$ transposed.
        (d)
        The $D_\sigma$ defect as represented by the tensor in~(c) can be unitarily moved by
        conjugating with $U_{D_\sigma}$.
        (e)
        The fusion rule $D_\sigma \times D_\epsilon = D_\sigma$ as it manifests in the defect
        tensors.
        $z$ denotes the $\sigma^z$ Pauli matrix.
        (f)
        $\trm_{D_\sigma}$ composed with the translation operator $\tro_{D_\sigma}$ that commutes
        with it.
    }\label{fig:D_sigma_collection}
\end{figure*}

Since we do not know how to represent the duality defect $D_\sigma$ in the classical Ising model, a
priori we also do not know how to insert it into a tensor network $\znet_{\hs,\vs}(A)$ [see
Fig.~\ref{fig:Z_and_T_collection}(a)].
However, because of the duality between the classical transfer matrix and the quantum Hamiltonian,
one would expect that if we had a transfer matrix $\trm_{D_\sigma}$ with a $D_\sigma$ line defect
looping through it, conjugating $\trm_{D_\sigma}$ with the unitary $U_{D_\sigma}$ would move the
defect.
We can make use of this intuition and the fusion rule $D_\sigma \times D_\sigma = D_\unity +
D_\epsilon$ to construct a tensor for $D_\sigma$.
As a reminder, in Fig.~\ref{fig:D_sigma_collection}(a) we have written the operator
$U_{D_\sigma}$ using the tensor network notation.
In the quantum Hamiltonian $H_{D_\sigma \times D_\sigma}$, having two duality defects $D_\sigma$ on
site $k$ appeared as the term $\sigma_{k-1}^z \sigma_k^x \sigma_{k+1}^z$.
The tensor network analog of this is an auxiliary spin in the network that, when written in the
parity eigenbasis, controls the coupling along a defect line.
The partition function $Z_{D_\unity + D_\epsilon}$ and the transfer matrix $\trm_{D_\unity +
D_\epsilon}$ that implement such defect are shown in Fig.~\ref{fig:Z_and_T_D_unity_plus_D_eps}.

We now have a tensor representing two duality defects $D_\sigma$ on the same site: the CNOT with
the two Hadamard gates.
We can move one of the defects away by conjugating with $U_{D_\sigma}$, thus obtaining the
tensor for a single $D_\sigma$ defect, as shown in Fig.~\ref{fig:D_sigma_collection}(b).
The explicit form of the tensor we obtain is shown in Fig.~\ref{fig:D_sigma_collection}(c).
We have omitted the calculation deriving this form as it is a long and uninformative exercise in
using the properties of Pauli matrices.

The form of the $D_\sigma$ tensor shown in Fig.~\ref{fig:D_sigma_collection}(c) is physically
intuitive though:
The Kramers-Wannier duality maps the Ising model to an equivalent model on the dual lattice (with a
degree of freedom in every plaquette).
Recall how the $D_\epsilon$ defect can be seen as separating two parts of the system, one
of which has been transformed with a spin-flip and the other one has not.
Similarly the $D_\sigma$ defect is the boundary separating a part of the system that has been
mapped with the Kramers-Wannier duality from the part that has not.
At this boundary we would expect spins on the dual lattice side [in
Fig.~\ref{fig:D_sigma_collection}(c) on the right] to represent domain walls between spins of
the original lattice [left, top and bottom in Fig.~\ref{fig:D_sigma_collection}(c)].
The tensor $Y_{ijk} = \unity_{ij} \, \delta_{0,k} + \sigma^y_{ij} \, \delta_{1,k}$ does exactly
that:
$Y_{ijk} \neq 0$ if and only if $k=0$ and $i=j$ (no domain wall between spins $i$ and $j$) or
$k=1$ and $i \neq j$ (a domain wall between spins $i$ and $j$).
The index $k$ of $Y_{ijk}$ (marked in the figure with a small arrowhead) thus directly represents a
domain wall and it is related to the free index on the right by a simple Hadamard rotation.
The $B$ matrices meanwhile provide the usual Ising couplings between the spins of the original
lattice.

The tensor for the duality defect $D_\sigma$ fulfills the property that it moves by one lattice
site under conjugation by $U_{D_\sigma}$, as illustrated in
Fig.~\ref{fig:D_sigma_collection}(d).
We can also observe the fusion rule $D_\sigma \times D_\epsilon = D_\sigma$ by multiplying the
$D_\sigma$ tensor with $V$ as in Fig.~\ref{fig:D_sigma_collection}(e).
The result is the same tensor $D_\sigma$ multiplied by two Pauli matrices $\sigma^z$ from above and
below.
This represents the same defect, because the Pauli matrices only provide a local change of basis.

Thus we propose the transfer matrix $\trm_{D_\sigma}$ for the twisted partition function
$Z_{D_{\sigma}}$ to be the one in Fig.~\ref{fig:D_sigma_collection}(f).
The validity of this this choice is ultimately confirmed by the numerical results shown below.
As with the symmetry defect $D_\epsilon$, the usual lattice translation $\tro$ moves the duality
defect $D_{\sigma}$ and we need to build a generalized translation operator $\tro_{D_\sigma} =
U_{D_\sigma} \tro$ that commutes with $\trm_{D_\sigma}$.
This, too, is shown in Fig.~\ref{fig:D_sigma_collection}(f).

\begin{figure}[htbp]
    \centering
    \includegraphics[width=1.0\linewidth]{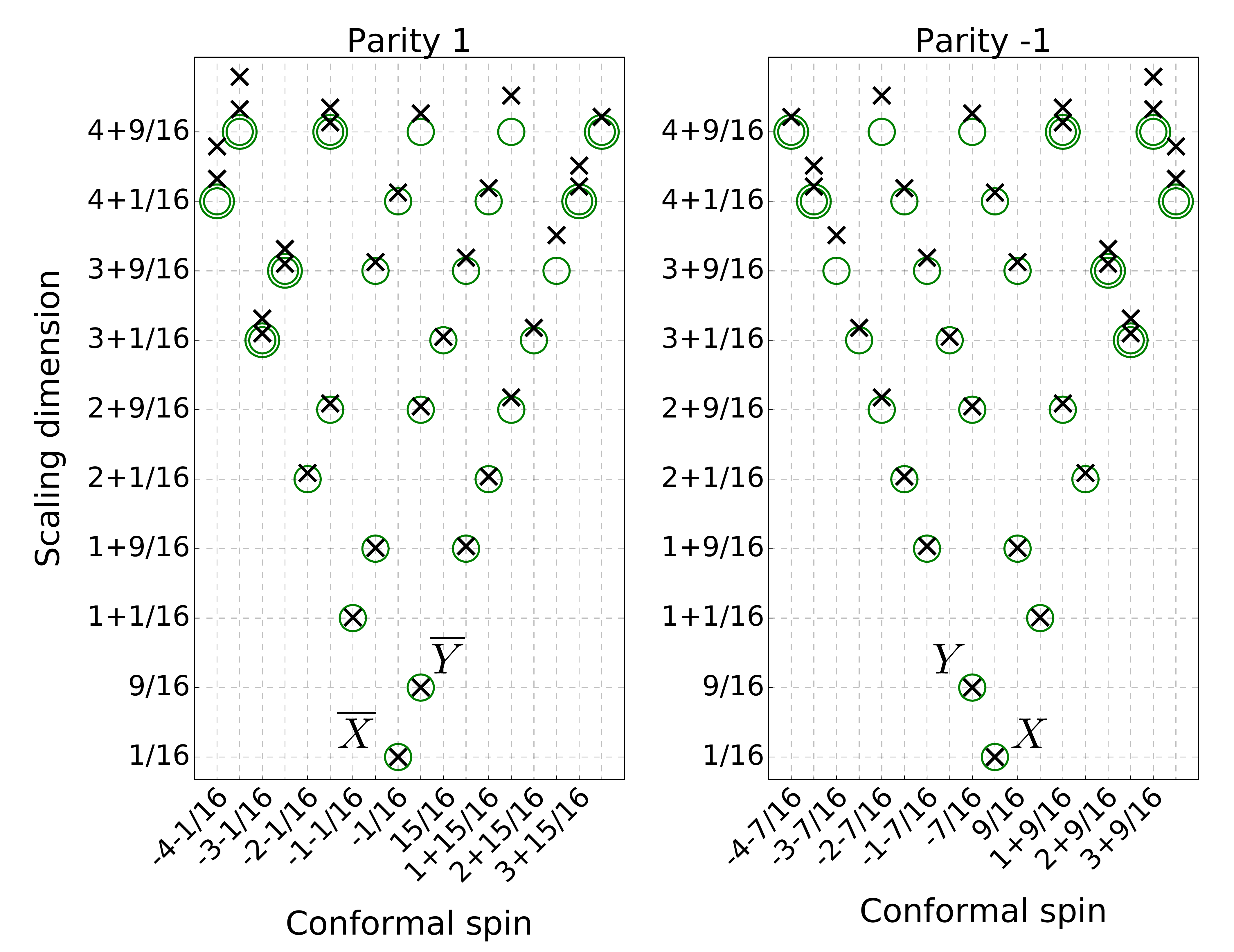}
    \caption{%
        The scaling dimensions (vertical axis) and conformal spins (horizontal axis) of the first
        scaling operators in $Z_{D_{\sigma}}$ obtained from exact diagonalization of a transfer
        matrix of $\hs=18$ sites.
        The scaling operators are again divided by their $\Integers_2$ charge.
        The crosses mark the numerical values that can be compared with the circles that are
        centered at the exact values.
        Several concentric circles denote the degeneracy $N_\alpha$ of that $(\Delta_\alpha,
        s_\alpha)$ pair.
    }\label{fig:ed_KW_results}
\end{figure}

In Fig.~\ref{fig:ed_KW_results}, we show scaling dimensions and conformal spins
$\{\Delta_{\alpha}, s_{\alpha}\}_{D_{\sigma}}$ for the operators in $Z_{D_{\sigma}}$, obtained by
diagonalizing $\tro_{D_\sigma} \cdot \trm_{D_\sigma}$ for 18 sites.
The results reproduce the expected conformal towers, confirming that our choice of tensor
$D_\sigma$ indeed represents the duality defect.
Again, the accuracy of the numerical estimates quickly deteriorates with increasing scaling
dimensions.
Next, we will discuss how to coarse-grain in the presence of a $D_\sigma$ defect, which will let us
reach larger system sizes and more accurate results.

\subsection{Coarse-graining}
Coarse-graining the symmetry defect $D_\epsilon$ was particularly simple because the global
$\Integers_2$ symmetry is explicitly realized in the individual tensors of the tensor network.
This is no longer the case for the Kramers-Wannier self-duality.
As a result, we need to coarse-grain the tensors for $D_\sigma$ as we would coarse-grain any other
line of impurity tensors (representing a generic conformal defect), that is, without being able to
exploit that they correspond to a topological defect.
For a sufficiently local coarse-graining scheme, such as TNR~\cite{evenbly_tensor_2015-2} (but also
TRG~\cite{levin_tensor_2007} and its generalizations, see Sec.~\ref{sec:discussion}), a line
defect is coarse-grained into a line defect at the next scale.
The details of how we coarse-grain the line defect using TNR are explained in
Appendix~\ref{app:tnr}.
The same appendix also shows how the translation operator $\tro_{D_\sigma}$ is coarse-grained in
this process.

\begin{figure}[htbp]
	\centering
	\includegraphics[width=1.0\linewidth]{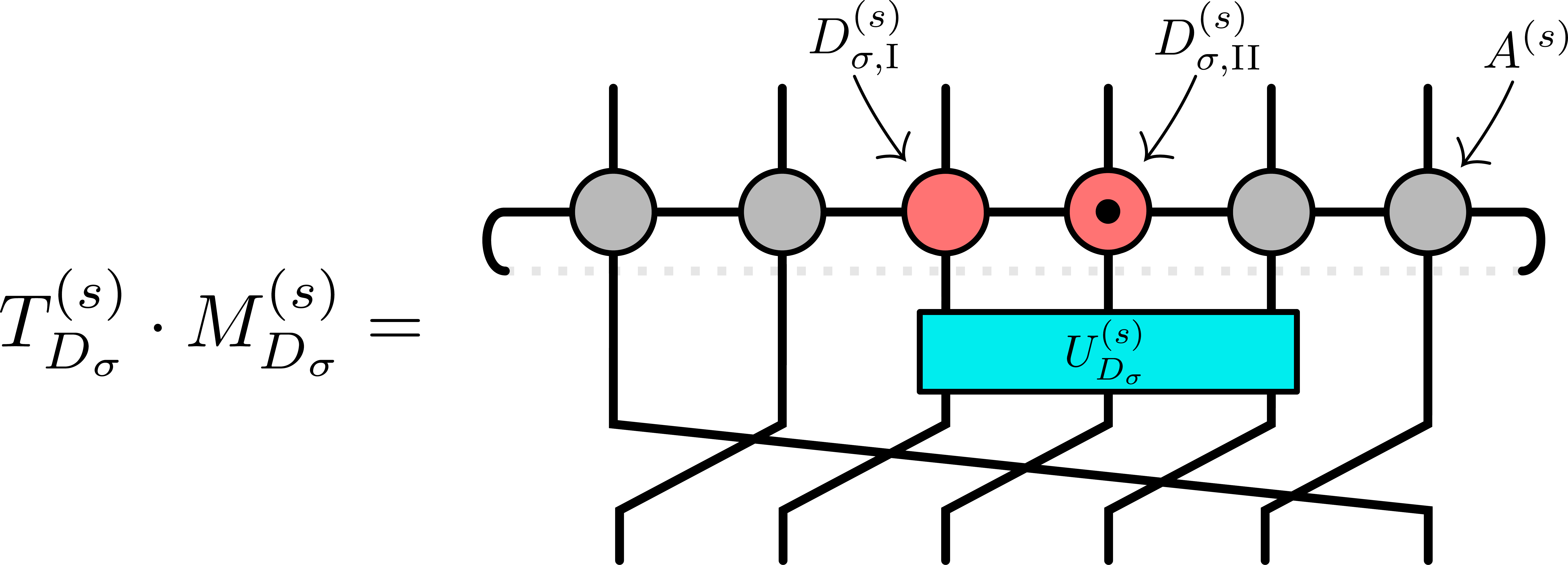}
	\caption{%
        The coarse-grained transfer matrix for the $D_\sigma$ defect composed with the generalized
        translation $\tro_{D_\sigma}^{(\iters)}$.
    }\label{fig:cT_T_sigma_m}
\end{figure}

Under TNR, the duality defect $D_\sigma$ is coarse-grained into a line defect with a width of two
tensors.
The coarse-grained transfer matrix $\trm_{D_\sigma}^{(\iters)}$ and translation operator
$\tro_{D_\sigma}^{(\iters)}$ are thus as shown in Fig.~\ref{fig:cT_T_sigma_m}.
The operator $\tro_{D_\sigma}^{(\iters)} \cdot \trm_{D_\sigma}^{(\iters)}$ can then be exactly
diagonalized and its eigenvalue spectrum yields the scaling dimensions and conformal spins
$\{\Delta_{\alpha}, s_{\alpha}\}_{D_{\sigma}}$ of the scaling operators with lowest dimensions in
$Z_{D_{\sigma}}$.
There is, however, a small technical subtlety in how to extract the conformal data from the
spectrum that is discussed in Appendix~\ref{app:Z_sigma_counting}.

\subsection{Numerical results}
\begin{figure}[htbp]
	\centering
	\includegraphics[width=1.0\linewidth]{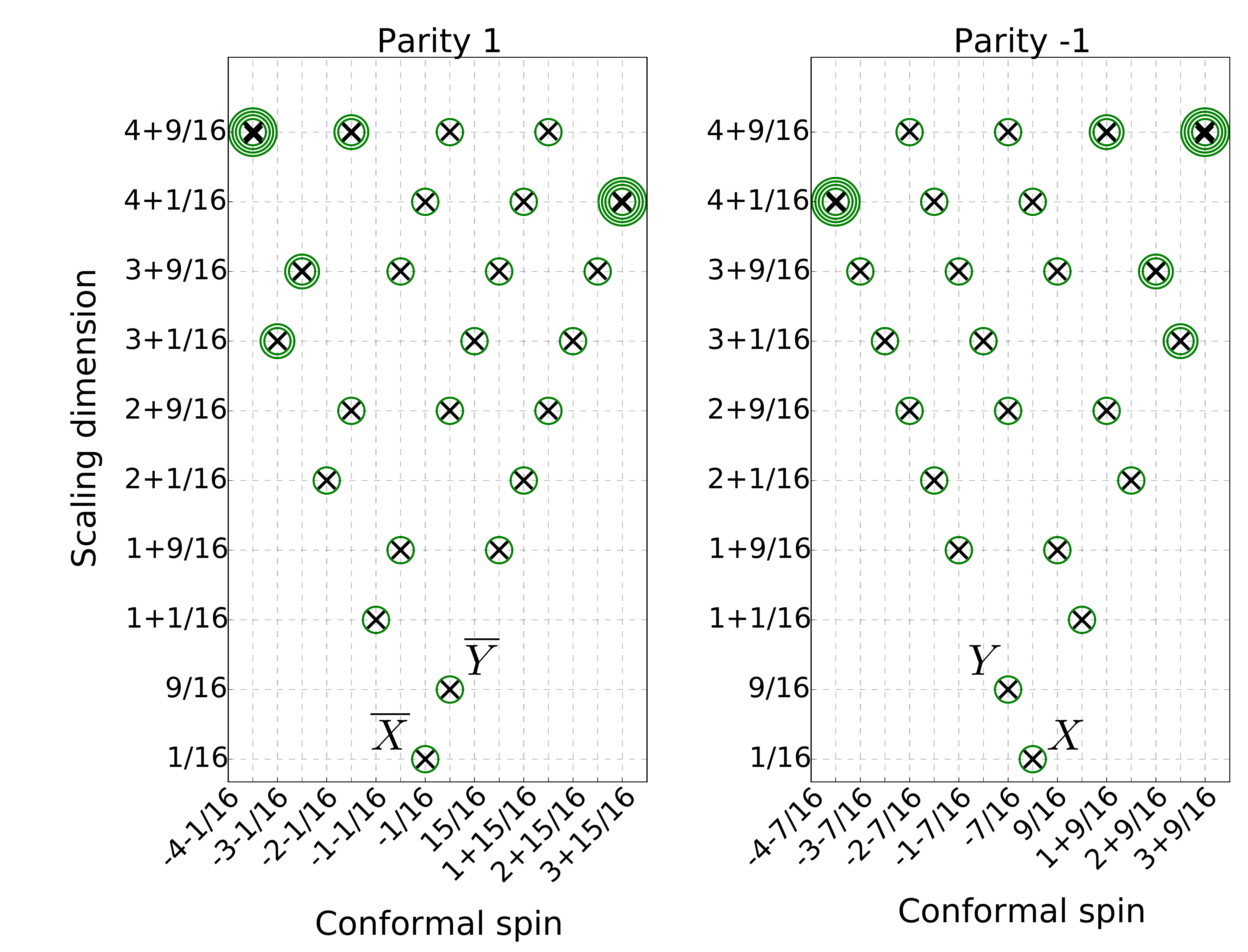}
	\caption{%
        The scaling dimensions (vertical axis) and conformal spins (horizontal axis) of the first
        scaling operators of the two-dimensional classical Ising model with a $D_\sigma$ defect as
        obtained with TNR\@.
        The crosses mark the numerical values that can be compared with the circles that are
        centered at the exact values.
        Several concentric circles denote the degeneracy $N_\alpha$ of that $(\Delta_\alpha,
        s_\alpha)$ pair.
        The keen-eyed reader may notice that high up in the conformal towers some of the momenta
        differ from those shown in Fig.~\ref{fig:ed_KW_results}.
        This is because the periodicity of the momenta in these results is lower, so that for
        instance the conformal spins $-4+\frac{7}{16}$ and $4+\frac{7}{16}$ are indistinguishable.
    }\label{fig:D_sigma_results}
\end{figure}

As discussed earlier, the primaries present in $Z_{D_{\sigma}}$ are the ones with
the conformal dimensions $(\frac{1}{2}, \frac{1}{16})$, $(0, \frac{1}{16})$, $(\frac{1}{16},
\frac{1}{2})$ and $(\frac{1}{16}, 0)$.
The first two have parity $+1$, the latter two $-1$.
Numerical results for the scaling dimensions and conformal spins of these primaries obtained with
the TNR method are shown in Table~\ref{tab:primary_data_sigma}.
Similar values for some of the first descendants are shown in Fig.~\ref{fig:D_sigma_results}.
The results are again in excellent agreement with the exact results even higher up in the conformal
towers.
These results were obtained by coarse-graining a transfer matrix of $2^5 \times (4 \times 2^5)$
$A^{(0)}$ tensors using TNR with bond dimensions $\chi = 22$ and $\chi' = 11$.
Again, a slightly larger system was used for obtaining the conformal spins, see
Appendix~\ref{app:coarse_momenta}.

\begin{table}[htbp]
    \begin{tabular}{cccccc}
        Primary & $\left(h, \overline{h}\right)$
        & $\Delta_\text{TNR}$ & $\Delta_\text{exact}$
        & $s_\text{TNR}$ & $s_\text{exact}$\\%
        \midrule

        $X$ & $\left(\sfrac{1}{16},0\right)$
        & $0.0626656$ & $0.0625$ & $\phantom{+}0.0624974$ & $\phantom{+}0.0625$\\%
        $\overline{X}$ & $\left(0,\sfrac{1}{16}\right)$
          & $0.0626656$ & $0.0625$ & $-0.0624974$ & $-0.0625$\\%
        $Y$ & $\left(\sfrac{1}{16},\sfrac{1}{2}\right)$
          & $0.5627685$ & $0.5625$ & $-0.4374828$ & $-0.4375$\\%
        $\overline{Y}$ & $\left(\sfrac{1}{2},\sfrac{1}{16}\right)$
          & $0.5627685$ & $0.5625$ & $\phantom{+}0.4374828$ & $\phantom{+}0.4375$

    \end{tabular}
    \caption{%
        The scaling dimensions $\Delta$ and conformal spins $s$ for the primaries of
        $Z_{D_{\sigma}}$ as obtained with TNR compared with the exact values.
        Note that, as with $D_\epsilon$, one can analytically deduce the possible values of the
        conformal spins in the two parity sectors by observing that
        $\left(\tro_{D_\sigma}\right)^{2n-1} = \frac{1}{\sqrt{2}}\left(\unity + i \Sigma^x
        \right)$, where $\Sigma^x$ is the global spin-flip operator.
        However, we choose to present the numerical values for the conformal spins, including the
        small numerical errors, to demonstrate the accuracy of our method.
    }\label{tab:primary_data_sigma}
\end{table}

\subsection{MERA}
\begin{figure}[tbp]
	\centering
    \includegraphics[width=1.0\linewidth]{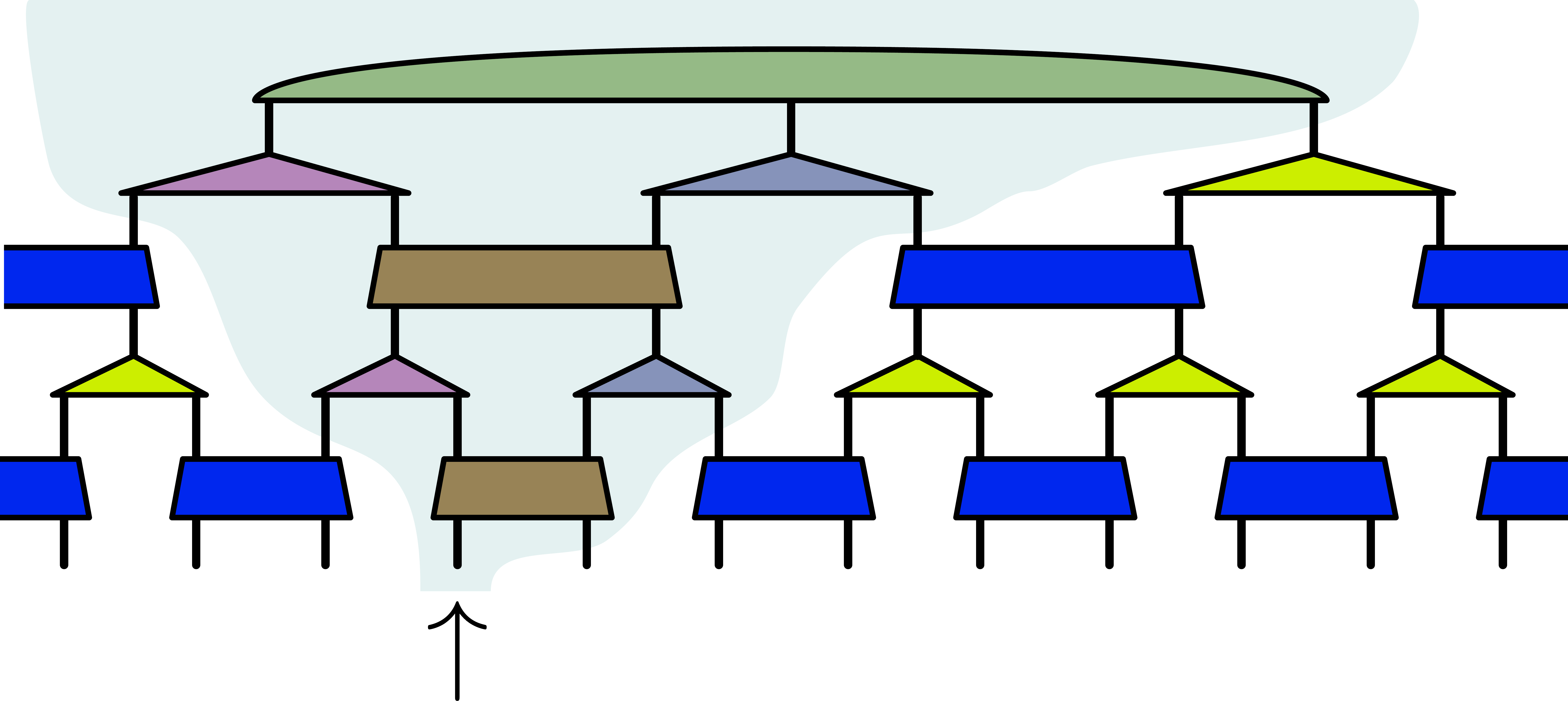}
	\caption{%
        A MERA for the ground state of the quantum spin chain that has a duality defect in it,
        produced by coarse-graining $Z_{D_{\sigma}}$ with TNR\@.
        The tensors in this MERA are the same ones that are used in coarse-graining the $D_\sigma$
        defect in Fig.~\ref{fig:TNR_sigma}.
        Even though our notation does not reflect it, the unitaries and isometries on different
        layers generally differ from each other.
        The arrow marks the position of the defect and the pale blue region is the causal cone of
        the defect site.
    }\label{fig:MERA_threelayer_sigma}
\end{figure}

As was the case with the $D_\unity$ and $D_\epsilon$ defects, coarse-graining the network for
$Z_{D_{\sigma}}$ with TNR produces a MERA for the ground state of the spin chain with $D_\sigma$
defect.
This MERA is shown in Fig.~\ref{fig:MERA_threelayer_sigma}.
As the figure shows, only the tensors within the causal cone of the defect are different from the
ones in the $D_\unity$ MERA in Fig.~\ref{fig:MERA_threelayer}.
Such a MERA is known as an impurity MERA~\cite{evenbly_algorithms_2014}.
It should be noted that unlike for a general impurity MERA the impurity tensors in
Fig.~\ref{fig:MERA_threelayer_sigma} can be moved with the unitary operator $U_{D_\sigma}$.

\section{Generic conformal defects}
\label{sec:generic}
This paper is devoted to the study of topological conformal defects using tensor network
techniques.
As discussed in the introduction, the main difference between a generic (i.e.\ non-topological)
conformal defect and a topological conformal defect is that in the latter case there is a local
unitary transformation that moves the location of the defect.
This allowed us to define a generalized translation operator $\tro_D$ whose eigenvalues yield the
conformal spins $s_{\alpha}$ associated with the defect $D$.

For a non-topological defect, the absence of a local unitary transformation that moves the defect
implies that we can no longer define the translation operator $\tro_D$, and therefore we cannot
extract conformal spins.
However, we can still build a tensor network representation of a non-topological defect $D$, and
thus of the corresponding partition function $Z_D$ on a torus and its transfer matrix $\trm_D$.
In addition, we can still coarse-grain and diagonalize the transfer matrix $\trm_D$ corresponding
to a number $n$ of sites much larger than what is accessible with exact diagonalization.
The spectrum of $\trm_D$ for a non-topological defect is no longer given by
Eq.~\eqref{eq:cardys_formula_no_f}, but it still provides information about the universal
properties of the system.

\subsection{Family of defects for the Ising model}
As a simple example, we consider a continuous family of conformal defects $D_\cdf$ of the
critical Ising model~\cite{oshikawa_boundary_1997,affleck_quantum_2008} where the coupling between
spins across the defect is proportional to a real number $\cdf \in [0,1]$.
The choice $\cdf=0$ corresponds to no coupling across the line and thus to open boundary
conditions, whereas $\cdf=1$ corresponds to periodic boundary conditions, and thus no defect.
Ignoring again finite-size corrections of higher order in $\hs^{-1}$, the eigenvalues
$\lambda_\alpha(\cdf)$ of $\trm_{D_\cdf}$ can be expressed as
\begin{IEEEeqnarray}{c}
    \frac{\lambda_\alpha(\cdf)}{\lambda_0(\cdf)} = e^{-\frac{2\pi}{\hs}\Delta_\alpha(\cdf)}.
\end{IEEEeqnarray}
At $\cdf = 1$, $\Delta_\alpha$ are the scaling dimensions of the Ising CFT\@.
For general $\cdf$, $\Delta_\alpha$ can be predicted analytically using a description in terms of
fermionic operators.
This will be discussed in detail in future work (Ref.~\onlinecite{hauru_inprep01}).
The result is that $\Delta_\alpha$ behave linearly in $\theta = \tan^{-1}\left(
\frac{1-\cdf}{1+\cdf} \right)$.

\subsection{Numerical results}
\begin{figure}[htbp]
	\centering
    \includegraphics[width=1.0\linewidth]{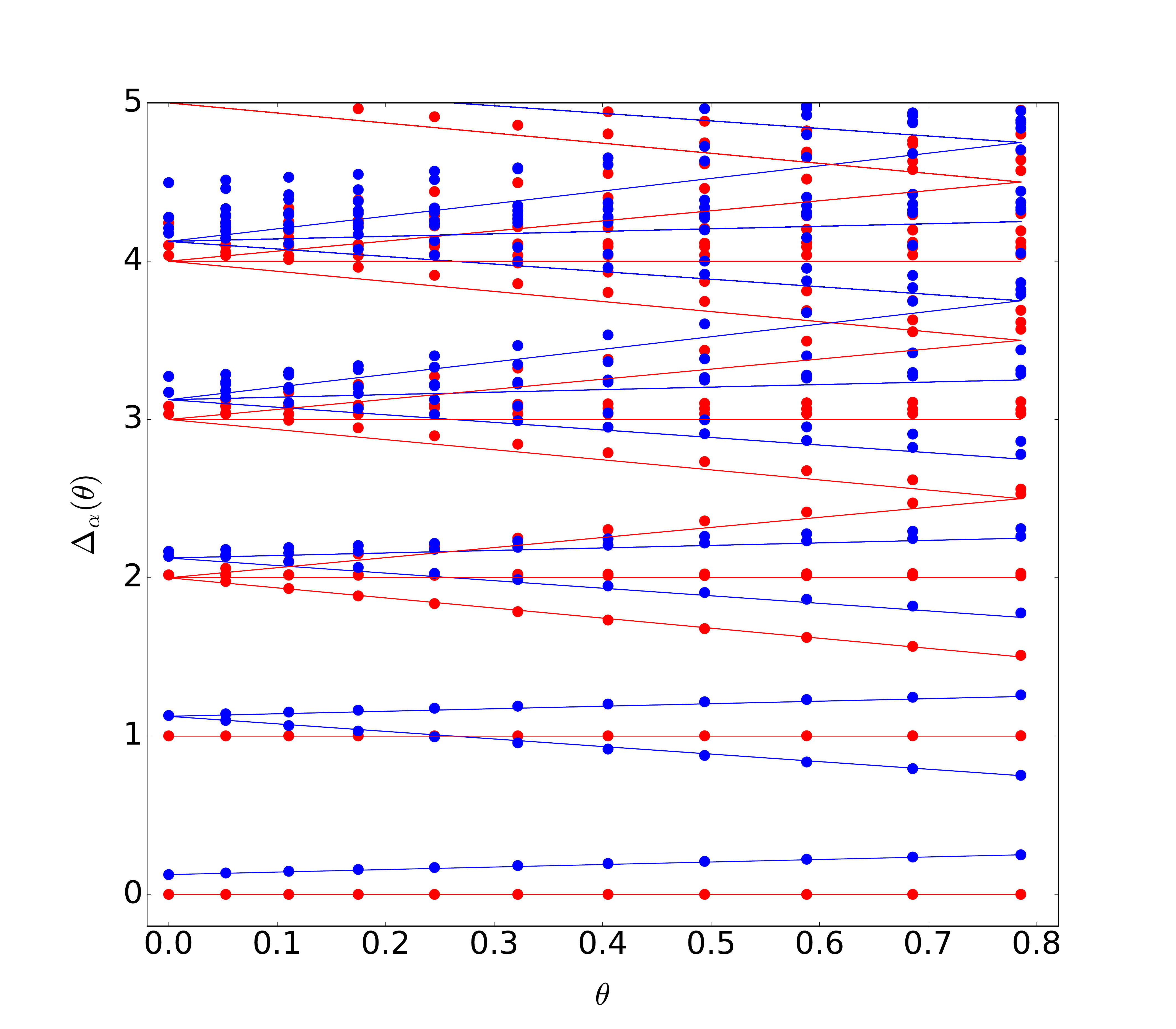}
	\caption{%
        Numerical estimates for
        $\Delta_\alpha(\theta) = -\frac{\hs}{2\pi}\log
        \left( \frac{\lambda_\alpha(\theta)}{\lambda_0(\theta)} \right)$
        obtained using exact diagonalization on a system of size $\hs = 18$.
        Here $\lambda_\alpha(\theta)$ are the eigenvalues of $\trm_{D_\cdf}$ and $\theta =
        \tan^{-1}\left( \frac{1-\cdf}{1+\cdf} \right)$.
        The lines are the analytic predictions coming from free fermion
        calculations~\cite{hauru_inprep01}.
        The blue and red colors mark the parity odd and parity even sectors, respectively.
        At the extreme left at $\theta=0$ is the case of periodic boundary conditions, where
        the $\Delta_\alpha$ are the scaling dimensions of the Ising CFT\@.
        At the extreme right at $\theta=\frac{\pi}{4}$ is the case of open boundary
        conditions.
        The numerical results clearly show qualitative agreement with the analytic values, but
        accuracy deteriorates significantly after the first few $\Delta_\alpha$'s.
    }\label{fig:cdf_ed_results}
\end{figure}

\begin{figure}[htbp]
	\centering
    \includegraphics[width=1.0\linewidth]{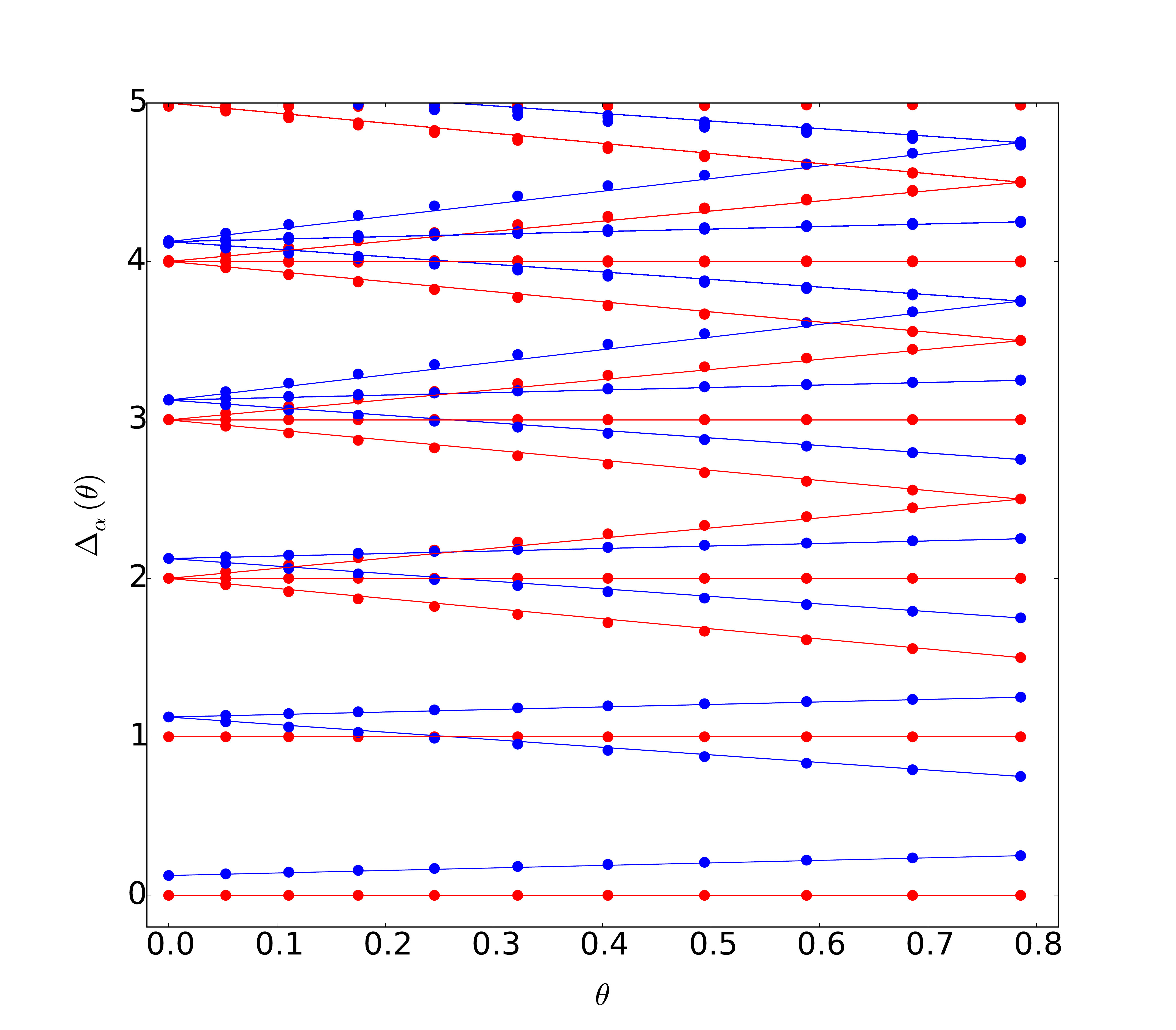}
	\caption{%
        Numerical estimates for $\Delta_\alpha(\theta) = -\frac{\hs}{2\pi}\log \left(
        \frac{\lambda_\alpha(\theta)}{\lambda_0(\theta)} \right)$ obtained by coarse-graining and
        then diagonalizing a transfer matrix for a system of roughly $4\,000$ spins.
        $\lambda_\alpha(\theta)$ are the eigenvalues of $\trm_{D_\cdf}$ and $\theta =
        \tan^{-1}\left( \frac{1-\cdf}{1+\cdf} \right)$.
        As in Fig.~\ref{fig:cdf_ed_results}, $\theta = 0$ corresponds to periodic boundary
        conditions, $\theta = \frac{\pi}{4}$ corresponds to open boundary conditions and the lines
        are the analytic predictions coming from free fermion calculations~\cite{hauru_inprep01}.
        The blue and red colors mark the parity odd and parity even sectors, respectively.
        The numerical and analytic values are seen to agree to a high precision, owing to the
        fact that the coarse-graining has significantly reduced finite-size effects.
    }\label{fig:cdf_results}
\end{figure}

Figure~\ref{fig:cdf_ed_results} shows estimates for $\Delta_\alpha$ as functions of $\theta$,
obtained by diagonalizing $\trm_{D_\cdf}$ for a system of size $\hs=18$, using exact
diagonalization.
The results are compared with the analytic values, which appear as lines.
Figure~\ref{fig:cdf_results} shows similar estimates obtained by diagonalizing a coarse-grained
transfer matrix $\trm_{D_\cdf}^{(\iters)}$, effectively reaching a system size of roughly
$4\,000$ spins.
Here we have used the same coarse-graining strategy as for the duality defect $D_{\sigma}$ in
Sec.~\ref{sec:D_sigma}, with bond dimension $\chi=22$ and $\chi'=11$.
Again, the results obtained using coarse-graining show greater accuracy with respect to the
analytic predictions, but even exact diagonalization gets many of the qualitative features of
the spectrum correct.

\section{Discussion}
\label{sec:discussion}
In this paper, we have explained how to compute accurate numerical estimates of the scaling
dimensions and conformal spins $\{\Delta_{\alpha }, s_{\alpha}\}_{D}$ associated to a topological
conformal defect $D$, or the scaling dimensions $\{\Delta_{\alpha}\}_{D}$ associated to a generic
(i.e., non-topological) conformal defect $D$.
For simplicity, we have focused on the topological conformal defects of the critical Ising model,
namely the symmetry defect $D_{\epsilon}$ and duality defect $D_{\sigma}$, and have briefly
considered also a family of non-topological, conformal defects.
In order to improve significantly on the numerical estimates readily available through exact
diagonalization, we have used a particular coarse-graining transformation: tensor network
renormalization (TNR).
Our numerical results clearly demonstrated that tensor network techniques are a useful tool to
characterize conformal defects in critical lattice models.

We conclude this paper with a short discussion on applying this approach to other lattice models,
on using other coarse-graining schemes for the same purpose, and on an alternative, more direct
route to extracting conformal data from critical lattice models based on building a lattice version
of the scaling operators of the theory.

\subsection{Conformal defects in other models}
The tensor network approach described in this paper can be applied to line defects on any critical
2D classical partition function on the lattice (or point defects on any critical 1D quantum lattice
system).
Let us recall what the requirements of the approach are.

The scaling dimensions $\{\Delta_{\alpha}\}_D$ associated to a conformal defect $D$ can be
extracted from the eigenvalues of the transfer matrix $M_D$ for the partition function $Z_D$ that
includes that defect.
Thus, we only need to have a lattice representation of the line defect $D$, from which to build
$Z_D$ and $M_D$.

In addition, for a topological conformal defect $D$, the associated conformal spins
$\{s_{\alpha}\}_D$ can be extracted from the eigenvalues of a generalized translation operator
$T_D$, built by composing a one-site translation with the application of the local unitary
transformation that moves the defect back to its initial position, as explained in
Secs.~\ref{sec:D_epsilon} and~\ref{sec:D_sigma} for the symmetry defect $D_{\epsilon}$ and
duality defect $D_{\sigma}$ of the critical Ising model.
Thus, in this case we also need to have a lattice representation of the local unitary
transformation that moves the location of the defect $D$.

Symmetry defects, associated to a global internal symmetry group $\mathcal{G}$, are a type of
topological conformal defect that is particularly easy to deal with within the tensor network
formalism.
Indeed, as we did in the case of the $\mathbb{Z}_2$ group of the Ising model, for each group
element $g\in \mathcal{G}$ we can create a line defect $D_g$ as a form of twisted boundary
conditions by inserting copies of a unitary representation $V_g$ along a vertical line of bond
indices.
This is remarkably simple when using $\mathcal{G}$-symmetric tensors to represent the tensor
network.
In this case, each tensor index is labeled by irreducible representations of $\mathcal{G}$, and
$V_g$ acts diagonally on the irreducible representations by placing a different complex phase on
each of the them.
As in the Ising model, the position of the line defect can be moved by a local unitary
transformation that acts on single sites.
Thus, we have all the required elements to extract both scaling dimensions and conformal spins for
any defect arising from an internal symmetry.
This is illustrated in Appendix~\ref{app:potts3} by presenting results for the symmetry defects of
the 3-state Potts model.

Duality defects are in general more difficult to characterize, but a lattice representation is
known in several models (see Ref.~\onlinecite{chui_integrable_2001}).
Finding a unitary transformation that moves the duality defect ought to also be possible after a
case-by-case analysis.
Then, we would be able to use the tools described in this paper.

\subsection{Why tensor network renormalization?}
In this paper, we have employed a particular choice of coarse-graining transformation, namely, the
tensor network renormalization (TNR) scheme, to coarse-grain the tensor network representation of
the generalized transfer matrix $M_D$ and translation operator $T_D$ corresponding to a topological
defect $D$, in order to extract accurate estimates of the associated scaling dimensions and
conformal spins $\{\Delta_{\alpha}, s_{\alpha}\}_D$ (or just $\{\Delta_{\alpha}\}_{D}$ for a
generic conformal defect).
However, we could have used many other coarse-graining schemes.

Indeed, any coarse-graining transformation that accurately preserves the spectra of $M_D$ and $T_D$
will allow us to extract the conformal data.
Examples of such coarse-graining transformations include tensor renormalization
group~(TRG)~\cite{levin_tensor_2007}, higher-order tensor renormalization
group~(HOTRG)~\cite{xie_coarsegraining_2012}, tensor entanglement-filtering
renormalization~(TEFR)~\cite{gu_tensorentanglementfiltering_2009}, the second renormalization group
(SRG)~\cite{xie_second_2009}, and higher-order second renormalization group
(HOSRG)~\cite{xie_coarsegraining_2012}.

TRG is the simplest option, and already produces a very significant gain of accuracy with respect
to exact diagonalization, since much larger systems can be considered for an equivalent
computational cost, without introducing significant truncation errors, so that the estimates for
$\{\Delta_{\alpha}, s_{\alpha}\}_D$ are less affected by non-universal, finite-size corrections.

The accuracy of TRG can be further increased, for the same bond dimension and similar computational
cost, in a number of ways.
Several improved algorithms, such as SRG and HOSRG, are based on computing an environment that
accounts for the rest of the tensor network during the truncation step of the coarse-graining.
The use of a global environment has the important advantage that it leads to a better truncation of
bond indices, and thus to more accurate results, compared to an equivalent scheme that does not
employ the environment (for instance, SRG compared to TRG, or HOSRG compared to HOTRG).

However, in the context of studying defects, the use of a global environment also has a second,
less favorable implication.
Since in order to truncate a given bond index we use a cost function that is aware of the whole
tensor network, the resulting coarse-grained tensors will notice the presence of the defect even
when the defect is away from those tensors (recall that in a critical systems correlations decay as
a power-law with the distance).
Consider a tensor network for the partition function $Z_D$ with line defect $D$, where the line
defect is initially characterized by a column of tensors that is inserted into the tensor network
for the partition function $Z$ in the absence of the defect, as we have done in this work.
Then under a coarse-graining transformation such as SRG or HOSRG, the coarse-grained tensor network
will consist of a collection of different tensors that depend on their distance to the defect.
In other words, the representation of the defect will spread throughout the whole tensor network,
instead of remaining contained in a (single or double) column of tensors, as was the case in this
paper.
As a matter of fact, for topological defects this can be prevented through a careful analysis,
since in some sense these defects can be made invisible to neighboring tensors (since their
location can be changed through local unitary transformations).
For generic conformal defects, however, one needs to consider a mixed strategy where the global
environment is used in order to coarse-grain the partition function $Z$ in the absence of a defect,
as well as the defect tensors in $Z_D$, but is not used in order to coarse-grain the rest of
tensors in $Z_D$, which are recycled from $Z$.

A direct comparison of computational resources required by several of these approaches has shown
that TNR provides significantly more accurate estimates than the above methods for
$\{\Delta_{\alpha}, s_{\alpha}\}$ from the transfer matrix $M$ and translation operator $T$ in the
absence of a defect, see e.g.\ Ref.~\onlinecite{evenbly_tensor_2015-2}.
The ultimate reason is that TNR, thanks to the use of disentanglers to remove short-range
correlations/entanglement in the partition function, provides a much more accurate description of
the partition function when using tensors with the same bond dimension.
We expect the same to be true for the estimation of $\{\Delta_{\alpha}, s_{\alpha}\}_D$ from the
transfer matrix $M_D$ and translation operator $T_D$ in the presence of a topological defect $D$
(or just $\{\Delta_{\alpha}\}_D$ from $M_D$ in the presence of a generic conformal defect $D$).

It is worth emphasizing, however, that a simpler algorithm such as TRG, which is easier to code,
already provides much better accuracy than exact diagonalization.

\subsection{Alternative approach to extracting conformal data from a critical lattice model}
In this paper, we have extracted the scaling dimensions and conformal spins $\{\Delta_{\alpha},
s_{\alpha}\}_D$ associated to a topological conformal defect $D$ by diagonalizing the transfer
matrix $M_D$ and translation operator $T_D$ of a corresponding partition function $Z_D$ on the
lattice.
This approach is based on generalizing, to the case of a line defect, the observation of
Ref.~\onlinecite{gu_tensorentanglementfiltering_2009} that the operator-state correspondence
of a CFT allows us to extract the scaling dimensions and conformal spins $\{\Delta_{\alpha},
s_{\alpha}\}$ of local scaling operators of the underlying CFT by diagonalizing the transfer matrix
$M$ and translation operator $T$ of the clean partition function $Z_D$ on the lattice.
Indeed, as discussed in Sec.~\ref{sec:ising_model}, the operator-state correspondence relates
the scaling operators $\phi_{\alpha}$ of the theory with the energy and momentum eigenvectors
$\ket{\alpha}$ of the Hamiltonian $H$ and momentum $P$ operators of the same theory (where the
transfer matrix $M$ and the translation operator $T$ can be thought of as the exponentials of $H$
and $P$, respectively), allowing the extraction of $\{\Delta_{\alpha}, s_{\alpha}\}$ directly from
(a properly normalized version of) the spectra of energies and momenta $\{E_{\alpha},
p_{\alpha}\}$.

An alternative, more direct way of extracting $\{\Delta_{\alpha}, s_{\alpha}\}$ from a lattice
system is also possible, by identifying a lattice version of the corresponding scaling operators
$\phi_{\alpha}$, and studying their transformation properties under changes of scale and rotations.
This alternative approach was recently made possible by the introduction of the tensor network
renormalization~(TNR)~\cite{evenbly_tensor_2015-2,evenbly_local_2015}.
The key of the approach is that, through the use of disentanglers that eliminate short-range
correlations / entanglement from the coarse-grained partition function $Z$, at criticality it is
possible to explicitly realize scale invariance: the tensor network before and after
coarse-graining is expressed in terms of the same critical fixed-point tensor.
As explained in Ref.~\onlinecite{evenbly_local_2015}, it is then possible to build a transfer
matrix $R$ \emph{in scale}, representing a lattice version of the dilation operator of the CFT,
whose eigenvectors correspond to a lattice version of the scaling operators $\phi_{\alpha}$, while
the eigenvalues are the exponential of the scaling dimensions $\Delta_{\alpha}$ (conformal spins
$s_{\alpha}$ are also extracted by analysis of two-point correlators).

This direct approach is computationally more challenging, since it requires ensuring that scale
invariance is explicitly realized during the coarse-graining, before building and diagonalizing the
scale transfer matrix $R$.
However, it also has some remarkable advantages.
On the one hand, it appears to provide even more accurate results for
$\{\Delta_\alpha, s_{\alpha}\}$ than the diagonalization of the space-time transfer matrix $M$
discussed in this paper, see Appendix in Ref.~\onlinecite{evenbly_local_2015}.
Even more important is the fact that, using the explicit lattice representation of the scaling
operators $\phi_{\alpha}$ obtained from the scale transfer matrix $R$, we can study the fusion of
two such operators into a third one, thus yielding the operator product expansion~(OPE)
coefficients of the CFT, which can not be obtained from the space-time transfer matrix $M$.
These possibilities extend to the presence of defects, as demonstrated in
Ref.~\onlinecite{evenbly_local_2015} for the symmetry defect $D_{\epsilon}$ of the Ising
model.

Finally, we also emphasize that in the context of quantum spin systems, conformal defects have
already been studied using MERA\@.
In that case, scale invariance was explicitly used to extract the scaling dimensions
$\Delta_{\alpha}$ attached to a conformal defect that represented an impurity, an open boundary or
an interface between two critical systems, see Refs.~\onlinecite{evenbly_boundary_2010}
and~\onlinecite{evenbly_algorithms_2014}.


\vspace{1cm}

\emph{Note added.}
A few weeks after this manuscript was posted on the arXiv, Aasen, Mong, and Fendley posted a
paper with independent, closely related work on topological defects of the classical
square-lattice Ising model~\cite{aasen_topological_2016}.
While the emphasis in our paper is on the use of tensor network methods,
Ref.~\onlinecite{aasen_topological_2016} is centered on constructing analytical models.
Thus, the two papers nicely complement each other.


\vspace{1cm}
We thank our anonymous referees for valuable feedback.
M.~Hauru thanks Andrew Tinits for helpful discussions on the numerical aspects of this work.
M.~Hauru is supported by an Ontario Trillium Scholarship.
G.~Vidal and G.~Evenbly acknowledge support by the Simons Foundation (Many Electron Collaboration).
G.~Vidal also acknowledges support by the John Templeton Foundation.
The authors also acknowledge support by Calcul Qu\'ebec and Compute Canada.
This research was supported in part by Perimeter Institute for Theoretical Physics.
Research at Perimeter Institute is supported by the Government of Canada through the Department of
Innovation, Science and Economic Development Canada and by the Province of Ontario through the
Ministry of Research, Innovation and Science.

\bibliography{defect_paper.bib}


\clearpage  

\appendix

\section{Tensor network renormalization}
\label{app:tnr}
This appendix gives a quick overview of the tensor network renormalization (TNR) scheme that we
use and shows how to adapt it to coarse-graining the topological defects $D_\epsilon$ and
$D_\sigma$ of the Ising model.
It also shows how the translation operators $\tro$, $\tro_{D_\epsilon}$ and $\tro_{D_\sigma}$ are
coarse-grained in the process.

\begin{figure}[htbp]
	\centering
	\includegraphics[width=1.0\linewidth]{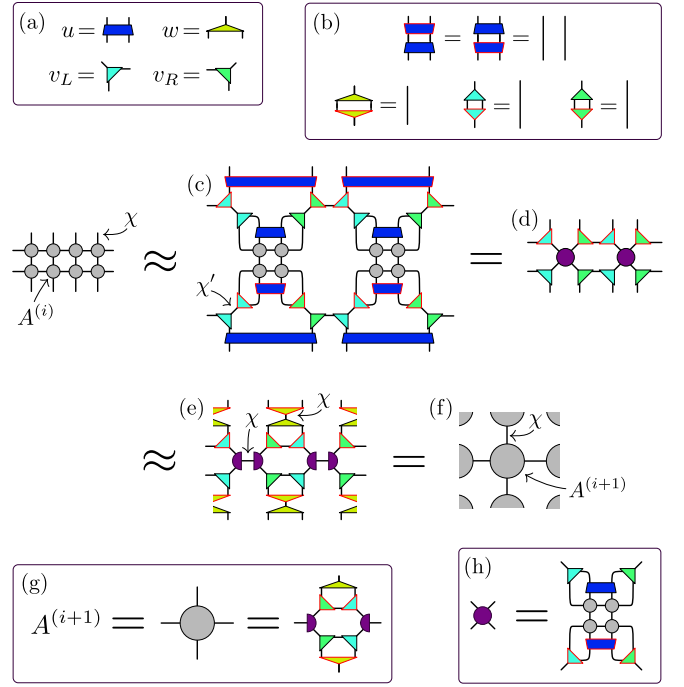}
	\caption{%
        An overview of the tensor network renormalization procedure that we use.
        The network is assumed to repeat as exactly similar in all directions.
        $\chi$ and $\chi'$ are the (maximum allowed) dimensions of the bonds they point at and all
        similar bonds in the network.
        The tensors with red borders are complex conjugates of the ones with the same shape but
        with black borders.
        (a) The tensors used in TNR\@.
        (b) The $u$ tensor is unitary whereas $v_L$, $v_R$ and $w$ are isometric.
        (c) Insertion of $u$'s, $v_L$'s and $v_R$'s as approximate partitions of unity.
        The fact that $v_L$ and $v_R$ are not unitary introduces a truncation error.
        $u$, $v_L$ and $v_R$ are optimized to minimize this error.
        (d) Some of the tensors have been contracted together to the one defined in (h).
        Because the network repeats itself in all directions the $u$'s and $u\dg$'s at the edges of
        the network have disappeared:
        they cancel with their counterparts in the next, similar blocks of tensors above and below
        the one shown.
        (e) Another approximate partition of unity has been inserted into the network as $ww\dg$
        and the purple tensor has been split into two using a truncated singular value
        decomposition, as is done in Levin \& Nave's tensor renormalization
        group~\cite{levin_tensor_2007}.
        (f) Tensors in (e) are contracted to form the new network of tensors $A^{(i+1)}$, defined
        in (g).
    }\label{fig:TNR}
\end{figure}

TNR is a coarse-graining transformation for tensor networks that is based on inserting approximate
partitions of unity into the network and optimizing them to minimize the truncation error.
It removes all short-range correlations during the coarse-graining and thus manages to realize a
proper renormalization group flow with the right fixed point structure.
Full details of the algorithm can be found in Refs.~\onlinecite{evenbly_tensor_2015-2}
and~\onlinecite{evenbly_algorithms_2015} and will not be repeated here.
As a summary and a reminder, Fig.~\ref{fig:TNR} shows the progression of a TNR coarse-graining
step.
It shows how a network $\znet_{\hs,\vs}\left(A^{(i)}\right)$ is coarse-grained into a new network
$\znet_{\frac{\hs}{2}, \frac{\vs}{2}}\left( A^{(i+1} \right)$, where each tensor $A^{(i+1)}$
corresponds to four $A^{(i)}$'s.
$\chi$ and $\chi'$ mark the dimensions of the bonds.
They control the accuracy of the approximations done in the algorithm but also the computational
cost:
Higher bond dimensions mean smaller truncation errors but require more memory and computation time.

\begin{figure}[htbp]
	\centering
	\includegraphics[width=0.5\linewidth]{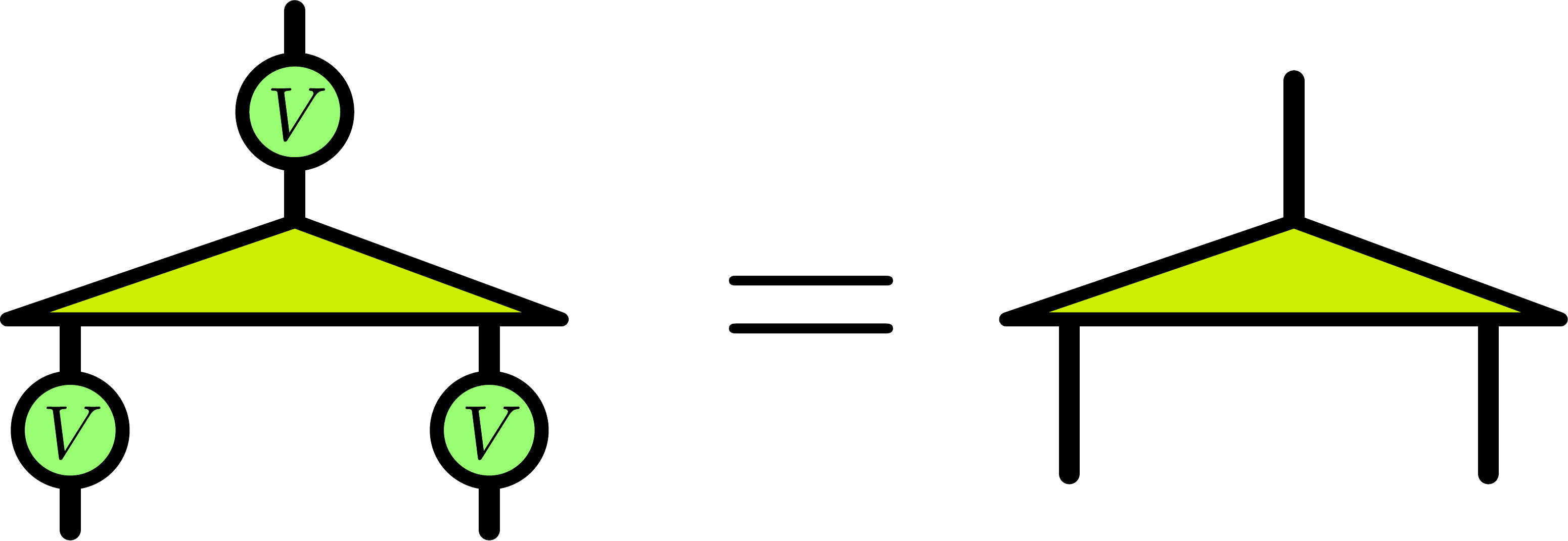}
	\caption{%
        The $\Integers_2$ invariance property of the $w$ tensor (in yellow).
        $V$ is the spin-flip matrix of the index it is on.
        Note that all the three $V$'s in this figure may in fact be different matrices, although
        our naming convention does not reflect this.
        Similar properties hold for all the tensors involved in coarse-graining the Ising model.
    }\label{fig:w_invar}
\end{figure}

All the tensors used in coarse-graining the Ising model are $\Integers_2$ invariant in
the following sense:
For every bond in the network there exists a unitary matrix representation $V$ of the non-trivial
element of $\Integers_2$, called the spin-flip matrix of that bond, such that for any tensor $t$ in
the network, multiplying all the legs of $t$ with the appropriate spin-flip matrix gives back $t$
again.
The spin-flip matrices for the different bonds are in general different, but we use $V$ or
$V^{(i)}$ to denote all of them: The bond they are on defines the correct $V$ unambiguously.
Figure~\ref{fig:w_invar} illustrates, as an example, the $\Integers_2$ invariance of the $w$
tensor.
Information on how such symmetry properties can be implemented and how to make computational use of
them can be found in Refs.~\onlinecite{singh_tensor_2010,singh_tensor_2011,singh_tensor_2012}.

\begin{figure}[htbp]
	\centering
	\includegraphics[width=0.9\linewidth]{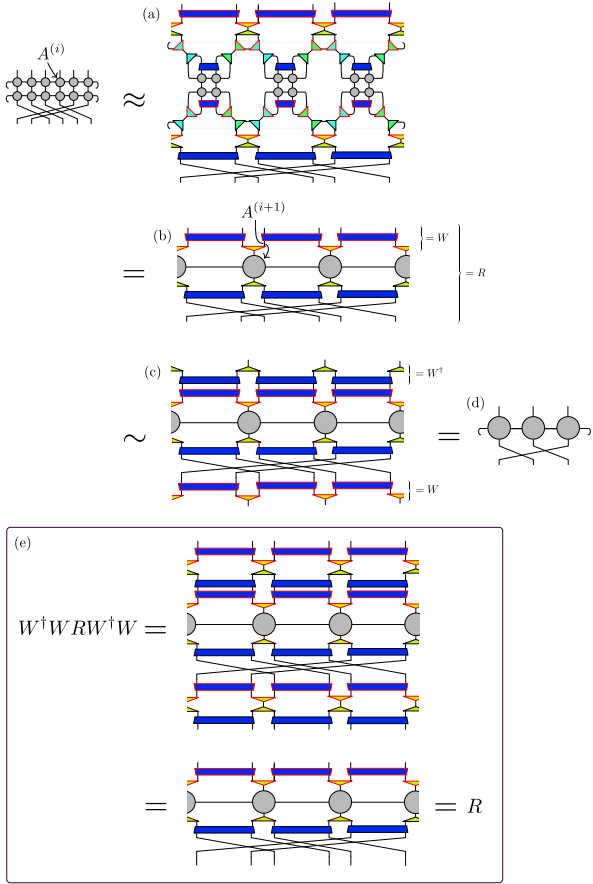}
	\caption{%
        One step of coarse-graining for the operator $\tro^{(i)} \cdot \trm^{(i)}$.
        The lattice translation by two sites gets coarse-grained into a translation by
        one site at the next scale.
        Periodic boundaries in the horizontal direction are assumed even when they are not
        explicitly shown.
        The tensors with red borders are complex conjugates of the ones with the same shape but
        with black borders.
        (a) The first step of the TNR algorithm has been applied to the tensors $A^{(i)}$.
        (b) A number of tensors are contracted to form $A^{(i+1)}$.
        We also define the operators $W$ and $R$.
        (c) The operator $R$ has been conjugated with $W$.
        $\sim$ denotes that the operators in (b) and (c) have the spectrum, or in other words that
        $W R W^\dagger$ has the same spectrum as $R$.
        This is true because, although $W$ is isometric and not unitary, it acts like a unitary on
        $M$, as shown in (e).
        (d) The unitaries and isometries cancel and we are left with the coarse-grained transfer
        matrix and translation operator.
    }\label{fig:cT_coarsing}
\end{figure}

If we coarse-grain a transfer matrix multiplied with a lattice translation by two sites, the
translation is coarse-grained into translation by one site at the next scale, as shown in
Fig.~\ref{fig:cT_coarsing}.
More generally, a translation by an even number of sites $n$ is coarse-grained into a
translation by $\frac{n}{2}$ sites at the next scale.
Because of this, the translation operator $\tro^{(\iters)}$ [Fig.~\ref{fig:cT_T_m}, back in
Sec.~\ref{sec:tensor_networks}] is a translation by $2^{\iters}$ sites at the original scale.
Thus the eigenvalues of $\tro^{(\iters)}$ are $\left(e^{\frac{2 \pi i}{2^\iters \hs_\iters}
s_\alpha} \right)^{2^\iters} = e^{\frac{2 \pi i}{\hs_\iters} s_\alpha}$ and the conformal spins are
determined modulo $\hs_\iters$.
To obtain the conformal spins with more possible values we can do an additional coarse-graining
step on $\tro^{(\iters)} \cdot \trm^{(\iters)}$, as explained in Appendix~\ref{app:coarse_momenta}.

\subsection{Coarse-graining the $D_\epsilon$ defect}

\begin{figure}[htbp]
	\centering
	\includegraphics[width=1.0\linewidth]{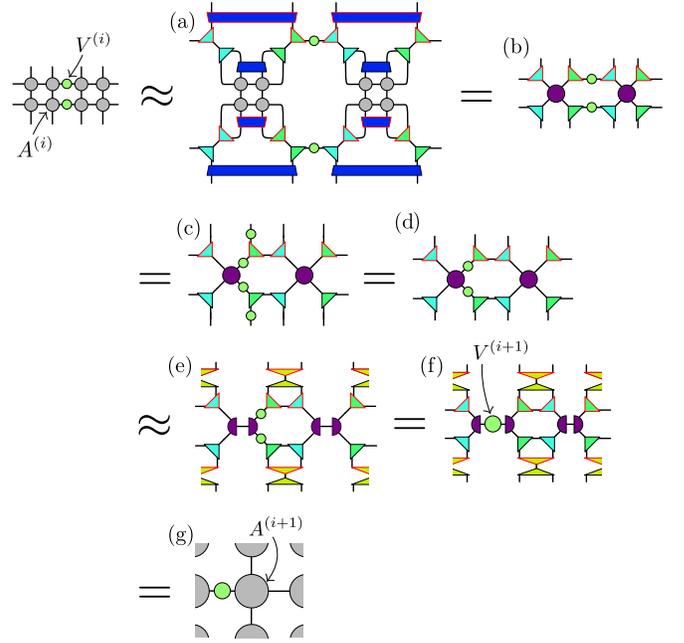}
	\caption{%
        Coarse-graining a system with a $D_\epsilon$ defect in it.
        By repeatedly using the $\Integers_2$ invariance of the different tensors and the fact that
        $\big( V^{(i)} \big)^2 = \unity$ we can take the string of spin-flip matrices $V^{(i)}$
        ``through'' the TNR procedure to the next scale.
        In the figure, the network is assumed to repeat as exactly similar in all directions.
        The tensors with red borders are complex conjugates of the ones with the same shape but
        with black borders.
        Steps (a), (b), (e) and (g) are the same steps as taken in the usual TNR algorithm [see
        Fig.~\ref{fig:TNR}].
        At steps (c) and (f) we have moved the spin-flip matrices to different legs using the
        $\Integers_2$ invariance of the tensors and the fact that $\left(V^{(i)}\right)^2 =
        \unity$.
        At step (d) the spin-flip matrices at the top and the bottom have canceled with similar
        matrices coming from the parts of the network above and below the one shown here.
    }\label{fig:TNR_epsilon}
\end{figure}

The TNR coarse-graining explained above can accommodate for a $D_\epsilon$ defect without any
changes.
As explained in Sec.~\ref{sec:D_epsilon}, a $D_\epsilon$ defect in a tensor network is a
realized by having spin-flip matrices on a string of bonds.
Figure~\ref{fig:TNR_epsilon} shows how, using the $\Integers_2$ invariance property of all the
tensors in the network, such a defect coarse-grains into a similar string of spin-flip matrices
at the next scale.
Hence, we can coarse-grain as if there was no defect, and at any scale $i$ we can insert a
$D_\epsilon$ defect into the system as a string of spin-flip matrices $V^{(i)}$, that are the
representations of the non-trivial element of $\Integers_2$ under which the matrix $A^{(i)}$ is
$\Integers_2$ invariant.

\begin{figure*}[!htbp]
	\centering
	\includegraphics[width=1.0\linewidth]{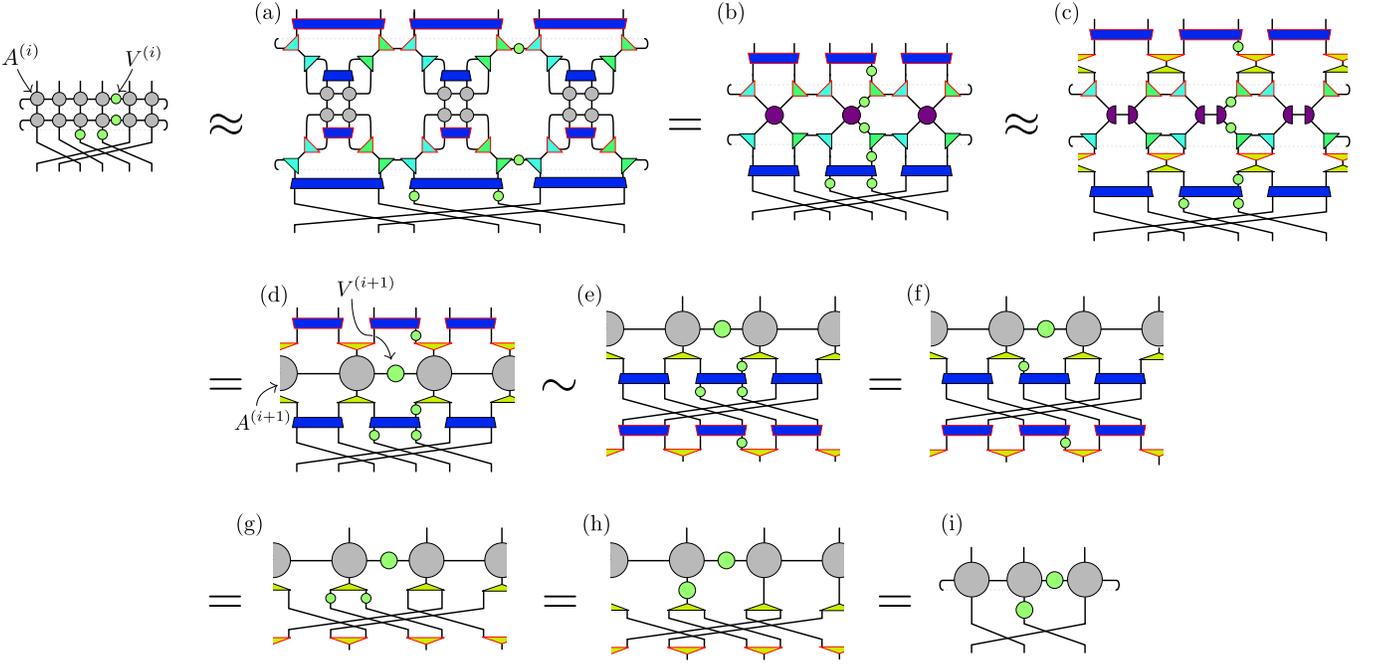}
	\caption{%
        One step of coarse-graining for the operator $\tro_{D_\epsilon}^{(i)} \cdot
        \trm_{D_\epsilon}^{(i)}$.
        The generalized translation by two sites gets coarse-grained into a generalized
        translation by one site at the next scale.
        Periodic boundaries in the horizontal direction are again assumed even when they are not
        explicitly shown.
        The tensors with red borders are complex conjugates of the ones with the same shape but
        with black borders.
        Steps (a-d) are a repetition of what was done in Fig.~\ref{fig:TNR_epsilon}, with some of
        the intermediate steps omitted.
        $\sim$ again denotes that the two operators in (d) and (e) have the same spectrum.
        That they indeed do have the same spectrum is not obvious, but can be shown
        with an argument exactly like the one used in Fig.~\ref{fig:cT_coarsing}(e).
        We have omitted this argument for the sake of brevity.
        In steps (f) and (h) we use the $\Integers_2$ invariance of $u$ and $w$ to move the
        spin-flip matrices in the network.
        In steps (g) and (i) the $u$ and $w$ tensors cancel with their complex conjugates because
        of their unitarity and isometricity.
    }\label{fig:cT_epsilon_coarsing}
\end{figure*}

In Sec.~\ref{sec:D_epsilon}, we introduced the generalized translation operator $\tro_{D_\epsilon}$
that is the proper notion of translation for a system with a $D_\epsilon$ defect.
Figure~\ref{fig:cT_epsilon_coarsing} shows how the generalized two-site translation operator
is coarse-grained into a generalized one-site translation at the next scale.
The procedure is the same as what we did with the usual lattice translation in
Fig.~\ref{fig:cT_coarsing}, except for the spin-flip matrices that get transferred to the next
scale.
Again, this generalizes to a generalized translation by an even number of sites $n$ that is
coarse-grained into a generalized translation by $\frac{n}{2}$ sites at the next scale.
Thus we know how to construct the generalized translation operator $\tro_{D_\epsilon}^{(i)}$ at any
scale with no additional numerical work needed.

\subsection{Coarse-graining the $D_\sigma$ defect}
\begin{figure}[!htpb]
	\centering
	\includegraphics[width=1.0\linewidth]{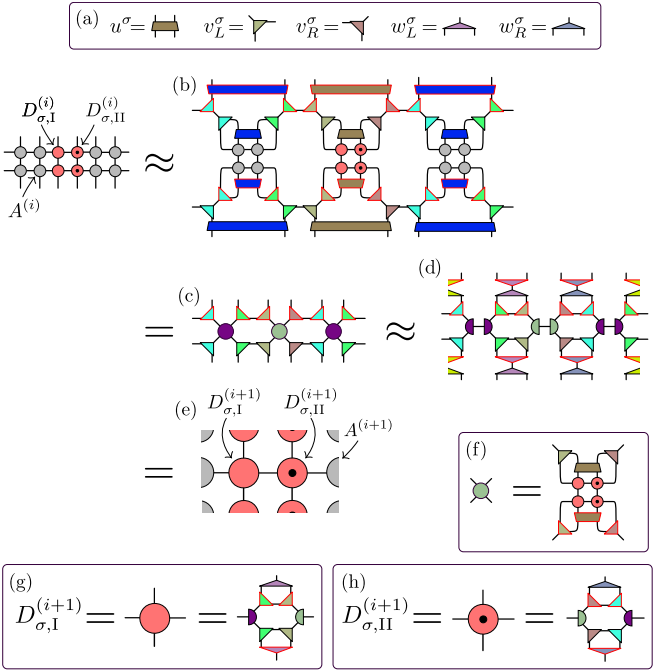}
	\caption{%
        The TNR procedure to coarse-grain a line defect of two tensors.
        The network is assumed to repeat as similar above and below the part shown, where as on the
        left and the right it is assumed to consist of tensors $A^{(i)}$.
        The tensors with red borders are complex conjugates of the ones with the same shape but
        with black borders.
        (a) The new tensors used in coarse-graining the defect.
        They have the same unitarity and isometricity properties as their counterparts in the usual
        scheme [see Fig.~\ref{fig:TNR}].
        (b -- d, f) The steps of the usual TNR algorithm are repeated for the new isometries and
        unitaries that coarse-grain the defect.  The isometries and unitaries are optimized to
        again minimize truncation errors, but for the environments that include the
        $D_{\sigma, \mathrm{I}}^{(i)}$ and $D_{\sigma, \mathrm{II}}^{(i)}$ tensors.
        (e) The different tensors are contracted together to form $D_{\sigma, \mathrm{I}}^{(i+1)}$
        and $D_{\sigma, \mathrm{II}}^{(i+1)}$, defined in (g) and (h), that represent the defect at
        next length scale.
    }\label{fig:TNR_sigma}
\end{figure}

Coarse-graining the symmetry defect $D_\epsilon$ is easy because of the $\Integers_2$ invariance of
the tensors in the TNR procedure.
The Kramers-Wannier symmetry of the critical Ising model is not similarly explicitly realized in
the individual tensors and does not help in coarse-graining the $D_\sigma$ defect.
Instead we treat the string of $D_\sigma$ tensors as we would treat any other string of impurity
tensors.
Coarse-graining such a string can be done using a variation of the usual TNR scheme.
The only change is that the unitaries and isometries that are in the vicinity of the $D_\sigma$
tensors need to be optimized for their respective environments~\cite{evenbly_algorithms_2015}.
The tensors elsewhere in the network are the same ones used in coarse-graining $D_\unity$.
This modified scheme is shown in Fig.~\ref{fig:TNR_sigma}.

In Fig.~\ref{fig:TNR_sigma}, we depict the $D_\sigma$ defect as a two-tensor-wide string.
This is because regardless of whether we start with a two-tensor or one-tensor-wide string it is
coarse-grained into a two-tensor-wide string at the next scale.\footnote{%
    This is related to the causal cone of the binary MERA that is two sites wide for certain
    operators~\cite{vidal_class_2008,evenbly_tensor_2015-1}.
}
For the first coarse-graining step we can simply choose $D_{\sigma, \mathrm{I}}^{(0)} = D_\sigma$
and $D_{\sigma, \mathrm{II}}^{(0)} = A$ and at all later scales the defect will consist of
two-tensor-wide string.

\begin{figure*}[!htbp]
	\centering
	\includegraphics[width=1.0\linewidth]{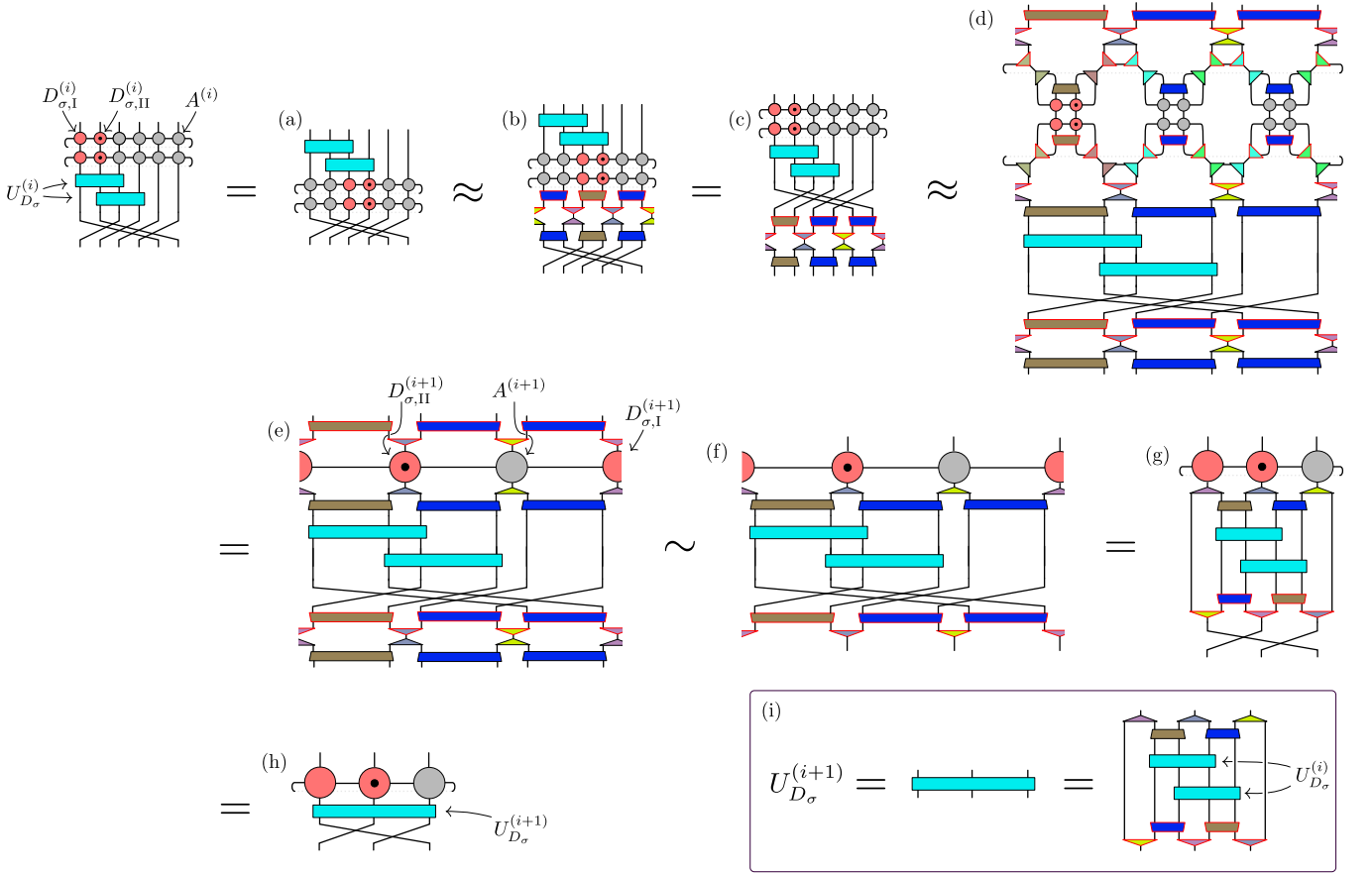}
	\caption{%
        One step of coarse-graining for the operator $\tro_{D_\sigma}^{(i)} \cdot
        \trm_{D_\sigma}^{(i)}$.
        The generalized translation by two sites gets coarse-grained into a generalized
        translation by one site at the next scale.
        Periodic boundaries in the horizontal direction are assumed even when they are not
        explicitly shown.
        All the tensors used in these diagrams are familiar from Fig.~\ref{fig:TNR_sigma} and
        Fig.~\ref{fig:TNR} and we have skipped a few intermediate steps that are just a
        repetition of what is done in those figures.
        (a) Because $U_{D_\sigma}^{(i)}$ moves the $D_\sigma$ defect we can move the
        unitaries to the other side of the transfer matrix as shown.
        (b) Unitaries and isometries have been inserted into the network.
        This is like the insertion in the first step of the TNR procedure in
        Fig.~\ref{fig:TNR_sigma}, but without the $v$ isometries and only on one side of the
        transfer matrix.
        Thus the truncation error introduced by this insertion is at most the truncation error
        introduced in the TNR step, because less projections are performed.
        (c) Using again the property that $U_{D_\sigma}^{(i)}$ moves the defect allows us to move
        the unitaries back to where they were originally.
        (d) The first step of the TNR procedure in Fig.~\ref{fig:TNR_sigma} has been applied
        to the tensors of the transfer matrix.
        (e) Some of the tensors have been contracted to form $A^{(i+1)}$,
        $D_{\sigma, \mathrm{I}}^{(i+1)}$ and $D_{\sigma,\mathrm{II}}^{(i+1)}$, as in
        Fig.~\ref{fig:TNR_epsilon}(e).
        Again, $\sim$ denotes that the operators in (e) and (f) have the same spectrum.
        This can be shown with an argument exactly analogous to the argument in
        Fig.~\ref{fig:cT_coarsing}(e).
        (g) Pair of unitaries $u u^\dagger$ have canceled and the graph of the network has been
        reorganized to make it easier to read.
        (h) The isometries and unitaries have been contracted to form $U_{D_\sigma}^{(i+1)}$,
        defined in (i).
        Note that the definition of $U_{D_\sigma}^{(i+1)}$ is nothing but a MERA ascending
        superoperator acting on two $U_{D_\sigma}^{(i)}$'s.
    }\label{fig:cT_sigma_coarsing}
\end{figure*}

The generalized translation operator $\tro_{D_\sigma}^{(i)}$ is coarse-grained as in
Fig.~\ref{fig:cT_sigma_coarsing}.
The principle is the same as for $\tro^{(i)}$ and $\tro_{D_\epsilon}^{(i)}$, but implementation is
not quite as simple because the defect-moving unitary $U_{D_\sigma}^{(i)}$ needs to be contracted
with what is essentially a MERA ascending superoperator~\cite{evenbly_algorithms_2009} to get the
$U_{D_\sigma}^{(i+1)}$ at the next scale.
The computational time to perform this coarse-graining of
$U_{D_\sigma}^{(i)}$ scales asymptotically as $\cO(\chi^{10})$, whereas all the other steps in the
TNR procedure (with or without a defect) are at most $\cO(\chi^7)$~\cite{evenbly_algorithms_2015},
making this step the computational bottleneck.

The argument in Fig.~\ref{fig:cT_sigma_coarsing} can be generalized to show that a
generalized translation by an even number of sites $n$ is coarse-grained to a generalized
translation by $\frac{n}{2}$ sites at the next scale, with the unitary $U^{(i+1)}$ being the one
defined in Fig.~\ref{fig:cT_sigma_coarsing}(i).
Analogously to how a one-tensor string of $D_\sigma$'s becomes a two-tensor string under
coarse-graining, the two-site operator $U_{D_\sigma}$ coarse-grains into a three-site operator
$U_{D_\sigma}^{(1)}$.
Because of this in Fig.~\ref{fig:cT_sigma_coarsing} $U_{D_\sigma}^{(i)}$ is shown as a
three-site operator that is then coarse-grained into a three-site operator $U_{D_\sigma}^{(i+1)}$.

\section{Increasing the period of the conformal spins}
\label{app:coarse_momenta}
The method we present in Sec.~\ref{sec:tensor_networks} and Appendix~\ref{app:tnr} yields
conformal spins modulo $\hs$, where $\hs$ is the number of tensors in the transfer matrix.
This appendix describes an additional coarse-graining step that can be taken to increase the period
to $2\hs$.
We explain how to do this for a model with no defect, but the method can easily be adapted
to accommodate for the presence of a defect.

\begin{figure}[htbp!]
	\centering
	\includegraphics[width=1.0\linewidth]{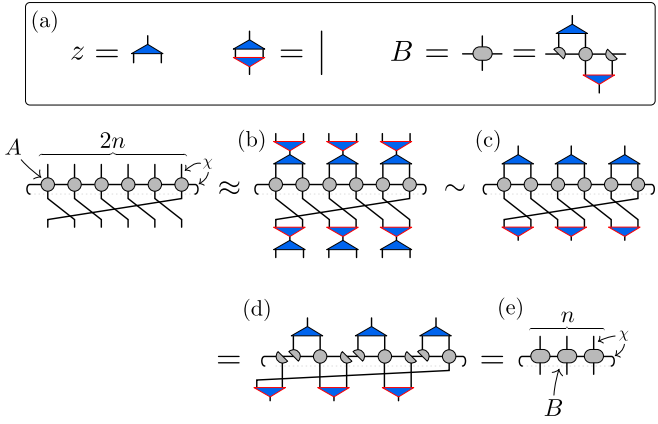}
	\caption{%
        The additional coarse-graining step to increase the period of the conformal spins.
        The tensors with red borders are complex conjugates of the ones with the same shape but
        with black borders.
        (a) $z$ is a coarse-graining isometry, with all of its legs having dimension $\chi$.
        $B$ is the tensor obtained at the end of the coarse-graining.
        (b) $zz\dg$ pairs have been inserted in the network as approximate partitions of unity.
        They can be optimized the same way the TNR isometries are optimized, see
        Ref.~\onlinecite{evenbly_algorithms_2015}.
        (c) $\sim$ denotes that the operators in (b) and (c) have the same spectrum.
        This can be shown with an argument exactly analogous to the argument in
        Fig.~\ref{fig:cT_coarsing}(e).
        (d) The $A$ tensors have been split into two using a truncated singular value
        decomposition, as in Levin \& Nave's tensor renormalization group~\cite{levin_tensor_2007}.
        (e) Some of the tensors have been contracted together into $B$, defined in (a).
        The bond dimension of $B$ is the same as the bond dimension of the original tensor $A$.
    }\label{fig:coarse_momenta}
\end{figure}

To get conformal spins with a period of $2\hs$ we want to diagonalize the operator $\tro \cdot
\trm$ of $2\hs$ tensors.
Assume, however, that we can numerically only afford to diagonalize a transfer matrix of $\hs$
tensors, i.e.\ a matrix with dimensions $\chi^\hs \times \chi^\hs$.
We can use an additional layer of isometries to coarse-grain $\tro \cdot \trm$ down to $\hs$
tensors (without changing the bond dimension).
The scheme we use to do this is shown in Fig.~\ref{fig:coarse_momenta}.

This additional coarse-graining does not utilize unitaries the same way TNR does and thus does not
remove all short-range correlations.
Because of this it causes a truncation error that is relatively large.
In practice we find that doing a single such coarse-graining does not qualitatively affect the
results:
Some accuracy is lost in the scaling dimensions but the conformal spins come out correctly to high
accuracy.
This lets us match the results obtained with and without the additional coarse-graining and pick
the best of both worlds:
Use the additional coarse-graining to get the values for the conformal spins and for the
scaling dimensions use values obtained without the additional coarse-graining.

\section{Results for the 3-state Potts model}
\label{app:potts3}
In this appendix, we present results for the critical 3-state Potts model on the square lattice.
It is defined by the partition function
\begin{IEEEeqnarray}{c}
    \label{eq:Z_potts3}
    Z = \sum_{\{s\}} \prod_{\langle i,j \rangle} e^{\beta \delta_{s_i, s_j}}.
\end{IEEEeqnarray}
$s_i$ are the local degrees of freedom which take three values, say $0$, $1$ and $2$, and
$\delta_{s_i, s_j}$ is a Kronecker $\delta$ that is $1$ only if neighboring degrees of freedom are
in the same state and $0$ otherwise.
The sum and the product are over all configurations and all nearest-neighbor pairs, respectively.
The model has a critical point at $\beta = \log(\sqrt{3} + 1)$ and the continuum limit at
criticality is described by a $c=\frac{4}{5}$ CFT\@.

If we permute the three different values of $s_i$ with the same permutation at every site the
Boltzmann weights remain unchanged.
Thus the 3-state Potts model has a global internal $S_3$ symmetry, $S_3$ being the symmetric group
for three elements.
For every element of $S_3$, of which there are $6$, there is topological defect for the CFT\@.
We will concentrate here on three elements $\unity$, $a$ and $a^2$ that form the $\Integers_3$
subgroup of $S_3$.
This is because $\Integers_3$ is Abelian, and manipulating symmetry preserving tensors is
computationally much less intensive for Abelian symmetries than for non-Abelian symmetries.

We call the defects related to these three group elements $D_\unity$, $D_a$ and $D_{a^2}$ and the
corresponding twisted partition functions $Z_{D_{\unity}}$, $Z_{D_a}$ and $Z_{D_{a^2}}$.
As discussed in Sec.~\ref{sec:discussion}, the tensor network methods used for the $D_\epsilon$
defect of the Ising model can be generalized to these symmetry defects of the Potts model.
Hence we can extract the conformal spins and scaling dimensions of the scaling operators
present in the partition functions $Z_{D_{\unity}}$, $Z_{D_a}$ and $Z_{D_{a^2}}$.

\begin{figure*}[!htbp]
	\centering

    \includegraphics[width=1.0\linewidth]{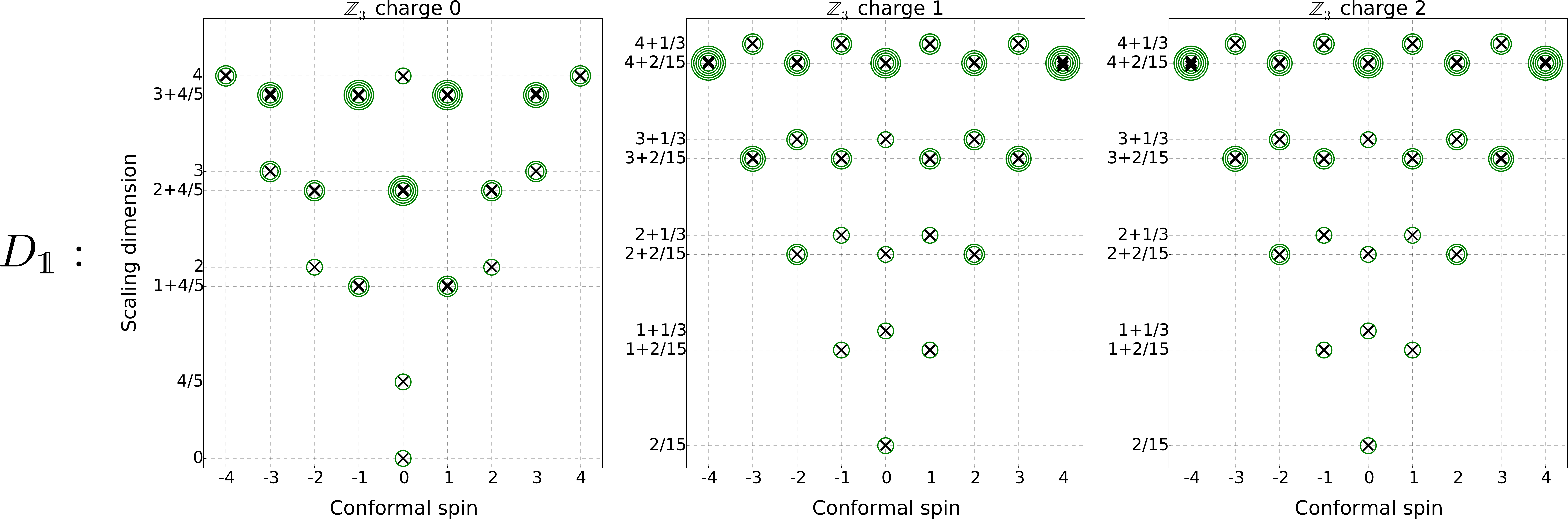}

    \vspace{20pt}
    \includegraphics[width=1.0\linewidth]{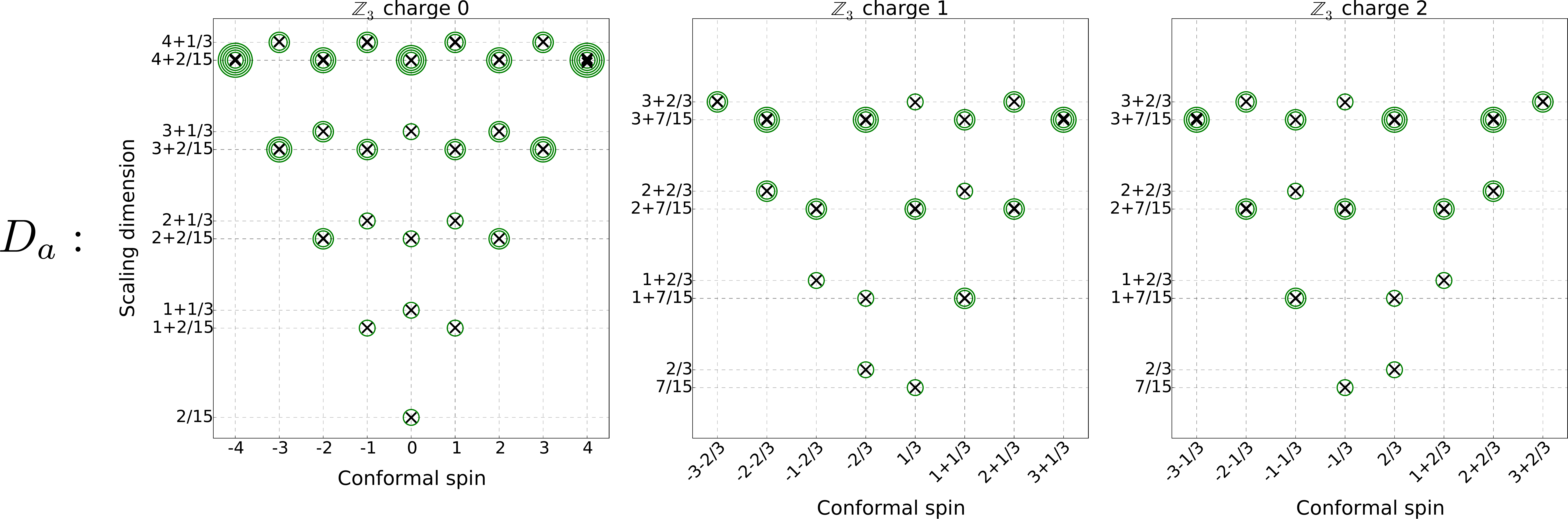}

    \vspace{20pt}
    \includegraphics[width=1.0\linewidth]{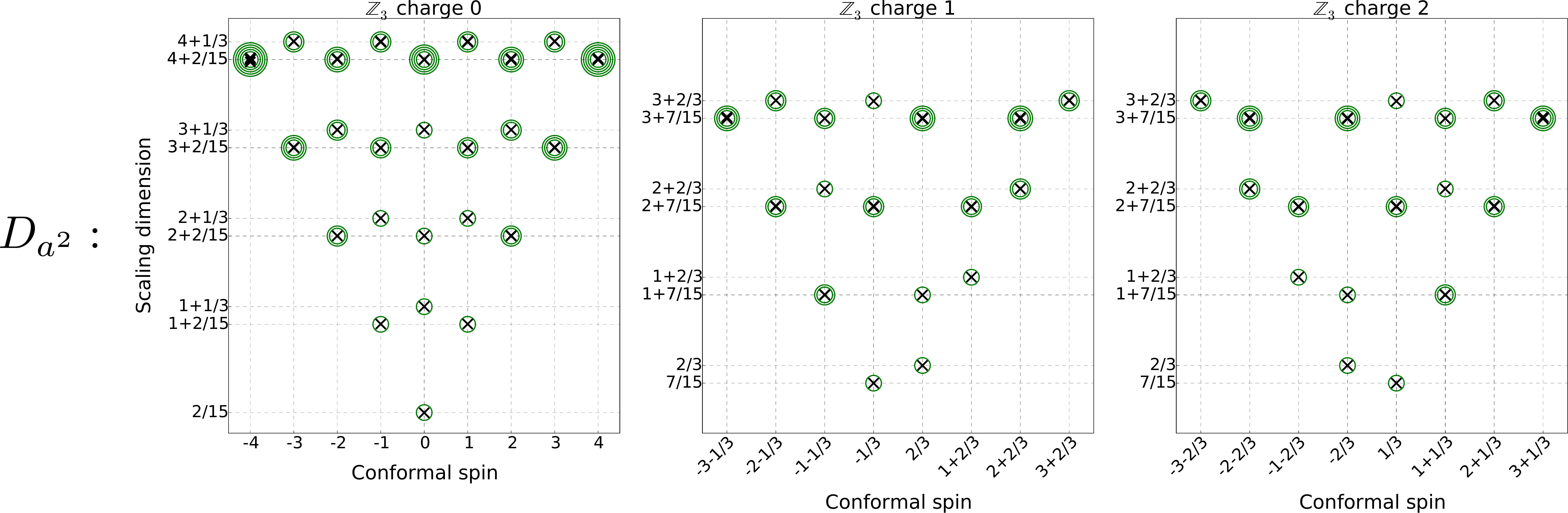}

	\caption{%
        The scaling dimensions and conformal spins of the first scaling operators of the
        square lattice 3-state Potts model with various defects as obtained with TNR\@.
        Every row of three plots includes the results for one of the defects we study:
        at the top the trivial defect $D_\unity$, in the middle the defect $D_a$ and at the bottom
        the defect $D_{a^2}$.
        Crosses mark the numerical values, circles mark the exact values.
        Several concentric circles denote the degeneracy $N_\alpha$ of that $(\Delta_\alpha,
        s_\alpha)$ pair.
        Although it is not clear from the figure, these degeneracies also come out correctly.
    }\label{fig:potts3_results}
\end{figure*}

The numerical results we obtain are shown in Fig.~\ref{fig:potts3_results}.
Because of the $\Integers_3$ symmetry each scaling operator comes with a $\Integers_3$ charge, and
we have organized the results by this charge.
The agreement with the exact values is again excellent.
To obtain these results we coarse-grained and diagonalized a transfer matrix of $2^6 \times 4
\times 2^6$ $A^{(0)}$ tensors ($\approx 32\,000$ of the original degrees of freedom).
The bond dimensions used in coarse-graining were $\chi=30$ and $\chi'=15$.
For the conformal spins, a slightly larger system was used, as explained in
Appendix~\ref{app:coarse_momenta}.

\section{Number of degrees of freedom in $Z_{D_{\sigma}}$}
\label{app:Z_sigma_counting}
In this appendix, we explain the correct way to count the degrees of freedom in a system with a
$D_\sigma$ defect.
This affects how we normalize the transfer matrix to extract conformal data.

We build the transfer matrix $\trm_{D_\sigma}^{(\iters)}$ from the tensors $A^{(\iters)}$,
$D_{\sigma,\mathrm{I}}^{(\iters)}$, and $D_{\sigma,\mathrm{II}}^{(\iters)}$ and diagonalize it.
We use the dependence of $\trm_{D_\sigma}^{(\iters)}$ on the system size to determine the free
energy term $\trms\hs f$ in the spectrum, as explained in Sec.~\ref{sec:ising_model}, and then
normalize this term away.
$\trms\hs$ is the number of spins included in the transfer matrix.
For the usual Ising model network $\znet_{\hs,\vs}(A)$ every bond corresponds to a spin and there
are twice as many spins as there are $A^{(0)}$ tensors.
However, the bonds on the right side of the $D_\sigma$ tensor are special:
they correspond to only half a spin.
This needs to taken into account when determining $f$.

That there is half a spin missing can be seen from the way the fusion rules manifest in the lattice
models or straight from the explicit form of the $D_\sigma$ tensor.
Perhaps the clearest way is to use the Jordan-Wigner transformation to map the quantum spin chain
of $\hs$ spins into a chain of $2\hs$ Majorana fermions.
There, the $D_\sigma$ defect is realized as one Majorana mode missing from the chain.

\end{document}